\documentclass[]{aa}
\usepackage[varg]{txfonts}
\usepackage{rotating}
\usepackage{natbib}
\usepackage{ae,aecompl}
\usepackage{graphicx}   
\usepackage{amsmath}    
\usepackage{amssymb}    
\usepackage{color,verbatim,url}
\usepackage{tabularx}
\usepackage{booktabs}
\usepackage{pdflscape}
\usepackage{mathtools}

\usepackage{hyperref}

\usepackage{breqn}

\definecolor{pink}{rgb}{0.858, 0.188, 0.478}
\definecolor{purple}{RGB}{76, 0,153}
\DeclarePairedDelimiter\set\{\}

\newcommand{\ie}{{\rm i.e.}}
\newcommand{\eg}{{\rm e.g.}}

\newcommand{\Nz}{{${\rm N}(z)$}}

\newcommand{\utok}{{ugriZY\!J\!H\!K_{\rm s}}}

\newcommand{\LCDM}{{$\Lambda{\rm CDM}$~}}

\begin{document}

\title{Photometric Redshift Calibration with Self Organising Maps}
\titlerunning{SOM Photometric Redshift Calibration}

\author{ Angus H. Wright\inst{1} 
\and Hendrik Hildebrandt\inst{1} 
\and Jan Luca van den Busch\inst{1} 
\and Catherine Heymans\inst{1,2} }
\authorrunning{Wright, Hildebrandt, v.d. Busch \& Heymans}
\institute{
Ruhr-Universit\"at Bochum, Astronomisches Institut, German Centre for Cosmological Lensing (GCCL), Universit\"atsstr.
150, 44801 Bochum, Germany. \\\email{awright@astro.rub.de} \and
Institute for Astronomy, University of Edinburgh, Royal Observatory, Blackford Hill, Edinburgh EH9 3HJ, UK.}

\date{Released 12/12/2121}

\graphicspath{{./figures/}}

\abstract{ 
Accurate photometric redshift calibration is central to the robustness of all cosmology constraints from cosmic shear
surveys. Analyses of the Kilo-Degree Survey, KiDS, re-weighted training samples from all overlapping spectroscopic
surveys to provide a direct redshift calibration. Using self-organising maps (SOMs) we demonstrate that this
spectroscopic compilation is sufficiently complete for KiDS, representing $99\%$ of the effective 2D cosmic shear
sample.  We use the SOM to define a $100\%$ represented  `gold’ cosmic shear sample, per tomographic bin.  Using mock
simulations of KiDS and the spectroscopic training set, we estimate the uncertainty on the SOM redshift
calibration, and find that photometric noise, sample variance, and spectroscopic selection effects (including redshift
and magnitude incompleteness) induce a combined maximal scatter on the bias of the redshift distribution
reconstruction ($\Delta \langle z \rangle=\langle z \rangle_{\rm est}-\langle z \rangle_{\rm true}$) of $\sigma_{\Delta
\langle z \rangle} \leq 0.006$ in all tomographic bins. Photometric noise and spectroscopic selection effects contribute
equally to the observed scatter.  We show that the SOM calibration is unbiased in the cases of noiseless photometry and
perfectly representative spectroscopic datasets, as expected from theory. The inclusion of both photometric noise and
spectroscopic selection effects in our mock data introduces a maximal bias of $\Delta \langle z \rangle =0.013\pm0.006$, 
 or $\Delta \langle z \rangle \leq 0.025$ at $97.5\%$ confidence, once quality flags have been applied to the SOM. 
The method presented here represents a significant improvement over the
previously adopted direct redshift calibration implementation for KiDS, owing to its diagnostic and quality assurance
capabilities. The implementation of this method in future cosmic shear studies will allow better diagnosis, examination,
and mitigation of systematic biases in photometric redshift calibration. 
}


\maketitle

\section{Introduction}
\label{sec: intro}
Comparisons between cosmological parameters from tomographic cosmic shear measurements
\citep[\eg~][]{hildebrandt/etal:2018,troxel/etal:2018,hikage/etal:2018} and the cosmic microwave background \citep[CMB;
][]{planck/cosmo:2018} reveal some tension between the amount and clustering strength of (predominantly dark) matter.
This is typically parameterised as ${S}_8=\sigma_8\sqrt{\Omega_m/0.3}$, where $\sigma_8$ is related to the
clustering amplitude of the dark matter power spectrum and $\Omega_m$ describes the overall energy density of matter.
Results from the recent Planck CMB measurements \citep{planck/cosmo:2018} suggest that the value of ${S}_8$ observed
at $z\sim1100$ is discrepant from that observed at low redshift by up to $3.2\sigma$ \citep[see the combined analyses of
cosmic shear surveys in][]{asgari/etal:2019}. 

This result presents the tantalising possibility that the highly successful \LCDM paradigm does not perfectly describe
the true nature of the universe \citep[see, \eg,][]{joudaki/etal:2017b}.  However naturally such a claim requires
significant evidence. Observations performed by different surveys within the tomographic cosmic shear community agree to
better than $1\sigma$ \citep[see, \eg, ][]{hildebrandt/etal:2018,joudaki/etal:2019}, with the results from both the Dark
Energy Survey \citep[DES; ][]{troxel/etal:2018} collaboration and the HyperSuprime Camera \citep[HSC;
][]{hikage/etal:2018} surveys finding no significant tension with respect to Planck, compared to the {mild} $2.3\sigma$ tension
reported by the Kilo Degree Survey \citep[KiDS; ][]{hildebrandt/etal:2018}. This begs the question as to whether the
reported CMB-cosmic shear tension is physical, or is simply reflecting systematic bias in the analysis methodologies of
one or more of these weak-lensing surveys. 

In an effort to explore the possible systematic biases within weak lensing analyses, members of both the DES and
KiDS collaborations have performed their own reanalysis of data from one-another's surveys.  \cite{troxel/etal:2018b}
utilise the DES analysis method on KiDS data and find a revised value of ${S}_8$ that is in closer agreement to
the results found by the fiducial DES analysis \citep{troxel/etal:2018}. More recently, \cite{joudaki/etal:2019} and
\cite{asgari/etal:2019} both performed a reanalysis of the DES data using the KiDS analysis methodology and found the
converse to be true; DES data was now in greater agreement with the fiducial results from KiDS
\citep{hildebrandt/etal:2018}. 

This difference in result as a function of methodology suggests a fundamental difference, possibly from unrecognised
systematic bias, in one or both of these analyses. \cite{hildebrandt/etal:2018} explored the influence of various model
and analytical choices on their cosmological constraints. In this analysis, they conclude that the only modification
that can be made to their analysis method which causes a decrease in the observed tension with the CMB
results from Planck is to utilise a different method of redshift calibration. Indeed, the approach to redshift
calibration is the most fundamental difference between the methodologies of the DES, HSC, and KiDS
collaborations, and therefore requires the most attention. 

The role of redshift calibration in cosmic shear tomography is pivotal. This is because the signal measured is
directly related to the strength of the gravitational distortion observed over redshift. If we estimate the true
redshift distribution of all of our sources to be systematically lower than they are in reality, then we incorrectly 
attribute the observed gravitational distortions as originating from an overall denser, more highly clustered
gravitational landscape than exists in reality. {\cite{hildebrandt/etal:2016} simulated the influence of redshift
uncertainties on cosmic shear analyses found that their cosmological estimates
were insensitive to redshift calibration biases of $\Delta \langle z \rangle \le 0.04$. However, with increasing amounts
of survey data and ever decreasing statistical uncertainties, biases at this level will be increasingly the dominant
source of error in future analyses. }

As a result of its importance, redshift calibration has been explored within (in particular) cosmic shear tomography for
many years. Three different redshift calibration methods are most prevalent in the literature.  These are estimation
via: cross correlation \citep[see, \eg, ][]{schneider/etal:2006,newman:2008,mcquinn/etal:2013,morrison/etal:2017};
stacking of individual redshift probability distributions
\citep{hildebrandt/etal:2012,hoyle/etal:2018,tanaka/etal:2018}; and direct calibration incorporating
spectroscopic redshift training samples, first presented by \cite{lima/etal:2008}, and implemented previously using
k-nearest-neighbour methods \citep[$k$NN; ][]{hildebrandt/etal:2016,hildebrandt/etal:2018} and unsupervised machine
learning \citep{buchs/etal:2019}. 

In this work we develop {and test} a new implementation of direct redshift calibration, also utilising unsupervised machine
learning methods. {We use this new method to measure the spectroscopic representation of KV450 photometric (cosmic shear) sources, and 
subsequently explore, via a suite of simulations, how sample variance, photometric noise, and spectroscopic
incompleteness influence photometric representation. We then use the simulations to estimate the influence of sample
variance, photometric noise, and spectroscopic incompleteness on the redshift reconstruction bias present in the method. }

The work presented in this paper is structured as follows. 
In Sect. \ref{sec: som method} we describe the theory behind {direct redshift calibration, and present the
theory behind our implementation of the same.} 
In Sects. \ref{sec: data} and \ref{sec: simulations} we describe the datasets and simulations utilized in this work. 
In Sect. \ref{sec: results} we present the main results of this work in four subsections. 
In Sect. \ref{sec: results I kv450} we estimate the representation of the KV450 photometric data
{ using currently available spectroscopic compilations}.  
In Sect. \ref{sec: results I mice2} we estimate the influence that systematic effects such as {sample variance,
photometric noise, and spectroscopic incompleteness} have on these estimates of representation. 
In Sect. \ref{sec: results II} we test the ability of our new implementation to calibrate redshift distributions, and
explore the influence of a variety of systematic effects on these reconstructed redshift distributions. 
{ We further compare our new method to previous implementations of the direct calibration. }
In Sect. \ref{sec: results III}, we present an additional set of calibrated redshift distributions for the
KiDS+VIKING-450 dataset, and propose an updated analysis methodology for the next iteration of KiDS cosmic shear
analyses. 
In Sect. \ref{sec: summary} we present a summary of our results, and our concluding remarks. 

This work also contains a number of technical Appendices (\ref{sec: som implementation}, \ref{sec: Clustering Results},
\ref{sec: spec training}, \ref{sec: SOM Bias}) which are relevant to how we implement our new direct photometric
redshift calibration. 

\section{Direct calibration with SOMs}\label{sec: som method}
The direct redshift calibration method was first proposed by \cite{lima/etal:2008}.  The method involves matching two
datasets, one with wide-field shear observations and one with deep spectroscopic observations, in high dimensional
multiband magnitude space. In the original description, and in previous cosmic shear analyses within KiDS
\citep{hildebrandt/etal:2017,hildebrandt/etal:2018}, this has been implemented using k-nearest-neighbour ($k$NN)
methods. {In this work, we present an updated direct calibration implementation using self organising maps
\citep{kohonen:1982}. 
The formulae describing the reweighting method are therefore identical to those presented by
\cite{lima/etal:2008,hildebrandt/etal:2017,hildebrandt/etal:2018}. Nonetheless, we reproduce them here for posterity.
We also briefly describe the previous KiDS direct calibration implementation, and subsequently present our updated
implementation utilising unsupervised machine learning methods.

In the recalibration method of \cite{lima/etal:2008}, the redshift distribution of an arbitrary set of photometric data,
$P$, is estimated via a given a set of spectroscopic data, $S$. To do this one first associates the photometric and
spectroscopic data in a way that maximises the spectroscopic redshift information. Typically this association involves
matching the two sets using colours/magnitudes $\mathbf{c}$, thereby creating $m\leq|S|$ associations\footnote{The
associations are defined with respect to the spectroscopic data, meaning that the maximal number of associations is
equal to the cardinality of set $S$. In the $k$NN method of association $m=|S|$ by definition, because the association
is performed by searching independently around each element of the spectroscopic set $s\in S$. In the SOM implementation
$m$ is equal to the number of SOM clusters containing spectroscopic data (see Appendix \ref{sec: Clustering Results}).} between
$S$ and $P$. Each of the $i\leq m$ associations produces a (possibly improper) subset of $S$ and $P$, which we define as
$\hat{S}_i = \set{S | \mathbf{c}_i}$ and $\hat{P}_i = \set{ P | \mathbf{c}_i}$ respectively.  

The goal of the redshift reconstruction procedure is to estimate the redshift distribution of the photometric data, 
$p\left(z|P\right)$ using a weighted combination of spectroscopic associations: 
\begin{equation}\label{eqn: template} 
p_{\rm S}^{\rm w}\left(z\right) = \sum_{i=1}^m\frac{w_i p\left(z|\hat{S}_i\right)}{\sum_{j=1}^m w_j},
\end{equation}
where $p\left(z|\hat{S}_i\right)$ is the spectroscopic redshift distribution of the $i^{\rm th}$
spectroscopic-to-photometric association, and $w_i$ is the reconstruction weight which maps the density of spectroscopic
sources to the photometric data. 
In the case of unweighted photometric data, the weights required to reconstruct the photometric redshift distribution,
$p\left(z|P\right)$, from the spectroscopic set $S$ is simply the ratio between the photometric and spectroscopic
set cardinalities of each association $w_i = |\hat{P}_i| / |\hat{S}_i|$. When the photometric sample is weighted by
some additional factor (such as a shear-measurement weight, $\hat{w}_j$ for $j\in P$), the weight formula changes to
a simple sum over these weights per association; 
\begin{equation}\label{eqn: weights} 
w_i=\frac{\sum_{j\in \hat{P}_i} \hat{w}_j}{|\hat{S}_i|}.
\end{equation}

As discussed in \cite{lima/etal:2008}, this reconstruction will yield an unbiased estimate $p_{\rm S}^{\rm
w}\left(z\right) = p\left(z|P\right)$ in the regime where $p\left(z|\hat{S}_i\right)=p\left(z|\hat{P}_i\right)$;
i.e. when the colour-redshift distributions of the spectroscopic and photometric subsets are exactly the same (including
selection effects, photometric noise, Poisson noise, etc). This is true for arbitrarily complex (e.g. broad)
$p_i\left(z|P,\mathbf{c}_i\right)$. However in practice, due to spectroscopic selection effects for example, the redshift
distribution of spectroscopic data are very different to those of photometrically defined data, 
$p\left(z|S\right)\neq p\left(z|P\right)$. Nonetheless, the recalibration remains valid provided that 
the colour-redshift relationship is unique: $p\left(z|\mathbf{c}_i\right)\to\delta\left(z|\mathbf{c}_i\right)$. In this 
limit $p\left(z|\hat{X}_i\right)=\delta\left(z|\hat{X}_i\right) \,\,\,\forall\,\,\, X \in P,S$ and so unbiased recovery is
again possible even given $p\left(z|S\right)\neq p\left(z|P\right)$. 

In \cite{hildebrandt/etal:2018} a $k$NN method is used to estimate, for the $i^{\rm th}$ spectroscopic galaxy, the 9-dimensional
{ hyper-spherical} volume, $V^S_i$, that contains precisely $|\hat{S}_i|=4$ spectroscopic nearest-neighbours. The (shear-contribution weighted)
number of photometric sources contained within the same volume, $N^P_i=\sum_{j\in \hat{P}_i} \hat{w}_j$, is then also
calculated, thereby allowing computation of the weights in Eqn. \ref{eqn: weights}. As stated earlier, in this work we perform a new 
association scheme utilising unsupervised machine learning. 
}

Self organising maps \citep[SOMs; ][]{kohonen:1982} 
are a form of unsupervised neural network which uses competitive learning of neurons to map a high dimensional 
manifold onto a low-dimensional grid. 
While SOMs were initially devised as a visualisation tool \citep{kohonen:1982}, they have found a range
of uses within the astronomical community over the last two decades \citep[see, e.g.,
][]{naim/etal:1997,davidzon/etal:2019}. The most notable implementation has been in the 
Complete Calibration of the Colour-Redshift Relation \citep[C3R2; ][]{masters/etal:2017,masters/etal:2019} 
project where the C3R2 team endeavour to utilise SOMs to identify unexplored parts of the $n$-dimensional
colour-redshift plane {\citep{masters/etal:2015}}, in an effort to subsequently observe spectra of such sources and thus,
as the name suggests, completely calibrate the colour-redshift relation for future weak lensing surveys such as Euclid
\citep{amendola/etal:2018,laureijs/etal:2011}. 

{ 
The importance of the C3R2 analysis for our work here lies in the use of SOMs to create a high-fidelity 
discrimination of the colour-redshift relation. As the SOM allows for a 
sophisticated mapping of the complex $n$-dimensional colour-magnitude manifold, it can be used as the basis of association
definitions $\hat{S}_i$ and $\hat{P}_i$. Therefore, a SOM trained on the
spectroscopic dataset $S$, and into which we subsequently map photometric data $P$, should generate a high fidelity
weighting $w_i \,\,\,\forall\,\,\, i\in [1,m]$, and therefore a high fidelity estimate of $p\left(z|P\right)$.  
}

Recently \cite{buchs/etal:2019} have presented an implementation of the direct redshift calibration method, utilising
multiple SOMs, also for the purpose of calibrating cosmic shear studies. Their implementation is designed primarily for
surveys observed in the manner of DES (that is, without comparable observations over both the wide-field and
deep-spectroscopic survey fields). This type of survey design makes directly mapping the wide-field and deep
spectroscopic surveys {challenging}, and as such \cite{buchs/etal:2019} are required to estimate the mapping of the
wide-field data onto the spectroscopy via a series of intermediate datasets. In their mock analysis, intermediate datasets are chosen to be
noiseless and/or fully representative with zero redshift uncertainty.
With these somewhat idealised conditions, \cite{buchs/etal:2019} recover the underlying redshift distributions
with maximal expected biases of $\sigma_{\Delta \langle z \rangle} \sim 0.007$ for DES-like wide-field observations and tomographic
binning. 

The results from C3R2 and \cite{buchs/etal:2019} suggest that a new implementation of the traditional direct redshift
calibration has merit.  Naturally though, they do not guarantee that SOMs will return an unequivocally superior
calibration method. The finite {binning} of the SOM manifold, for example, presents a limitation that is clearly not
present when performing a $k$NN matching of every spectroscopic source individually. Such a {binning} creates
discreteness in the final mapping, which could lead to a degradation of the final weighting. 
Conversely, the same continuity of the $k$NN method (which we just described as a positive) can also lead to pitfalls.
For example, if the colour-magnitude space of the spectroscopic sample is not adequately representative of the
photometric sample, then the $k$NN matching will be forced to extend to sources well beyond what we might consider the
local region of the $n$-dimensional manifold. In the SOM implementation, such regions without spectroscopic
representation are directly visible and so the misidentification of photometric sources can be kept to a minimum. 
These are but two examples of possible differences between the $k$NN and SOM direct calibration methods, and
demonstrate why comprehensive testing of the two methods is necessary. 

For all of the SOM analysis presented in this work, we utilise a branched version of the widely used and tested {\tt
kohonen} package \citep{wehrens/kruisselbrink:2018,wehrens/lutgarde:2007} within R \citep{R}. The original package 
version is available from the Comprehensive R Archive Network (CRAN). 
Our branched version of the CRAN package, available at \url{https://github.com/AngusWright/kohonen.git}, contains 
modifications for better plotting, and has been used for all SOM visualisations here. 
All of the scripts required to run the SOM direct calibration, and produce the figures here, are made public at 
\url{https://github.com/AngusWright/SOM\_DIR.git}. 

\section{Dataset}\label{sec: data}
In this work we will explore the performance of the SOM calibration method using a series of simulations, and then use
the SOM to calibrate the KiDS+VIKING-450 dataset presented in \cite{wright/etal:2018b,hildebrandt/etal:2018}. Our
simulations are built to resemble the KiDS+VIKING-450 dataset, which can be split into two fundamental sections: 
the photometric survey which contains shape measurements for cosmic shear, and the spectroscopic compilation which
contains redshift estimates from spectroscopy. 
\subsection{Photometric Survey}
A comprehensive description of the combined full KiDS dataset is provided in \cite{wright/etal:2018b}. The
dataset comprises of KiDS optical imaging probing the $3000$ to $9000$ \AA ngstrom range in 4 bands ($ugri$). 
Imaging is taken with the OmegaCAM instrument, mounted at the Cassegrain focus of ESO's VLT Survey
Telescope \citep[VST; ][]{dejong/etal:2017} on Cerro Paranal, Chile.  The imaging used here comprises of $454$
distinct $\sim1$ deg$^2$ pointings of the camera, which (after masking) covers $360.3$ deg$^2$ on-sky.

These optical data are then combined with the infrared imaging from the VISTA Kilo degree INfrared Galaxy
\citep[VIKING; ][]{edge/etal:2013,venemans/etal:2015} survey, probing the NIR wavelengths between $8000$ and $24000$
\AA ngstroms. These data are taken using the Visible and InfraRed CAMera (VIRCAM) on ESO's 4m VISTA telescope, also located
on Cerro Paranal, Chile. The imaging is taken in 5 near-IR bands ($ZYJHK_{\rm s}$), using 16
individual HgCdTe detectors, each with a $0.2\times0.2$ square degree angular size, but which jointly span a
$\sim1.2$ square degree field of view. These detectors are designed for dedicated near-IR observations, which allows
for a vastly improved efficiency in even the bluest ($Z$) band compared to observations taken in a similar range using 
optical detectors. 

The combined KiDS+VIKING dataset is extremely well matched in terms of depth and coverage. Photometry in every band 
is measured using the Gaussian Aperture and PSF \citep[GAaP][]{kuijken:2008,kuijken/etal:2015} method, with additional
methodological details described in \cite{wright/etal:2018b}. \cite{wright/etal:2018b} also present  
statistics for the photometric detection of sources in the combined dataset, demonstrating that over $80\%$ of
KiDS+VIKING lensing sources have finite detections in all bands from $g$-$K_{\rm s}$. 

The full KiDS+VIKING-450 dataset (after masking) comprises of $447$ distinct $\sim1$ deg$^2$ pointings of the camera,
which (after masking) covers $341.3$ deg$^2$ on-sky \citep{wright/etal:2018b}.

\subsection{Spectroscopic Surveys}
Spectroscopic data utilised for direct calibration in the KiDS survey originates from 5 distinct redshift surveys:
zCOSMOS \citep[the bright selection presented in ][ and a non-public deep compliment]{lilly/etal:2009}, the DEEP2
Redshift Survey \citep[the colour-selected equatorial fields; ][]{newman/etal:2013}, VIMOS VLT Deep Survey \citep[VVDS; ][]{lefevre/etal:2013}, GAMA G15-Deep
\citep{kafle/etal:2018} and ESO-GOODS CDFS \citep{popesso/etal:2009,balestra/etal:2010,vanzella/etal:2008}. These
surveys were chosen for two reasons.  Firstly, they each probe (at least partially) the colour-magnitude range of
photometric sources utilised for KiDS cosmic shear.  Secondly, they are selected because they all either overlap with
the KiDS and VIKING photometry directly (G15-Deep, zCOSMOS\footnote{In the zCOSMOS field observations performed by
VISTA have been undertaken extensively for the UltraVISTA survey \citep{mccracken/etal:2012}. UltraVISTA observations 
do not include data taken in the VISTA $Z$-band, and so we construct a VISTA-$Z$-like dataset
from other deep $z^\prime$ data in the field, taken with the MegaCAM instrument on the Canada-France-Hawaii-Telescope 
\citep[CFHT; ][]{bielby/etal:2012,hudelot/etal:2012}. Using this deep data we are able to construct a $Z$-band substitute
that, for the photometric depth and redshift coverage probed by KiDS data, has similar colour properties for all
galaxies at all redshifts to better than $|z^\prime - Z| < 0.1$.}) or have dedicated KiDS- and VIKING-like observations
(VVDS, CDFS, DEEP2).

\begin{figure}
\centering
\includegraphics[width=\columnwidth]{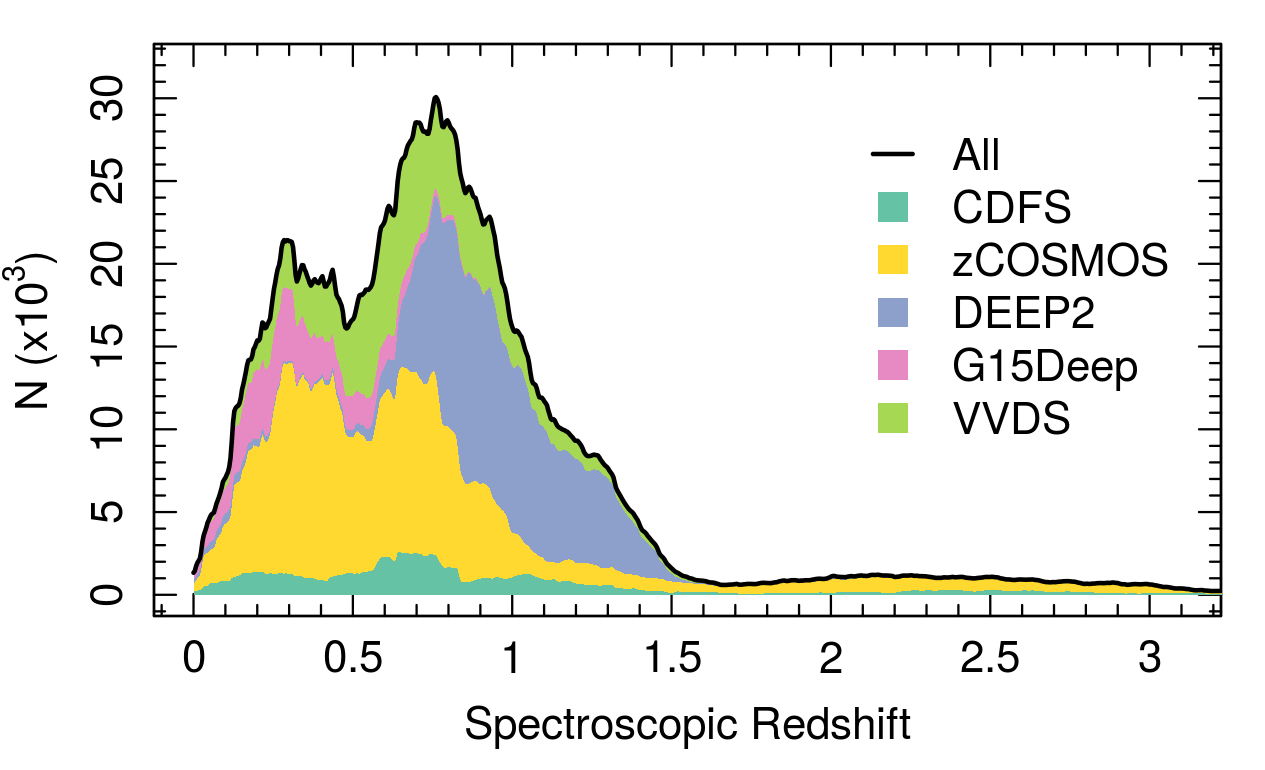}
\caption{The spectroscopic redshift distribution of our combined spectroscopic calibration dataset. The figure shows
the redshift distribution as a kernel density estimate (KDE), constructed using a rectangular $\delta z=0.1$ kernel. The 
KDE is weighted such that lines are interpretable as the instantaneous counts per $\delta z$. 
The KDE is coloured by the fractional contribution from each of our 5 datasets
to the total, which is shown by the black line. 
}\label{fig: specz dist}
\end{figure}

Statistics for the various spectroscopic datasets are provided in Table \ref{tab: specz samp}. The full spectroscopic 
compilation is described in detail in \cite{hildebrandt/etal:2018}. The table shows the size of the individual 
spectroscopic datasets {in area and number of spectra. We can see that
the three largest of our spectroscopic datasets are zCOSMOS, DEEP2, and VVDS; combined they make up more than $85\%$ of
the spectroscopy used for our calibration. The table is complemented by Figure \ref{fig: specz dist}, which shows the
spectroscopic redshift distribution of the combined sample, coloured by the survey from which each source
originated. The figure demonstrates the different selections that have been applied to the various spectroscopic
datasets. zCOSMOS is a complicated combination of multiple spectroscopic campaigns, containing a bright low-redshift
($z\sim0.35$) population, a fainter middle-redshift ($z\sim0.75$) population, and finally a population of very
high-redshift ($z>1.7$) sources. Conversely, DEEP2 is a single population of colour-selected targets, which show a clear
singular population between $0.7\lesssim z \lesssim 1.5$.} 

\begin{table}
\centering
\caption{Spectroscopic redshift samples used for the direct redshift calibration in KiDS.  }\label{tab: specz samp}
\begin{tabular}{ccc}
\hline
\hline
Survey & Area & No. of \\ 
       & $[\rm deg^2]$ & spec-$z$ \\
\hline
zCOSMOS  & $0.7$ & $9930$ \\
DEEP2    & $0.8$ & $6919$ \\
VVDS     & $1.0$ & $4688$ \\
G15Deep  & $1.0$ & $1792$ \\
CDFS     & $0.1$ & $2044$ \\
\hline
\end{tabular}
\end{table}

\section{Simulations}\label{sec: simulations}
In this section we describe the construction of our KiDS-like simulated datasets which we use 
to verify the performance of the SOM direct calibration methodology 
on a cosmological dataset such as KiDS. We therefore must construct a simulated dataset that mimics the
complexity of KiDS and the spectroscopic compilation in terms of extraction depth, photometric depth, wavelength
coverage, spectroscopic target selection, and shear estimation precision. 

Our simulations are constructed from the MICE2 simulation which is presented in detail in
\cite{Fosalba/etal:2015,Crocce/etal:2015,Fosalba/etal:2015b,Carretero/etal:2015,Hoffmann/etal:2015}. MICE2 is based on an N-body dark
matter simulation, which is used to derive an all-sky lensing mock catalogue between $0.1\leq z \leq 1.4$. The lensing
catalogue contains source positions, morphological information, lensing convergence measurements, and model magnitudes in the
$\utok$-bands. From these products, we are able to construct multiple realisations of high-fidelity KiDS-like mock
photometric and spectroscopic catalogues. Given its construction, the MICE2 mocks present an excellent starting point for
our analysis. {Our simulations are constructed using the pipeline of \cite{vandenbusch/etal:2019}, which is
publicly available at \url{https://www.github.com/KiDS-WL/MICE2_mocks.git}. We detail parts of the mock catalogue
generation here, including a description of spectroscopic selections and sample definitions applied to the simulation. }

{ 
Prior to selection of photometric and spectroscopic sources using MICE2, we apply the recommended evolution corrections
to the model magnitudes:
\begin{equation}\label{eqn: evolution}
m_{\rm e} = m_{\rm model} - 0.8 (\arctan(1.5 z_{\rm true}) - 0.1489), 
\end{equation}
and apply the required flux magnifications \citep{Fosalba/etal:2015b} as determined by the lensing convergence, $\kappa$:
\begin{equation}\label{eqn: magnification}
m_{\rm e,m} = m_{\rm e} - 2.5\log{(1+2\kappa)}. 
\end{equation}
We then derive effective photometric apertures from the on-sky bulge and disk effective radii and the bulge fractions.
The photometric aperture of each source, $a^f_{\rm ap}$ and $b^f_{\rm ap}$, is then approximated, per filter $f$, as a
function of the effective radius of the two-component light-profile, $R_{\rm eff}$, the intrinsic two-component profile
axis ratio $b_{\rm intr}/a_{\rm intr}$, and the filter $f$ PSF FWHM $\sigma^f_{\rm PSF}$:
\begin{align}\label{eqn: aperture}
a^f_{\rm ap}=&\sqrt{\left(2.5 R_{\rm eff}\right)^2+\left(\sigma^f_{\rm PSF}\right)^2}\\
b^f_{\rm ap}=&\sqrt{\left(2.5 R_{\rm eff}\frac{b_{\rm intr}}{a_{\rm intr}}\right)^2+\left(\sigma^f_{\rm PSF}\right)^2}.
\end{align}

With these aperture parameters and the documented per-filter point-source magnitude limits \citep[see
][]{kuijken/etal:2019}, we generate KiDS-like photometry for all sources in the MICE2 octant. We use the magnitude
limits to calculate a true signal-to-noise (SN) for every source in every band of the simulated catalogue. These SN estimates
incorporate realistic estimates of aperture noise, using the apertures calculated above, and all flux uncertainties
encoded in the point source magnitude limits (shot noise, image noise, etc).  
With these SN estimates, we calculate an `observed' flux for every source in each band, and compute a final uncertainty
from these observed fluxes. 
}

{ 
For the photometric sample definition, we first subset a section of the MICE2 octant into a KiDS sized patch of area
$341.3$ square degrees (i.e. the post-masking area of KV450). All simulated galaxies that lie within this footprint
have KiDS-like photometric noise realisations, as described above. All sources are then matched in 9-band
magnitude-space to the actual KiDS photometric data (via $k$NN within a maximum 1 magnitude Euclidean radius), and
inherit the nearest-neighbour shear calibration weight\footnote{Individual galaxy shear estimates in KiDS are made using the
\emph{Lens}fit algorithm \citep{miller/etal:2007}, which produces an inverse variance weight per galaxy, which is highly
magnitude dependant.  Therefore matching simulated sources to KiDS galaxies in magnitude space is able to reproduce the
shear-weight distribution of KiDS photometric galaxies well.}. Unmatched sources are assigned zero weight, as by
definition they do not appear in the KiDS source sample. This latter step has the added benefit of implicitly encoding
any unrecognised colour and magnitude dependant selection effects present in the KiDS data. The final simulated
photometric source catalogue is then selected as being all sources with non-zero calibration weight. 

We simulate our 3 primary spectroscopic datasets (zCOSMOS, DEEP2, and VVDS), using the evolution corrected
photometry in the Johnson $BVRI$-bands, to simulate the selection of spectroscopic targets from deep imaging. We start
by selecting all galaxies within distinct patches (\ie not part of one-another nor the KV450 area), each of which is the
same size as the spectroscopic survey being modelled (\ie as shown in Table \ref{tab: specz samp}). Their
various magnitude- and colour-based selection functions are applied, except typically with minor adjustments to better
reproduce the observed redshift distributions of the surveys:
\begin{itemize}
\item zCOSMOS: $15 < I < 22.5$ (i.e. the bright selection only);
\item DEEP2: $18.5<R<24.0$ and $(B-R < 2.0 (R-I)-0.4$ or $R-I > 1.1$ or $ B-R < 0.2)$; and 
\item VVDS: $18.5 < I < 24.0$.  
\end{itemize}
After each of these selections is applied, we further trim each sample using documented (per-survey) spectroscopic
failure/incompleteness functions \citep[in both magnitude and redshift space, as described
in][]{newman/etal:2013,lilly/etal:2009,lefevre/etal:2013}. {This process allows us to encode spectroscopic
incompleteness, albeit imperfectly: we are unable to explicitly incorporate incompleteness as a function of colour in this
implementation.} Finally, if the number density of remaining spectra is higher 
than in the data, we perform a sparse sampling of each spectroscopic subset to reproduce the number of observed
spectra (to within $\sim 3$ percent). 
}

As was shown in Table \ref{tab: specz samp}, our three main spectroscopic datasets correspond to $85\%$ of the
full spectroscopic dataset {(although part of this is the zCOSMOS faint selection). We have opted not to include
the G15DEEP, zCOSMOS-faint, and CDFS compilations in our simulations: G15Deep is small and relatively low-impact given our lensing
sample (see Sect. \ref{sec: results I kv450}), and both the CDFS and zCOSMOS-faint samples are complex combination of
tens of individual survey
datasets. These post-hoc combinations have naturally complex selection functions which are difficult to faithfully
reconstruct in our simulations. Excluding these datasets from the simulated compilation ensures that we do not
accidentally over-estimate the depth of our true spectroscopic data, and thus overestimate the performance of our 
redshift calibration methodology.
}

In cases where we test the effects of sample variance\footnote{ Sample variance here is used in the standard
cosmological context, meaning the variance introduced in astronomical observations of finite area due to large scale
structures along the line-of-sight. This is distinct from shot noise and cosmic variance, the latter of which relates to
the variance induced by differing realisations of the observable universe.}, we perform the spectroscopic catalogue
creation 100 times in 100 sets of completely independent fields {(lines-of-sight);} i.e. all 100 fields of all 3
surveys are independent of one-another. We also produce 100 realisations of the photometric noise in one realisation of
the spectroscopic fields, to test the influence of photometric noise within the spec-z distributions. 

{ 
Finally, we note that the limited redshift range of the MICE2 simulation places a limitation on the interpretability 
of the results with respect to redshift calibration for real cosmic shear surveys. However this limitation is somewhat
common to the literature \citep[see, e.g., ][]{buchs/etal:2019,alarcon/etal:2019}, and will only be resolved with the
construction of larger, deeper simulations of cosmic shear samples. To this end, simulations such as the Euclid
Flagship\footnote{\url{https://sci.esa.int/web/euclid/-/59348-euclid-flagship-mock-galaxy-catalogue}} and the Legacy
Survey of Space and Time (LSST) DC2 
\citep{korytov/etal:2019} represent an obvious testing-ground for this (and other) redshift calibration methods in
the future. 
}


\section{Results}\label{sec: results}
\subsection{Suitability of current spectroscopy for direct photometric redshift calibration}\label{sec: results I kv450} 
In this section we explore the question of whether the currently available spectroscopic compilations, used primarily by
the KiDS consortium, are of sufficient depth and diversity for use in direct {photometric redshift calibration, so as to
not cause significant biases in reconstructed redshift distributions}. We explore this question of representation using the
real KV450 photometric and spectroscopic datasets. {We also measure the representation of KV450 photometric sources
using the individual spectroscopic surveys, in an effort to quantify the influence that any one calibration dataset may
have over the final calibrated redshift distribution estimates.}

\begin{figure*}
\centering
\includegraphics[scale=0.7]{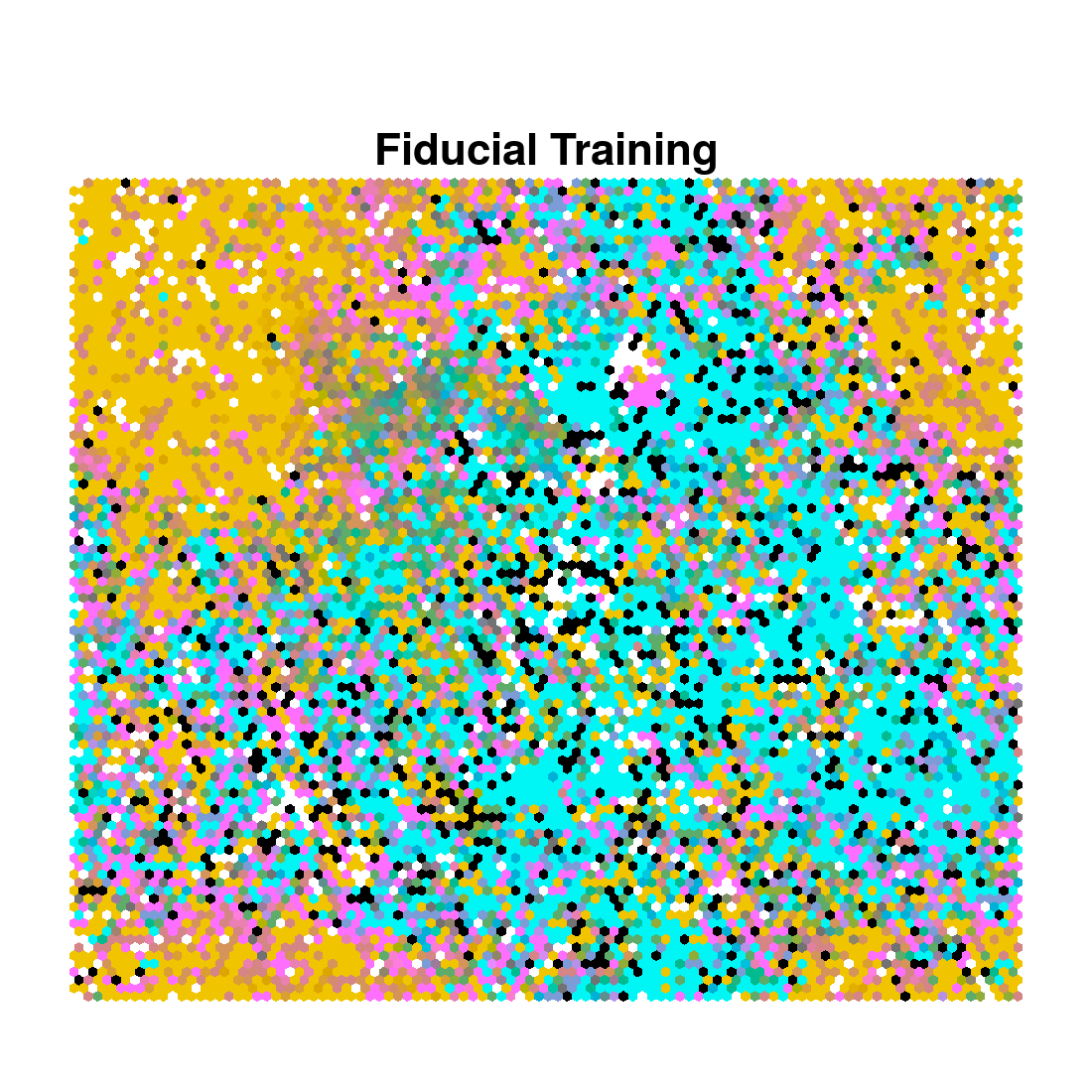}\includegraphics[scale=0.7]{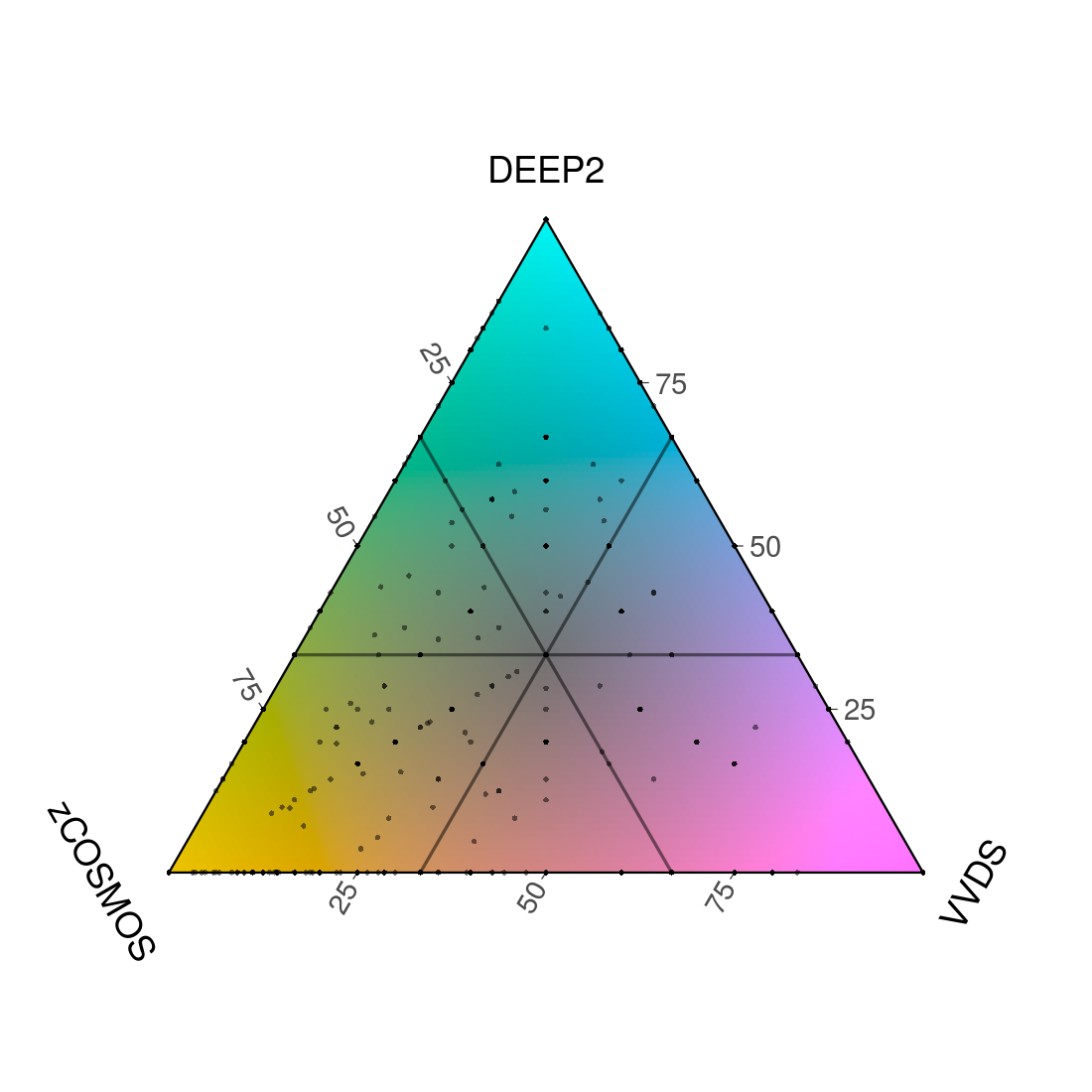}
\caption{The distribution of the 3 primary KV450 spectroscopic samples within the SOM. The figure on the left shows the 
SOM coloured by the fractional contribution of each of the 3 main spectroscopic samples from KV450. The ternary colour bar 
is shown on the right. {The makeup of individual cells is annotated within the colour-bar as points.} SOM {
cells }that are filled entirely by sources from DEEP2, for example, are blue. Conversely {cells} that are filled
by equal mixtures from all 3 samples are grey. {Cells }which contain spectroscopic data from other surveys (which
are not shown) are coloured white. {Cells }which contain photometric galaxies but no spectroscopy from any survey
are coloured black. The figure highlights the complementarity between the DEEP2 and zCOSMOS data, as well as the breadth
of coverage of the VVDS data.}\label{fig: som samples}
\end{figure*}

\begin{table*}
\centering
\caption{Representation of photometric galaxies within the true KV450 data and spectroscopic compilation, 
for variously defined spectroscopic samples, determined using our `full-sample' SOM and variable {cell} clustering 
per tomographic bin (see Appendix \ref{sec: Clustering Results}). The 
table shows the overall sample size of each spectroscopic sample (``training size''), the 
fraction of SOM {cells }containing photometric sources which also contain spectroscopy ($f_{\rm pix}$), 
and the change in the effective number density of sources for cosmic shear studies, $n_{\rm eff}$, that we get when only
using photometric sources which are represented within each particular spectroscopic sample ($n^\prime_{\rm eff}/n_{\rm
eff}$).  The statistics are shown for the overall source samples (``All''), and for each of the individually 
defined tomographic bins (note that the ``All'' cases are not averages/summations of the binned values; see Sect. 
\protect \ref{sec: results I kv450}). The best performing individual spectroscopic dataset (\ie\ the middle section) 
is shown in bold (for each column), as is the worst performing joint spectroscopic compilation (\ie\ the lower section). 
The table demonstrates the complementarity of our spectroscopic compilation: our 3 main spectroscopic samples
(zCOSMOS, VVDS, and DEEP2) each uniquely dominate the description of the photometric data in 
the different tomographic bins ($1+2$, $3$, and $4+5$ respectively).}\label{tab: coverage}
\begin{tabular}{c|c|c|cccccc}
\hline
\hline
Spectroscopic& Training    & $f_{\rm pix}$ & \multicolumn{6}{c}{$\rm n^\prime_{\rm eff}/n_{\rm eff} (\%)$}\\
Compilation  &  Size (all) & (all, $\%$)   & All & bin1 & bin2 & bin3 & bin4 & bin5 \\
& & &  $z_{\rm B}\in(0.1,1.2]$ &$(0.1,0.3]$ & $(0.3,0.5]$ & $(0.5,0.7]$ &$(0.7,0.9]$ &$(0.9,1.2]$ \\
\hline
Full Sample & $ 25373 $& $ 91.9 $&  $ 99.5 $ &  $ 82.7 $ &  $ 83.9 $ &  $ 84.6 $ &  $ 82.7 $ &  $ 94.0 $ \\
\hline
CDFS only    & $ 2044 $& $ 17.4 $&  $ 67.3 $ &  $ 57.1 $ &  $ 58.7 $ &  $ 53.2 $ &  $ 40.2 $ &  $ 54.6 $ \\
zCOSMOS only & $ 9930 $& $ {\bf 48.5} $&  $ 79.7 $ &  $ {\bf 74.9} $ &  $ {\bf 75.3} $ &  $ 65.8 $ &  $ 60.3 $ &  $ 63.2 $ \\
DEEP2 only   & $ 6919 $& $ 43.7 $&  $ 73.8 $ &  $ 17.8 $ &  $  5.5 $ &  $ 35.2 $ &  $ {\bf 68.8} $ &  $ {\bf 89.5} $ \\
G15DEEP only & $ 1792 $& $ 10.1 $&  $ 42.0 $ &  $ 63.1 $ &  $ 63.6 $ &  $ 44.1 $ &  $ 19.8 $ &  $ 14.0 $ \\
VVDS only    & $ 4688 $& $ 34.7 $&  $ {\bf 81.4} $ &  $ 54.9 $ &  $ 72.8 $ &  $ {\bf 70.7} $ &  $ 57.3 $ &  $ 70.2 $ \\
\hline
without CDFS    & $ 23329 $& $ 89.1 $&  $ 98.9 $ &  $ 81.5 $ &  $ 82.6 $ &  $ 82.2 $ &  $ 81.0 $ &  $ 93.0 $ \\
without zCOSMOS & $ 15443 $& $ 77.6 $&  $ 97.8 $ &  $ {\bf 76.4} $ &  $ 80.0 $ &  $ 80.8 $ &  $ 78.9 $ &  $ 92.6 $ \\
without DEEP2   & $ 18454 $& $ {\bf 73.4} $&  $ {\bf 93.4} $ &  $ 81.6 $ &  $ 83.0 $ &  $ 80.4 $ &  $ {\bf 72.2} $ &  $ {\bf 80.8} $ \\
without G15DEEP & $ 23581 $& $ 90.6 $&  $ 99.5 $ &  $ 80.1 $ &  $ 83.6 $ &  $ 84.3 $ &  $ 82.7 $ &  $ 94.0 $ \\
without VVDS    & $ 20685 $& $ 84.7 $&  $ 98.1 $ &  $ 81.3 $ &  $ {\bf 79.5} $ &  $ {\bf 77.5} $ &  $ 80.1 $ &  $ 92.9 $ \\
\hline
\end{tabular}
\end{table*}

{ 
In order to estimate the representation of the KV450 photometric dataset, we first train a SOM using the full spectroscopic
compilation. In our fiducial case, we train a $101\times101$ hexagonal-cell SOM with toroidal topology, using the
combination of $36$ colours and $1$ magnitude; the $r$-band. Specific details regarding these
construction and training parameters, and how they influence our results, can be found in Appendix  
\ref{sec: som implementation}}. We then propagate our full photometric and spectroscopic datasets into this trained SOM, 
producing like-for-like groupings between spectroscopic and photometric {sources; specific details of this process
are presented in Appendix \ref{sec: Clustering Results}.} Once we have constructed like-for-like groupings within the
spectroscopic and photometric catalogue, we can then directly measure the number of photometric sources which are
without a spectroscopic counterpart. 

{ 
This direct measurement of representation is used to construct a subsample of the photometric catalogue
which is represented by the spectroscopy. This subsample} with guaranteed representation we define to be
the `gold sample'. 

A visual representation of the propagation of the spectroscopic data into our trained SOM is provided in Figure
\ref{fig: som samples}. The figure shows our trained SOM coloured using a ternary {scale, shown as the large
triangle}. It demonstrates the fractional
contribution of each of our primary spectroscopic surveys to the individual SOM cells: blue for DEEP2, yellow for
zCOSMOS, and pink for VVDS, with a continuum scale for intermediate mixtures of the three catalogues. We also show
the {cells }which are filled by other spectroscopic datasets (but not any of the 3 primary sets) in white, 
and {cells }that have no spectroscopic data in black. 
The figure showcases a few interesting features of the KV450 spectroscopic compilation. Firstly, the complementarity of
the three primary spectroscopic datasets is clear; {cells }are overwhelmingly either blue, yellow, or pink, rather than
intermediate colours (green, purple, brown, grey). This is an indication that our spectroscopic datasets have little overlap in
multidimensional colour-space. 

We quantify the spectroscopic representation of KV450 photometric sources in Table \ref{tab: coverage}. The table shows
the coverage statistics of the SOM, for various splits of the spectroscopic compilation. Starting with the `Full
Sample', shown by all the non-black {cells }in Figure \ref{fig: som samples}, the table shows the
overall size of the spectroscopic sample (`Training Size'; $25373$ galaxies for the full sample), as well as the
fraction of SOM {cells }that these sources occupy ($f_{\rm pix}$; $91.9\%$ for the full sample)\footnote{The $f_{\rm pix}$ value
indicates the percentage of {cells }in the SOM which are occupied by at least 1 spectroscopic galaxy and 1
photometric galaxy. There is no weighting based on the number/weight of photometric sources.}. {As can be seen in} Figure
\ref{fig: som samples}, $\sim 8\%$ of SOM {cells }are unoccupied by spectra. 
However this is not indicative of the fraction of photometric sources which are missing spectra, as photometric counts
vary strongly across the SOM. Furthermore, all photometric sources do not hold the same weight in cosmic shear
estimates, owing to the shape measurement weighting described in Sect. \ref{sec: simulations}. {To correctly quantify the
photometric representation, we choose a statistic that correctly accounts for
the weights of the missed photometric sources}. \cite{heymans/etal:2012} define the effective number
density of weighted photometric sources in cosmic shear studies, $n_{\rm eff}$, as: 
\begin{equation}\label{eqn: neff}
n_{\rm eff} = \frac{1}{A}\frac{\left[\Sigma_{i\in p} w_i\right]^2}{\left[\Sigma_{i\in p} w_i^2\right]}; 
\end{equation}
where $w_i$ is the lensing weight assigned to each photometric source \citep[see][for details]{miller/etal:2007}, and $A$ is the effective survey area. 
This statistic can be calculated for all of photometric sources ($p$), and the 
subset of photometric sources which reside within SOM groupings which contain spectroscopy (\ie the `gold' sample; $p^\prime \subseteq p$). 
{ We can then accurately compute fractional change in $n_{\rm eff}$: $n^\prime_{\rm eff}/n_{\rm eff}$ caused by the
requirement of spectroscopic representation}. 

Table \ref{tab: coverage} shows the fractional changes in $n_{\rm eff}$ (as percentages) when going from the full to the
gold sample for each combination of the spectroscopic data. We start by showing these values for the entire
photometric catalogue (`All'), without tomographic binning. Looking first to the 
case of the full spectroscopic sample, we can see that while roughly $8\%$ of our SOM {cells }contain no
spectroscopy, these {cells }contain just $0.5\%$ of the total lensing weight of the photometric catalogue.  The cells
lie in unimportant parts of the colour space (for the cosmic shear sample), and so contribute negligibly {to the
$n_{\rm eff}$. }

We also show the fractional change in $n_{\rm eff}$ for the five tomographic bins defined in
\cite{hildebrandt/etal:2018}. The redshift limits of each of these bins are annotated in the table header. 
In these split statistics, both the spectroscopic and photometric
dataset are selected such that they have photometric redshifts within the tomographic bin. This has the effect of
decreasing the spectroscopic training size for each bin by a factor of roughly five\footnote{Note that the individual bin
representations need not sum/average to the value in the `All' case, as sources from different tomographic bins can
occupy the same cells.}. As a result, the representation
statistics also decrease. However this process is critical so as to not bias the resulting groupings (see Appendix
\ref{sec: SOM Bias}). In these tomographic bins, we see that the photometric data representation is between $83\%$ (in
the first and fourth tomographic bins) and as high as $94\%$ in the {fifth and highest redshift} tomographic bin.
This result is counterintuitive, as a naive expectation would suggest {that poorer high-$z$ spectroscopic success would
translate to a dearth of representation at high photometric redshifts. However in practice volume effects and the choice
of tomographic bins in KiDS means that the highest redshift bins contain the most spectra: $N_{\rm
spec}=\{2715,3031,4971,6058,6145\}$ for the $5$ tomographic bins, respectively. Furthermore, the increasing rate of
spectroscopic failure is somewhat counter-balanced by the highly redshift and magnitude dependent shear-measurement efficiency,
encoded by the shear measurement weight. Prior to consideration of the shape measurement weights, the highest
tomographic bin has more than $10\%$ misrepresentation. This indicates that spectroscopic failures and
shape-measurement failures are correlated; sources for which it is difficult to measure shapes are also difficult to
redshift.} As a result, the highest tomographic bin is actually the best
represented in KV450. 

We further explore the makeup of our spectroscopic catalogue by splitting it into subsets. 
{ This is motivated by the possibility that one or more of the individual spectroscopic surveys within our
compilation may be affected by unrecognised systematic effects. Recent re-observation and redshifting of
VVDS high-confidence redshifts by the LEGA-C collaboration \citep{straatman/etal:2018}, for example, suggests that the
outlier rate for high-confidence redshifts may be {higher than the expected $\lesssim 3\%$} rate. Should future
studies verify that any of our spectroscopic compilations have serious systematic effects present, then our Table
\ref{tab: coverage} may be used to infer the impact that this may have over current and future KiDS analyses.} 

We look at the representation of each individual spectroscopic dataset (the `only' rows), and
of the full compilation minus each individual dataset (the `without' rows).  
The representation statistics here are all calculated from the
SOM trained on the full spectroscopic dataset; subsequent samples are simply 
propagated into the pre-trained SOM and the coverage statistics are then
calculated. 

Table \ref{tab: coverage} highlights that while the 
zCOSMOS dataset occupies the most individual SOM {cells }of any individual catalogue ($f_{\rm pix} = 48.5\%$), it is the removal of the 
DEEP2 dataset which causes the greatest reduction in {cell} coverage compared to the full compilation (from $91.9\%$ to
$73.4\%$). This indicates that while zCOSMOS is the largest of the KV450 spectroscopic datasets, DEEP2 is the most
unique; the sources within DEEP2 occupy the greatest fraction of systematically different SOM {cells }to those of any
other survey. 

Looking at the change to $n_{\rm eff}$, the story changes slightly. We see that, when considering all photometric
sources together without tomographic binning, it is VVDS which (despite being less than half the size of zCOSMOS) most
effectively describes the lensing weighted KV450 photometric data (when each spectroscopic dataset is considered alone;
$n^\prime_{\rm eff}/n_{\rm eff} = 81.4\%$). Per tomographic bin, however, the best-representation is split between our three main datasets:
zCOSMOS best describes the two lowest tomographic bins ($0.1 < z \leq 0.5$; $74.9\%$ and $75.3\%$), VVDS best describes
the middle bin ($0.5 < z \leq 0.7$; $70.7\%$), and DEEP2 best describes the highest tomographic bins ($0.7 < z \leq 1.2$;
$68.8\%$ and $89.5\%$). This result supports the hypothesis of \cite{hildebrandt/etal:2018}, who argued that the removal
of DEEP2 from the spectroscopic compilation would preferentially affect the photometric representation of the higher
tomographic bins, effectively pulling them to lower mean redshifts, and thus causing a bias in the estimated value of
${S}_8$. 

Interestingly, the same trends are largely true for the compilations without each of these datasets. In the
tomographically binned cases, all bins other than the $2^{\rm nd}$ are most heavily misrepresented when the most
uniquely represented dataset is removed. In the $2^{\rm nd}$ bin, removal of VVDS (rather than zCOSMOS) causes the
greatest decrease in $n_{\rm eff}$. Overall, it is DEEP2 that is the most important dataset: while removal of all other
datasets triggers a maximal reduction in $n_{\rm eff}$ of $7.1\%$ (VVDS in bin 3), removal of DEEP2 sees a reduction of over
$10\%$ in both bins 4 and 5. This is more than twice the decrease seen when removing the next most important dataset in bin 4
(from $78.9\%$ without zCOSMOS to $72.2\%$ without DEEP2), and nearly $10$ times smaller than the decrease seen when
removing the next most important dataset in bin 5 (also zCOSMOS; from $92.6\%$ to $80.8\%$). 
Overall, these statistics indicate that, for the calibration of the KiDS tomographic cosmic shear dataset, our three
primary KV450 datasets are equal parts individually important and unique. {As a result, for a coherent redshift
offset in all tomographic bins to be seen, a conspiracy of unknown systematic biases in at least two major spectroscopic
surveys would be needed. }

\subsection{Influence of sample variance, noise, and selection biases}\label{sec: results I mice2} 
Our second set of results regards the sensitivity of our photometric representation estimates to the presence of a
number of systematic effects present in spectroscopic surveys: sample variance, photometric noise, and selection biases. 
To test the influence of these systematics we use our MICE2 simulations, for which we are able to generate many
realisations of lines-of-sight (to analyse sample variance) and noise realisations (because we know the true fluxes). 

\begin{figure}
\centering
\includegraphics[width=\columnwidth]{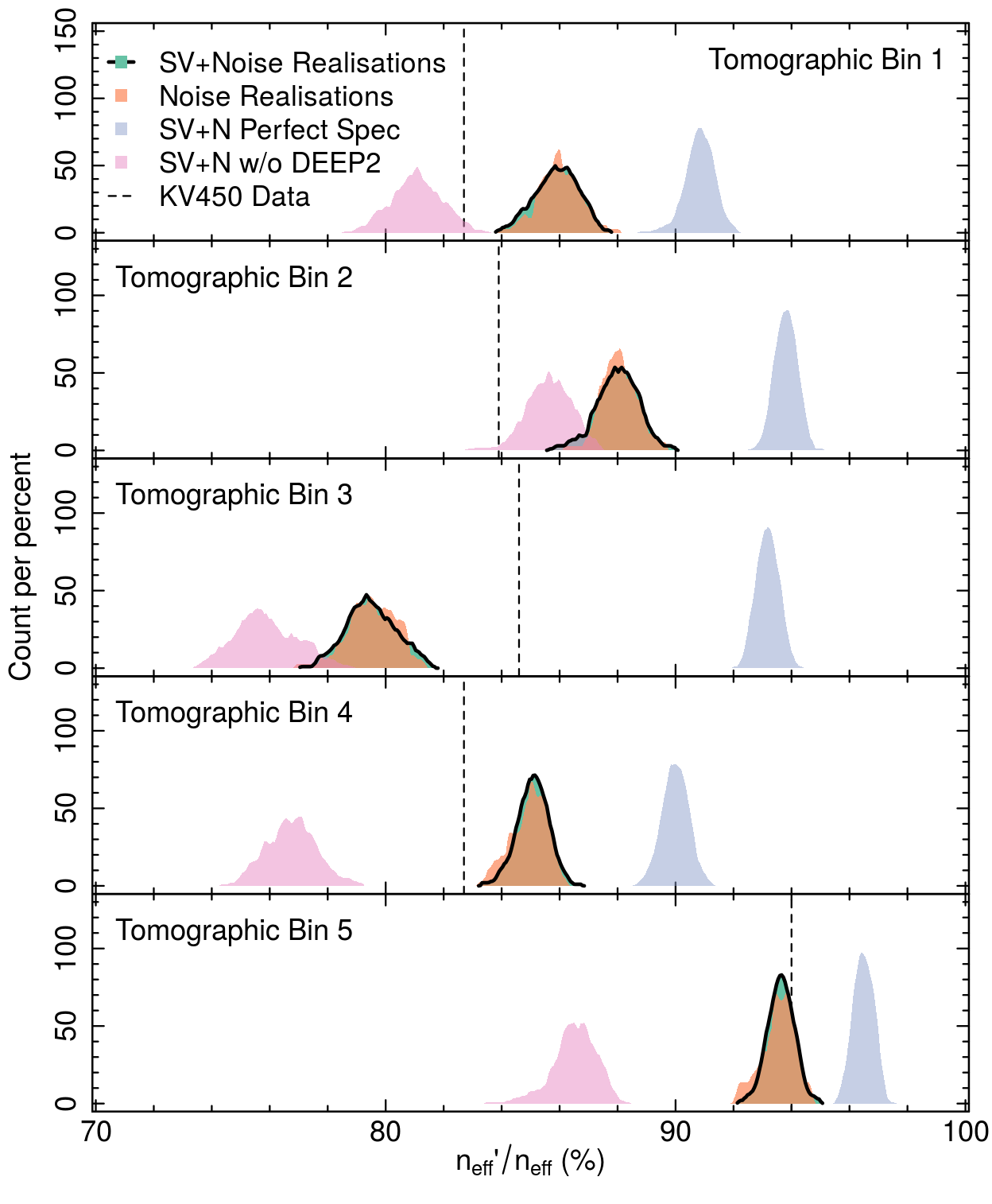}
\caption{The change in the value of $n^\prime_{\rm eff}/n_{\rm eff}$ with 100 different lines of sight (testing both
noise and sample variance; green), 100 different noise realisations of a single line-of-sight (testing the importance of
photometric noise; orange), 100 perfectly sampled spectroscopic catalogues (testing spectroscopic selection effects;
purple), and 100 lines-of-sight excluding DEEP2 (testing the similarities to simulations and data; pink). The
representations seen in the real KV450 data are also shown (black dashed lines). The
distributions show that simulation is a {reasonable match to the observed representations, being typically within
$\pm5\%$ of the representations seen in the data.} We see that photometric noise dominates our observed
misrepresentation, and that the MICE2 KV450-like spectroscopic compilation is typically $\sim5\%$ less representative of
the full photometric sample, when compared to a perfectly sampled spectral catalogue (with the exception of bin 3). 
Thus the majority of the under-representation is caused by Poisson sampling and photometric noise.
}\label{fig: cv pdfs}
\end{figure}

Figure \ref{fig: cv pdfs} shows the representation of the cosmic shear sample, $n^\prime_{\rm eff}/n_{\rm eff}$, for  
100 realisations of different spectroscopic lines-of-sight within our MICE2 simulations (green with black outline). Each
line-of-sight is unique, {both} per realisation and per spectroscopic survey (DEEP2, zCOSMOS, and VVDS), and so samples a
unique part of the MICE2 octant. We use a single realisation of the photometric catalogue for these tests, to
exclusively probe the impact of sample variance in the spectroscopic catalogues on our representation estimates. As the
different lines-of-sight naturally contain independent galaxies, the {width of} our green histograms contain the
influence of both sample
variance and photometric noise. We therefore also show 100 realisations of {a single set of spectroscopic
lines-of-sight} (orange), so as to demonstrate exclusively
the influence of photometric noise. {These histograms} suggest that the joint effect of sample variance and
photometric noise is to perturb our estimated $n^\prime_{\rm eff}/n_{\rm eff}$ at the $\sim1$ percent level, and 
(more interestingly) that this scatter is overwhelmingly driven by photometric noise rather than sample variance; the 
{ width of the green histograms is equivalent to the width of the orange}. 

The figure also shows the observed representations for the KV450 dataset (vertical black dashed lines). These lines show that the
simulation is a {reasonable reflection of the representation seen in the data, being within $\sim 5$ percent of the
data representation in all bins.  
The simulations also show the same behaviour as the data with regard to the inclusion or exclusion of individual
spectroscopic samples; we show the influence of removing DEEP2 (pink) causes a similar pathological
reduction in representation, per tomographic bin, for the simulations as in the data (Sect. \ref{sec:
results I kv450}).} 

We demonstrate how the estimated value of $n^\prime_{\rm eff}/n_{\rm eff}$ changes if we were able to
use a perfectly representative spectroscopic compilation. To do this we construct a spectroscopic sample of the same
size as our full spectroscopic compilation, but which is sparse sampled (100 times) from the photometric dataset itself.
The results are shown in purple in Figure \ref{fig: cv pdfs}. Surprisingly, the perfectly representative spectroscopic
sample is typically only $\sim 5\%$ better than that of our standard spectroscopic compilation. The exception here is
again the third tomographic bin. 
This suggests that the decrease in KV450 representation is not driven {predominantly} by systematically missing spectra in the
multidimensional colour space, but rather the dearth of spectra overall. {Such a circumstance, however, is unlikely to be
true (with this spectroscopic compilation) for stage-III cosmic-shear surveys like Euclid and LSST, which will extend to
considerably higher redshifts than KiDS. These surveys will likely require additional dedicated programs, such as C3R2,
to compile samples of spectra capable of calibrating their highest redshift sources.}

\subsection{Accuracy of the SOM Direct Calibration}\label{sec: results II}
We now utilise our simulations to {explore the bias and scatter in our photometric redshift calibration method
by comparing the true and estimated redshift distributions from our mocks. In the following analyses we assume that 
all spectroscopy within our spectroscopic sample are recovered with $100\%$ accuracy. For a discussion of the influence
of catastrophic spectroscopic failures on our analysis, we direct the interested reader to Appendix \ref{sec:
catastrophic specz}.} 

We estimate the tomographic redshift distributions for each of our $100$ spectroscopic lines-of-sight, as described in
Sect. \ref{sec: som method}, keeping the photometric dataset static. {We explore the performance of our method with 
both noisy and noiseless photometry, and for both perfectly representative and systematically incomplete spectroscopic
compilations. Measured redshift distribution biases for each of our simulations are given in Table \ref{tab:
reconstruction bias}. }

\begin{table*}
\caption{Biases in the mean redshift estimation ($\Delta\langle z \rangle=\langle z \rangle_{\rm est}-\langle z
\rangle_{\rm true}$), per tomographic bin, for our various runs
of the MICE2 simulations. We show the results using both KV450-like and perfectly representative spectroscopic data,
using both noisy and exact photometry.  Values shown are the mean biases over 100 different lines of sight
(MICE2), as well as the stdev population scatters from the same. Entries with both bias and scatter less than
$1\times 10^{-3}$ are simply shown with a null symbol $\emptyset$. Conversely, entries with large biases ($\Delta\langle
z \rangle > 0.01$) are highlighted via boldface. The results demonstrate that the SOM method is unbiased in the absence
of photometric noise, even in the presence of sample variance and spectroscopic selection effects. {Photometric
noise at the level of KV450 introduces colour redshift degeneracies which subsequently introduce a maximal bias of
$\Delta\langle z \rangle = 0.023$ in some tomographic bins. Basic quality cuts (`QC1') are able to reduce the maximal
bias to $\Delta\langle z \rangle = 0.013$, {or $\Delta\langle z \rangle \leq 0.025$ at $97.5\%$ confidence}, at the cost
of $\Delta n_{\rm eff}=\{2.0,0.3,2.4,0.6,0.1\}\%$ in the 5 tomographic bins respectively. More stringent quality cuts
(`QC2') reduce the biases to $\Delta\langle z \rangle = 0.010$, but at further cost to the effective number density
($\Delta n_{\rm eff}=\{15.9,13.6,23.3,26.2,21.1\}\%$). The results computed when using the previous $k$NN association
are shown in the final row.} }\label{tab: reconstruction bias}
\centering
\begin{tabular}{lc|r@{\hskip 0in}lr@{\hskip 0in}lr@{\hskip 0in}lr@{\hskip 0in}lr@{\hskip 0in}l}
\hline
\hline
\multicolumn{2}{c}{ Dataset} & \multicolumn{10}{c}{ Reconstruction Bias $(\Delta\langle{\rm z}\rangle)$} \\
 Type & Phot & \multicolumn{2}{c}{bin1} & \multicolumn{2}{c}{bin2} &
    \multicolumn{2}{c}{bin3} & \multicolumn{2}{c}{bin4} & \multicolumn{2}{c}{bin5} \\
\hline
perfect & exact  & 
$    0.001 $ &$   \pm <\!10^{-3} $ & 
$          $ &$   \emptyset $ & 
$          $ &$   \emptyset $ & 
$          $ &$   \emptyset $ & 
$          $ &$   \emptyset $ \\ 
KV450 & exact & 
$    0.002$ & $   \pm 0.001 $ &
$    0.002$ & $   \pm 0.002 $ &
$    0.003$ & $   \pm 0.001 $ &
$    0.003$ & $   \pm 0.001 $ &
$   -0.001$ & $   \pm 0.001 $ \\
Perfect & noisy & 
$         0.004$ & $   \pm 0.003 $ &
$    <\!10^{-3}$ & $   \pm 0.002 $ &
$         0.006$ & $   \pm 0.003 $ &
$         0.004$ & $   \pm 0.003 $ &
$         0.003$ & $   \pm 0.003 $ \\
KV450 & noisy & 
$    0.009 $ & $   \pm 0.005 $  &
$    0.004 $ & $   \pm 0.006 $  &
$\bf 0.023 $ & $\bf\pm 0.006 $  &
$\bf 0.012 $ & $\bf\pm 0.004 $  &
$   -0.007 $ & $   \pm 0.005 $ \\
KV450 & noisy+QC1 & 
$    <\!10^{-3}$ & $\pm 0.005 $      &
$    0.002 $ & $   \pm 0.006 $  &
$\bf 0.013 $ & $\bf\pm 0.006 $  &
$\bf 0.011 $ & $\bf\pm 0.004 $  &
$   -0.006 $ & $   \pm 0.005 $\\
KV450 & noisy+QC2 & 
$    0.002 $ & $   \pm 0.005 $ &
$    0.003 $ & $   \pm 0.006 $ &
$    0.007 $ & $   \pm 0.005 $ &
$    0.009 $ & $   \pm 0.004 $ &
$   -0.006 $ & $   \pm 0.004 $\\
\hline
&&\multicolumn{10}{c}{\em using $k$NN association} \\
KV450 & noisy & 
$\bf 0.047 $ & $\bf\pm 0.005 $ &
$\bf 0.025 $ & $\bf\pm 0.004 $ &
$\bf 0.032 $ & $\bf\pm 0.005 $ &
$   -0.004 $ & $   \pm 0.004 $ &
$\bf-0.013 $ & $\bf\pm 0.004 $\\
\hline
\hline

\end{tabular}

\end{table*}

First we focus on the simulations run without photometric noise (`exact'). The results indicate that in all
circumstances {(\ie\ with both perfect and biased spectroscopic compilations) the SOM direct
photometric redshift calibration method is unbiased. All tomographic bins, in the case of complete and incomplete
spectroscopy, show biases $\Delta \langle z \rangle = \langle z\rangle_{\rm est} - \langle z\rangle_{\rm true} \leq
0.003$. In the case of both perfectly representative spectra and noiseless photometry, all but the first tomographic bin
shows both bias and scatter less than $0.001$; these entries we simply mark with a null symbol ($\emptyset \equiv
<10^{-3} \pm <10^{-3}$). Introducing incomplete spectroscopy causes the bias to increase very slightly, but nonetheless
remains at a level that we would consider negligible for current weak lensing surveys.} 

Once we add photometric noise, we see that the results degrade. {As a baseline for comparison, we include in Table
\ref{tab: reconstruction bias} the biases measured using the $k$NN association using our noisy and systematically incomplete spectroscopic
compilation. The $k$NN method returns a maximal biases of $\Delta \langle z \rangle = 0.047\pm0.005$, {or $\Delta \langle
z \rangle \leq 0.057$ at $97.5\%$ confidence}, in the first
tomographic bin. The highest tomographic bin exhibits bias of $\Delta \langle z \rangle = -0.013\pm0.004$, {or $\Delta
\langle z \rangle \geq -0.021$ at $97.5\%$ confidence}. 
While the observed $k$NN biases are non-negligible, it is worth noting that they agree well with the estimated redshift
uncertainties presented by \cite{hildebrandt/etal:2018}. They estimated the uncertainty on their $k$NN direct
calibration via a spatial bootstrap analysis, and found biases of $\sigma_{\Delta \langle z \rangle} \in
\{0.039,0.023,0.026,0.012,0.011\}$. Furthermore, the biases estimated for the $k$NN method are incoherent (\ie the sign of
the bias changes for different tomographic bins) thereby limiting the impact that they would have on cosmological conclusions.

For our updated SOM implementation, we find that the method is still largely unbiased
($\Delta \langle z \rangle \leq 0.006$) in the case of perfectly representative spectroscopic data, as is expected from
the theory of \cite{lima/etal:2008}. In the presence of biased spectroscopy, however, the biases increase to a value of 
$\Delta \langle z \rangle = 0.023\pm0.006$, {or $\Delta \langle z \rangle \leq 0.035$ at $97.5\%$ confidence}, in the
third tomographic bin. The magnitude of this bias remains unchanged when performing our redshift calibration with data
detected in all photometric bands, indicating that our treatment of non-detections is not the cause of this bias.
However we can leverage additional information, encoded by our new direct calibration, to improve these results and
minimise systematic bias. 

A primary strength of our direct calibration implementation is the ability to perform diagnostic and quality checks on
the resulting calibration. We can then perform some simple quality control checks on the spectroscopic-to-photometric
groupings  as a means of minimising
the bias introduced by photometric noise in our calibration. As a demonstration, we perform two sets of quality control
checks on our simulations with noisy photometry and systematically incomplete spectroscopy. First, we flag and remove
spectroscopic-to-photometric groupings which are catastrophic outliers in the distribution of photo-$z$ vs. SOM$z$ (\ie
the mean redshift of the SOM grouping): 
\begin{equation}
\frac{|\langle z_{\rm spec} \rangle - \langle Z_{\rm B} \rangle|}{
{\rm nMAD}(\langle z_{\rm spec} \rangle - \langle Z_{\rm B} \rangle)} > 5. 
\end{equation}
This quality cut (`QC1') effectively flags and removes regions of colour-colour space where template-fitting photo-$z$ and
machine learning photo-$z$ catastrophically disagree. This simple quality control step removes $\{2.0,0.3,2.4,0.6,0.1\}\%$ of the
photometric $n_{\rm eff}$ in each of the tomographic bins, and reduces the maximal bias to $\Delta \langle z \rangle =
0.013\pm0.006$, {or $\Delta \langle z \rangle \leq 0.025$ at $97.5\%$ confidence} (see Table \ref{tab:
reconstruction bias}). We can then apply additional, stricter, layers of quality cuts (`QC2') to further reduce the
bias. This layer of quality control flags and removes regions of colour-colour space where the average photo-$z$ of the
spectroscopic and photometric sources disagree: 
\begin{equation}
{|\langle Z_{\rm B} \rangle_{\rm spec} - \langle Z_{\rm B}\rangle_{\rm phot}|} > 0.02. 
\end{equation}
Such measures reduce the maximal bias to $\Delta \langle z \rangle = 0.009\pm0.004$ (now in the fourth tomographic bin),
but at the cost of decreased photometric $n_{\rm eff}$: $\Delta n_{\rm eff} = \{15.9,13.6,23.3,26.2,21.1\}\%$. 

These quality controls steps are not designed to be final, but are merely a demonstration of the refinement which is
possible using our updated direct calibration implementation. Determination of the best possible quality metrics should
ideally be performed on simulations beyond those presented here, which extend to higher redshifts. 
}


\subsection{Gold-Sample Tomographic Redshift Distributions for KV450}\label{sec: results III}
Here we present the application of the SOM direct photometric redshift calibration method to the KV450 dataset, and
derive tomographic redshift distributions for the KV450 gold-sample. Recall that the gold sample is the subset of the
full photometric sample $p$ which is represented by spectroscopic data, per tomographic bin, within our SOM ($p^\prime$,
such that $p^\prime \subseteq p$). {Importantly, for this gold selection we have also implemented the `QC1' quality
cuts described in Section \ref{sec: results II}.} As the gold sample is not the same set of photometric galaxies as was used in
previous KV450 cosmic shear analyses, we must note clearly that the gold-sample redshift distributions can not be
directly applied to these previous analyses. A re-analysis of KV450 cosmic shear using {various} SOM-defined `gold'
samples will be presented in a forthcoming paper.  

\begin{figure*}
\centering
\includegraphics[width=0.8\textwidth]{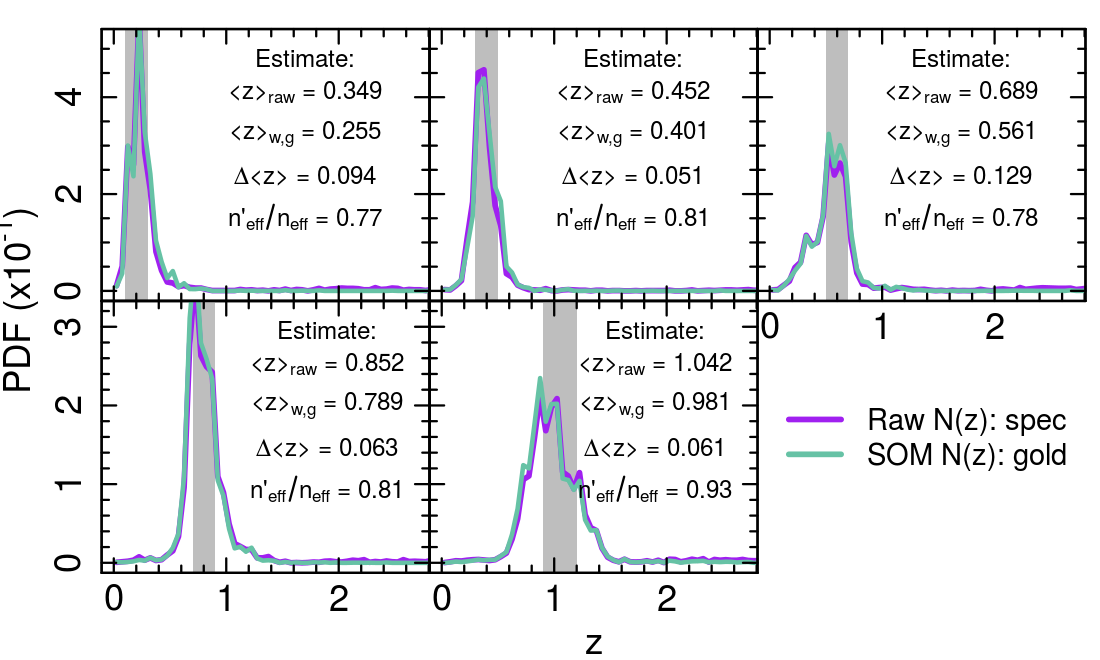}
\caption{Our new KV450 redshift distribution estimates, for the `gold' sample; a reduced photometric sample of galaxies
with $100\%$ representation in the spectroscopic sample, {and which satisfies the quality cuts `QC1'.} The figure
shows the reconstructed redshift distributions (green) alongside the purely tomographically binned spectroscopic data
(purple). The figure is annotated with the mean redshift estimates for the purely tomographically binned sample
($\langle z \rangle_{\rm raw}$) and the weighted gold sample ($\langle z \rangle_{\rm w,g}$), the difference that our
reweighting has had on the spectroscopic means ($\Delta \langle z \rangle$), and the fractional number of galaxies in
the gold sample compared to the original KV450 cosmic
shear sample $n^\prime_{\rm eff}/n_{\rm eff}$. }\label{fig: kv450 nz}
\end{figure*}

Figure \ref{fig: kv450 nz} shows the estimated gold-sample redshift distributions for KV450. Each panel shows one
tomographic bin (the tomographic selection is shown by the grey shaded region), and contains two lines. The first is 
the unweighted \Nz\ distribution of the tomographically binned spectroscopic sample (`raw'; purple). The second is the 
weighted \Nz\ estimate of the photometric gold-sample (`w,g'; green). The panels are each annotated with the raw and
weighted mean redshifts, $\langle z \rangle$, the difference between the two ($\Delta \langle z \rangle$), and the
fractional loss of galaxies in the gold sample, 
of $n^\prime_{\rm eff}/n_{\rm eff}$, for each bin (which were also shown in Table \ref{tab: coverage}). 

The redshift distributions demonstrate that the weighting shifts the distribution means by {between $|\Delta z| =
0.05$ and $|\Delta z| = 0.13$;} significant shifts for the case of KV450-like cosmic shear studies. {More importantly, we
note that shifts are coherent, as the raw redshift distributions are consistently at higher redshift than their
reweighted counterparts. This observation
simply demonstrates the importance of the reweighting process for cosmic shear studies; a simple null test. In the 
first two} tomographic bins the weighted PDF is less peaked than the raw, indicating that the reweighting is increasing
the significance of sources in the wings of the distributions. In the highest {four tomographic bins, however, the
reverse is true; the gold selection and reweighting truncates the wings of the distributions. }

\section{Summary}\label{sec: summary}
We present an updated implementation of direct redshift calibration, utilising unsupervised machine learning methods. We
verify, via a suite of simulations, the suitability of currently available spectroscopic datasets for direct
calibration, and subsequently the fidelity of the direct calibration methodology as a whole. Testing using both data
from the Kilo Degree Survey (KiDS) and simulated data from MICE2, we demonstrate via our SOMs that currently available
spectroscopic compilations are sufficiently complete for use in KiDS, representing $99\%$ of the effective 2D cosmic
shear sample. The representations decrease slightly when performing tomographic binning, to $\sim 84\%$ in the first
four tomographic bins, and to $94\%$ in the highest tomographic bin. {Calibration of stage-III cosmic shear surveys
with this archival spectroscopic dataset, however, would likely result in much lower levels of representation
(particularly at high-redshift).} We demonstrate using mock simulations of KiDS and the
spectroscopic training set that these measured representation fractions are robust to the effects of photometric noise,
sample variance, and spectroscopic incompleteness. We use this SOM-based selection to define a $100\%$ represented `gold’ cosmic shear
sample, per tomographic bin.  Using our mock simulations, we demonstrate that the mean redshift of the `gold’ sample
can be recovered by the SOM perfectly in the absence of photometric noise, agnostic to the effects of sample
variance and spectroscopic incompleteness. Photometric noise does not introduce bias when analysing perfectly
representative spectroscopic compilations. Under photometric noise and spectroscopic incompleteness, however, we find
maximal biases of {$\Delta \langle z \rangle = 0.023\pm0.006$, or $\Delta \langle z \rangle \leq 0.035$ at $97.5\%$
confidence.} The observed scatter in $\Delta \langle z \rangle$ in each tomographic bin, $\sigma_{\Delta \langle z \rangle} \leq 0.006$,
is found to be driven equally by photometric noise and spectroscopic selection effects. {
With limited quality control (which induces a reduction in the effective source number density, $\Delta
n_{\rm eff}$, of $\sim 5\%$) these biases can be mitigated, 
to the maximal bias of: $\Delta \langle z \rangle = 0.013\pm0.006$, or $\Delta \langle z \rangle \leq 0.025$ at
$97.5\%$ confidence. With more restrictive quality control ($\Delta n_{\rm eff} \sim 20\%$), the maximal bias can be
reduced to $\Delta \langle z \rangle = 0.009\pm0.004$, or $\Delta \langle z \rangle \leq 0.017$ at $97.5\%$ confidence. 
} Finally, we apply our new SOM photometric redshift calibration to the KiDS+VIKING-450 data, deriving `gold' redshift
distributions for use in future KV450 cosmological reanalyses.


\begin{acknowledgements}
We thank the anonymous referee for their comments, which have improved the clarity and content of this paper.
We acknowledge support from the European Research Council under grant numbers 770935 (AWH, HH, JvdB) and 647112 (CH).
HH is also supported by Heisenberg grants (Hi1495/5-1) of the Deutsche Forschungsgemeinschaft.  CH also acknowledges
support from the Max Planck Society and the Alexander von Humboldt Foundation in the framework of the Max
Planck-Humboldt Research Award endowed by the Federal Ministry of Education and Research. We thank C. Morrison for
useful discussions. This work is based on observations made with ESO Telescopes at the La Silla Paranal Observatory
under programme IDs 100.A-0613, 102.A-0047, 179.A-2004, 177.A-3016, 177.A-3017, 177.A-3018, 298.A-5015.  The MICE
simulations have been developed at the MareNostrum supercomputer (BSC-CNS) thanks to grants AECT-2006-2-0011 through
AECT-2015-1-0013. Data products have been stored at the Port d'Informació Científica (PIC), and distributed through the
CosmoHub webportal (cosmohub.pic.es).
\end{acknowledgements}


\bibpunct{(}{)}{;}{a}{}{,}
\bibliographystyle{aa}
\bibliography{library}

\newcommand{\noopsort}[1]{}
\begin{thebibliography}{63}
\expandafter\ifx\csname natexlab\endcsname\relax\def\natexlab#1{#1}\fi

\bibitem[{{Alarcon} {et~al.}(2019){Alarcon}, {S{\'a}nchez}, {Bernstein}, \&
  {Gazta{\~n}aga}}]{alarcon/etal:2019}
{Alarcon}, A., {S{\'a}nchez}, C., {Bernstein}, G.~M., \& {Gazta{\~n}aga}, E.
  2019, arXiv e-prints, arXiv:1910.07127

\bibitem[{{Amendola} {et~al.}(2018){Amendola}, {Appleby}, {Avgoustidis},
  {Bacon}, {Baker}, {Baldi}, {Bartolo}, {Blanchard}, {Bonvin}, {Borgani},
  {Branchini}, {Burrage}, {Camera}, {Carbone}, {Casarini}, {Cropper}, {de
  Rham}, {Dietrich}, {Di Porto}, {Durrer}, {Ealet}, {Ferreira}, {Finelli},
  {Garc{\'{\i}}a-Bellido}, {Giannantonio}, {Guzzo}, {Heavens}, {Heisenberg},
  {Heymans}, {Hoekstra}, {Hollenstein}, {Holmes}, {Hwang}, {Jahnke},
  {Kitching}, {Koivisto}, {Kunz}, {La Vacca}, {Linder}, {March}, {Marra},
  {Martins}, {Majerotto}, {Markovic}, {Marsh}, {Marulli}, {Massey}, {Mellier},
  {Montanari}, {Mota}, {Nunes}, {Percival}, {Pettorino}, {Porciani},
  {Quercellini}, {Read}, {Rinaldi}, {Sapone}, {Sawicki}, {Scaramella},
  {Skordis}, {Simpson}, {Taylor}, {Thomas}, {Trotta}, {Verde}, {Vernizzi},
  {Vollmer}, {Wang}, {Weller}, \& {Zlosnik}}]{amendola/etal:2018}
{Amendola}, L., {Appleby}, S., {Avgoustidis}, A., {et~al.} 2018, Living Reviews
  in Relativity, 21, 2

\bibitem[{{Asgari} {et~al.}(2019){Asgari}, {Tr{\"o}ster}, {Heymans},
  {Hildebrandt}, {van den Busch}, {Wright}, {Choi}, {Erben}, {Joachimi},
  {Joudaki}, {Kannawadi}, {Kuijken}, {Lin}, {Schneider}, \&
  {Zuntz}}]{asgari/etal:2019}
{Asgari}, M., {Tr{\"o}ster}, T., {Heymans}, C., {et~al.} 2019, A\&A submitted

\bibitem[{{Balestra} {et~al.}(2010){Balestra}, {Mainieri}, {Popesso},
  {Dickinson}, {Nonino}, {Rosati}, {Teimoorinia}, {Vanzella}, {Cristiani},
  {Cesarsky}, {Fosbury}, {Kuntschner}, \& {Rettura}}]{balestra/etal:2010}
{Balestra}, I., {Mainieri}, V., {Popesso}, P., {et~al.} 2010, \aap, 512, A12

\bibitem[{{Bielby} {et~al.}(2012){Bielby}, {Hudelot}, {McCracken}, {Ilbert},
  {Daddi}, {Le F{\`e}vre}, {Gonzalez-Perez}, {Kneib}, {Marmo}, {Mellier},
  {Salvato}, {Sanders}, \& {Willott}}]{bielby/etal:2012}
{Bielby}, R., {Hudelot}, P., {McCracken}, H.~J., {et~al.} 2012, \aap, 545, A23

\bibitem[{{Buchs} {et~al.}(2019){Buchs}, {Davis}, {Gruen}, {DeRose}, {Alarcon},
  {Bernstein}, {S{\'a}nchez}, {Myles}, {Roodman}, {Allen}, {Amon}, \&
  et~al.}]{buchs/etal:2019}
{Buchs}, R., {Davis}, C., {Gruen}, D., {et~al.} 2019, arXiv e-prints,
  arXiv:1901.05005

\bibitem[{{Carretero} {et~al.}(2015){Carretero}, {Castander}, {Gazta{\~n}aga},
  {Crocce}, \& {Fosalba}}]{Carretero/etal:2015}
{Carretero}, J., {Castander}, F.~J., {Gazta{\~n}aga}, E., {Crocce}, M., \&
  {Fosalba}, P. 2015, \mnras, 447, 646

\bibitem[{{Crocce} {et~al.}(2015){Crocce}, {Castander}, {Gazta{\~n}aga},
  {Fosalba}, \& {Carretero}}]{Crocce/etal:2015}
{Crocce}, M., {Castander}, F.~J., {Gazta{\~n}aga}, E., {Fosalba}, P., \&
  {Carretero}, J. 2015, \mnras, 453, 1513

\bibitem[{{Davidzon} {et~al.}(2019){Davidzon}, {Laigle}, {Capak}, {Ilbert},
  {Masters}, {Hemmati}, {Apostolakos}, {Coupon}, {de la Torre}, {Devriendt},
  {Dubois}, {Kashino}, {Paltani}, \& {Pichon}}]{davidzon/etal:2019}
{Davidzon}, I., {Laigle}, C., {Capak}, P.~L., {et~al.} 2019, arXiv e-prints,
  arXiv:1905.13233

\bibitem[{{de Jong} {et~al.}(2017){de Jong}, {Verdois Kleijn}, {Erben},
  {Hildebrandt}, {Kuijken}, {Sikkema}, {Brescia}, {Bilicki}, {Napolitano},
  {Amaro}, {Begeman}, {Boxhoorn}, {Buddelmeijer}, {Cavuoti}, {Getman}, {Grado},
  {Helmich}, {Huang}, {Irisarri}, {La Barbera}, {Longo}, {McFarland},
  {Nakajima}, {Paolillo}, {Puddu}, {Radovich}, {Rifatto}, {Tortora},
  {Valentijn}, {Vellucci}, {Vriend}, {Amon}, {Blake}, {Choi}, {Conti}, {Gwyn},
  {Herbonnet}, {Heymans}, {Hoekstra}, {Klaes}, {Merten}, {Miller}, {Schneider},
  \& {Viola}}]{dejong/etal:2017}
{de Jong}, J. T.~A., {Verdois Kleijn}, G.~A., {Erben}, T., {et~al.} 2017, \aap,
  604, A134

\bibitem[{{Edge} {et~al.}(2013){Edge}, {Sutherland}, {Kuijken}, {Driver},
  {McMahon}, {Eales}, \& {Emerson}}]{edge/etal:2013}
{Edge}, A., {Sutherland}, W., {Kuijken}, K., {et~al.} 2013, The Messenger, 154,
  32

\bibitem[{Everitt(1974)}]{everitt/1974}
Everitt, B. 1974, Cluster analysis / [by] Brian Everitt (Heinemann Educational
  [for] the Social Science Research Council London), vi, 122 p. :

\bibitem[{{Fosalba} {et~al.}(2015{\natexlab{a}}){Fosalba}, {Crocce},
  {Gazta{\~n}aga}, \& {Castand er}}]{Fosalba/etal:2015}
{Fosalba}, P., {Crocce}, M., {Gazta{\~n}aga}, E., \& {Castand er}, F.~J.
  2015{\natexlab{a}}, \mnras, 448, 2987

\bibitem[{{Fosalba} {et~al.}(2015{\natexlab{b}}){Fosalba}, {Gazta{\~n}aga},
  {Castander}, \& {Crocce}}]{Fosalba/etal:2015b}
{Fosalba}, P., {Gazta{\~n}aga}, E., {Castander}, F.~J., \& {Crocce}, M.
  2015{\natexlab{b}}, \mnras, 447, 1319

\bibitem[{Hartigan(1975)}]{hartigan/1975}
Hartigan, J.~A. 1975, Clustering Algorithms, 99th edn. (USA: John Wiley \&
  Sons, Inc.)

\bibitem[{{Heymans} {et~al.}(2012){Heymans}, {Van Waerbeke}, {Miller}, {Erben},
  {Hildebrandt}, {Hoekstra}, {Kitching}, {Mellier}, {Simon}, {Bonnett},
  {Coupon}, {Fu}, {Harnois D{\'e}raps}, {Hudson}, {Kilbinger}, {Kuijken},
  {Rowe}, {Schrabback}, {Semboloni}, {van Uitert}, {Vafaei}, \&
  {Velander}}]{heymans/etal:2012}
{Heymans}, C., {Van Waerbeke}, L., {Miller}, L., {et~al.} 2012, \mnras, 427,
  146

\bibitem[{{Hikage} {et~al.}(2018){Hikage}, {Oguri}, {Hamana}, {More},
  {Mandelbaum}, {Takada}, {K{\"o}hlinger}, {Miyatake}, {Nishizawa}, {Aihara},
  {Armstrong}, {Bosch}, {Coupon}, {Ducout}, {Hsieh}, {Komiyama}, {Lanusse},
  {Leauthaud}, {Medezinski}, {Mineo}, {Miyazaki}, {Murata}, {Murayama},
  {Shirasaki}, {Sif{\'o}n}, {Simet}, {Speagle}, {Spergel}, {Strauss},
  {Sugiyama}, {Tanaka}, \& {Wang}}]{hikage/etal:2018}
{Hikage}, C., {Oguri}, M., {Hamana}, T., {et~al.} 2018, ArXiv e-prints
  [\eprint[arXiv]{1809.09148}]

\bibitem[{{Hildebrandt} {et~al.}(2016){Hildebrandt}, {Choi}, {Heymans},
  {Blake}, {Erben}, {Miller}, {Nakajima}, {van Waerbeke}, {Viola},
  {Buddendiek}, {Harnois-D{\'e}raps}, {Hojjati}, {Joachimi}, {Joudaki},
  {Kitching}, {Wolf}, {Gwyn}, {Johnson}, {Kuijken}, {Sheikhbahaee}, {Tudorica},
  \& {Yee}}]{hildebrandt/etal:2016}
{Hildebrandt}, H., {Choi}, A., {Heymans}, C., {et~al.} 2016, \mnras, 463, 635

\bibitem[{{Hildebrandt} {et~al.}(2012){Hildebrandt}, {Erben}, {Kuijken}, {van
  Waerbeke}, {Heymans}, {Coupon}, {Benjamin}, {Bonnett}, {Fu}, {Hoekstra},
  {Kitching}, {Mellier}, {Miller}, {Velander}, {Hudson}, {Rowe}, {Schrabback},
  {Semboloni}, \& {Ben{\'{\i}}tez}}]{hildebrandt/etal:2012}
{Hildebrandt}, H., {Erben}, T., {Kuijken}, K., {et~al.} 2012, \mnras, 421, 2355

\bibitem[{{Hildebrandt} {et~al.}(2018){Hildebrandt}, {K{\"o}hlinger}, {van den
  Busch}, {Joachimi}, {Heymans}, {Kannawadi}, {Wright}, {Asgari}, {Blake},
  {Hoekstra}, {Joudaki}, {Kuijken}, {Miller}, {Morrison}, {Tr{\"o}ster},
  {Amon}, {Archidiacono}, {Brieden}, {Choi}, {de Jong}, {Erben}, {Giblin},
  {Mead}, {Peacock}, {Radovich}, {Schneider}, {Sif{\'o}n}, \&
  {Tewes}}]{hildebrandt/etal:2018}
{Hildebrandt}, H., {K{\"o}hlinger}, F., {van den Busch}, J.~L., {et~al.} 2018,
  arXiv e-prints, arXiv:1812.06076

\bibitem[{{Hildebrandt} {et~al.}(2017){Hildebrandt}, {Viola}, {Heymans},
  {Joudaki}, {Kuijken}, {Blake}, {Erben}, {Joachimi}, {Klaes}, {Miller},
  {Morrison}, {Nakajima}, {Verdoes Kleijn}, {Amon}, {Choi}, {Covone}, {de
  Jong}, {Dvornik}, {Fenech Conti}, {Grado}, {Harnois-D{\'e}raps}, {Herbonnet},
  {Hoekstra}, {K{\"o}hlinger}, {McFarland}, {Mead}, {Merten}, {Napolitano},
  {Peacock}, {Radovich}, {Schneider}, {Simon}, {Valentijn}, {van den Busch},
  {van Uitert}, \& {Van Waerbeke}}]{hildebrandt/etal:2017}
{Hildebrandt}, H., {Viola}, M., {Heymans}, C., {et~al.} 2017, \mnras, 465, 1454

\bibitem[{{Hoffmann} {et~al.}(2015){Hoffmann}, {Bel}, {Gazta{\~n}aga},
  {Crocce}, {Fosalba}, \& {Castander}}]{Hoffmann/etal:2015}
{Hoffmann}, K., {Bel}, J., {Gazta{\~n}aga}, E., {et~al.} 2015, \mnras, 447,
  1724

\bibitem[{{Hoyle} {et~al.}(2018){Hoyle}, {Gruen}, {Bernstein}, {Rau}, {De
  Vicente}, {Hartley}, {Gaztanaga}, {DeRose}, {Troxel}, {Davis}, {Alarcon},
  {MacCrann}, {Prat}, {S{\'a}nchez}, {Sheldon}, {Wechsler}, {Asorey}, {Becker},
  {Bonnett}, {Carnero Rosell}, {Carollo}, {Carrasco Kind}, {Castander},
  {Cawthon}, {Chang}, {Childress}, {Davis}, {Drlica-Wagner}, {Gatti},
  {Glazebrook}, {Gschwend}, {Hinton}, {Hoormann}, {Kim}, {King}, {Kuehn},
  {Lewis}, {Lidman}, {Lin}, {Macaulay}, {Maia}, {Martini}, {Mudd},
  {M{\"o}ller}, {Nichol}, {Ogando}, {Rollins}, {Roodman}, {Ross}, {Rozo},
  {Rykoff}, {Samuroff}, {Sevilla-Noarbe}, {Sharp}, {Sommer}, {Tucker}, {Uddin},
  {Varga}, {Vielzeuf}, {Yuan}, {Zhang}, {Abbott}, {Abdalla}, {Allam}, {Annis},
  {Bechtol}, {Benoit-L{\'e}vy}, {Bertin}, {Brooks}, {Buckley-Geer}, {Burke},
  {Busha}, {Capozzi}, {Carretero}, {Crocce}, {D'Andrea}, {da Costa}, {DePoy},
  {Desai}, {Diehl}, {Doel}, {Eifler}, {Estrada}, {Evrard}, {Fernandez},
  {Flaugher}, {Fosalba}, {Frieman}, {Garc{\'{\i}}a-Bellido}, {Gerdes},
  {Giannantonio}, {Goldstein}, {Gruendl}, {Gutierrez}, {Honscheid}, {James},
  {Jarvis}, {Jeltema}, {Johnson}, {Johnson}, {Kirk}, {Krause}, {Kuhlmann},
  {Kuropatkin}, {Lahav}, {Li}, {Lima}, {March}, {Marshall}, {Melchior},
  {Menanteau}, {Miquel}, {Nord}, {O'Neill}, {Plazas}, {Romer}, {Sako},
  {Sanchez}, {Santiago}, {Scarpine}, {Schindler}, {Schubnell}, {Smith},
  {Smith}, {Soares-Santos}, {Sobreira}, {Suchyta}, {Swanson}, {Tarle},
  {Thomas}, {Tucker}, {Vikram}, {Walker}, {Weller}, {Wester}, {Wolf}, {Yanny},
  {Zuntz}, \& {DES Collaboration}}]{hoyle/etal:2018}
{Hoyle}, B., {Gruen}, D., {Bernstein}, G.~M., {et~al.} 2018, \mnras, 478, 592

\bibitem[{{Hudelot} {et~al.}(2012){Hudelot}, {Cuillandre}, {Withington},
  {Goranova}, {McCracken}, {Magnard}, {Mellier}, {Regnault}, {Betoule},
  {Aussel}, {Kavelaars}, {Fernique}, {Bonnarel}, {Ochsenbein}, \&
  {Ilbert}}]{hudelot/etal:2012}
{Hudelot}, P., {Cuillandre}, J.-C., {Withington}, K., {et~al.} 2012, VizieR
  Online Data Catalog, 2317

\bibitem[{{Joudaki} {et~al.}(2019){Joudaki}, {Hildebrandt}, {Traykova},
  {Chisari}, {Heymans}, {Kannawadi}, {Kuijken}, {Wright}, {Asgari}, {Erben},
  {Hoekstra}, {Joachimi}, {Miller}, {Tr{\"o}ster}, \& {van den
  Busch}}]{joudaki/etal:2019}
{Joudaki}, S., {Hildebrandt}, H., {Traykova}, D., {et~al.} 2019, arXiv
  e-prints, arXiv:1906.09262

\bibitem[{{Joudaki} {et~al.}(2016){Joudaki}, {Mead}, {Blake}, {Choi}, {de
  Jong}, {Erben}, {Fenech Conti}, {Herbonnet}, {Heymans}, {Hildebrandt},
  {Hoekstra}, {Joachimi}, {Klaes}, {K{\"o}hlinger}, {Kuijken}, {McFarland},
  {Miller}, {Schneider}, \& {Viola}}]{joudaki/etal:2017b}
{Joudaki}, S., {Mead}, A., {Blake}, C., {et~al.} 2016, ArXiv e-prints
  [\eprint[arXiv]{1610.04606}]

\bibitem[{{Kafle} {et~al.}(2018){Kafle}, {Robotham}, {Driver}, {Deeley},
  {Norberg}, {Drinkwater}, \& {Davies}}]{kafle/etal:2018}
{Kafle}, P.~R., {Robotham}, A.~S.~G., {Driver}, S.~P., {et~al.} 2018, \mnras,
  479, 3746

\bibitem[{Kohonen(1982)}]{kohonen:1982}
Kohonen, T. 1982, Biological Cybernetics, 43, 59

\bibitem[{{Korytov} {et~al.}(2019){Korytov}, {Hearin}, {Kovacs}, {Larsen},
  {Rangel}, {Hollowed}, {Benson}, {Heitmann}, {Mao}, {Bahmanyar}, {Chang},
  {Campbell}, {DeRose}, {Finkel}, {Frontiere}, {Gawiser}, {Habib}, {Joachimi},
  {Lanusse}, {Li}, {Mandelbaum}, {Morrison}, {Newman}, {Pope}, {Rykoff},
  {Simet}, {To}, {Vikraman}, {Wechsler}, {White}, \& {(The LSST Dark Energy
  Science Collaboration}}]{korytov/etal:2019}
{Korytov}, D., {Hearin}, A., {Kovacs}, E., {et~al.} 2019, \apjs, 245, 26

\bibitem[{{Kuij\-ken}(2008)}]{kuijken:2008}
{Kuij\-ken}, K. 2008, \aap, 482, 1053

\bibitem[{{Kuij\-ken} {et~al.}(2015){Kuij\-ken}, {Heymans}, {Hildebrandt},
  {Nakajima}, {Erben}, {de Jong}, {Viola}, {Choi}, {Hoekstra}, {Miller}, {van
  Uitert}, {Amon}, {Blake}, {Brouwer}, {Buddendiek}, {Conti}, {Eriksen},
  {Grado}, {Harnois-D{\'e}raps}, {Helmich}, {Herbonnet}, {Irisarri},
  {Kitching}, {Klaes}, {La Barbera}, {Napolitano}, {Radovich}, {Schneider},
  {Sif{\'o}n}, {Sikkema}, {Simon}, {Tudorica}, {Valentijn}, {Verdoes Kleijn},
  \& {van Waerbeke}}]{kuijken/etal:2015}
{Kuij\-ken}, K., {Heymans}, C., {Hildebrandt}, H., {et~al.} 2015, \mnras, 454,
  3500

\bibitem[{{Kuijken} {et~al.}(2019){Kuijken}, {Heymans}, {Dvornik},
  {Hildebrandt}, {de Jong}, {Wright}, {Erben}, {Bilicki}, {Giblin}, {Shan},
  {Getman}, {Grado}, {Hoekstra}, {Miller}, {Napolitano}, {Paolilo}, {Radovich},
  {Schneider}, {Sutherland }, {Tewes}, {Tortora}, {Valentijn}, \& {Verdoes
  Kleijn}}]{kuijken/etal:2019}
{Kuijken}, K., {Heymans}, C., {Dvornik}, A., {et~al.} 2019, \aap, 625, A2

\bibitem[{{Laigle} {et~al.}(2019){Laigle}, {Davidzon}, {Ilbert}, {Devriendt},
  {Kashino}, {Pichon}, {Capak}, {Arnouts}, {de la Torre}, {Dubois},
  {Gozaliasl}, {Le Borgne}, {Lilly}, {McCracken}, {Salvato}, \&
  {Slyz}}]{laigle/etal:2019}
{Laigle}, C., {Davidzon}, I., {Ilbert}, O., {et~al.} 2019, \mnras, 486, 5104

\bibitem[{{Laigle} {et~al.}(2016){Laigle}, {McCracken}, {Ilbert}, {Hsieh},
  {Davidzon}, {Capak}, {Hasinger}, {Silverman}, {Pichon}, {Coupon}, {Aussel},
  {Le Borgne}, {Caputi}, {Cassata}, {Chang}, {Civano}, {Dunlop}, {Fynbo},
  {Kartaltepe}, {Koekemoer}, {Le F{\`e}vre}, {Le Floc'h}, {Leauthaud}, {Lilly},
  {Lin}, {Marchesi}, {Milvang-Jensen}, {Salvato}, {Sanders}, {Scoville},
  {Smolcic}, {Stockmann}, {Taniguchi}, {Tasca}, {Toft}, {Vaccari}, \&
  {Zabl}}]{laigle/etal:2016}
{Laigle}, C., {McCracken}, H.~J., {Ilbert}, O., {et~al.} 2016, \apjs, 224, 24

\bibitem[{{Laureijs} {et~al.}(2011){Laureijs}, {Amiaux}, {Arduini},
  {Augu{\`e}res}, {Brinchmann}, {Cole}, {Cropper}, {Dabin}, {Duvet}, {Ealet},
  \& et~al.}]{laureijs/etal:2011}
{Laureijs}, R., {Amiaux}, J., {Arduini}, S., {et~al.} 2011, ArXiv e-prints
  [\eprint[arXiv]{1110.3193}]

\bibitem[{{Le F{\`e}vre} {et~al.}(2013){Le F{\`e}vre}, {Cassata}, {Cucciati},
  {Garilli}, {Ilbert}, {Le Brun}, {Maccagni}, {Moreau}, {Scodeggio}, {Tresse},
  {Zamorani}, {Adami}, {Arnouts}, {Bardelli}, {Bolzonella}, {Bondi},
  {Bongiorno}, {Bottini}, {Cappi}, {Charlot}, {Ciliegi}, {Contini}, {de la
  Torre}, {Foucaud}, {Franzetti}, {Gavignaud}, {Guzzo}, {Iovino}, {Lemaux},
  {L{\'o}pez-Sanjuan}, {McCracken}, {Marano}, {Marinoni}, {Mazure}, {Mellier},
  {Merighi}, {Merluzzi}, {Paltani}, {Pell{\`o}}, {Pollo}, {Pozzetti},
  {Scaramella}, {Tasca}, {Vergani}, {Vettolani}, {Zanichelli}, \&
  {Zucca}}]{lefevre/etal:2013}
{Le F{\`e}vre}, O., {Cassata}, P., {Cucciati}, O., {et~al.} 2013, \aap, 559,
  A14

\bibitem[{{Lilly} {et~al.}(2009){Lilly}, {Le Brun}, {Maier}, {Mainieri},
  {Mignoli}, {Scodeggio}, {Zamorani}, {Carollo}, {Contini}, {Kneib}, {Le
  F{\`e}vre}, {Renzini}, {Bardelli}, {Bolzonella}, {Bongiorno}, {Caputi},
  {Coppa}, {Cucciati}, {de la Torre}, {de Ravel}, {Franzetti}, {Garilli},
  {Iovino}, {Kampczyk}, {Kovac}, {Knobel}, {Lamareille}, {Le Borgne}, {Pello},
  {Peng}, {P{\'e}rez-Montero}, {Ricciardelli}, {Silverman}, {Tanaka}, {Tasca},
  {Tresse}, {Vergani}, {Zucca}, {Ilbert}, {Salvato}, {Oesch}, {Abbas},
  {Bottini}, {Capak}, {Cappi}, {Cassata}, {Cimatti}, {Elvis}, {Fumana},
  {Guzzo}, {Hasinger}, {Koekemoer}, {Leauthaud}, {Maccagni}, {Marinoni},
  {McCracken}, {Memeo}, {Meneux}, {Porciani}, {Pozzetti}, {Sanders},
  {Scaramella}, {Scarlata}, {Scoville}, {Shopbell}, \&
  {Taniguchi}}]{lilly/etal:2009}
{Lilly}, S.~J., {Le Brun}, V., {Maier}, C., {et~al.} 2009, \apjs, 184, 218

\bibitem[{{Lima} {et~al.}(2008){Lima}, {Cunha}, {Oyaizu}, {Frieman}, {Lin}, \&
  {Sheldon}}]{lima/etal:2008}
{Lima}, M., {Cunha}, C.~E., {Oyaizu}, H., {et~al.} 2008, \mnras, 390, 118

\bibitem[{{Masters} {et~al.}(2015){Masters}, {Capak}, {Stern}, {Ilbert},
  {Salvato}, {Schmidt}, {Longo}, {Rhodes}, {Paltani}, {Mobasher}, {Hoekstra},
  {Hildebrandt}, {Coupon}, {Steinhardt}, {Speagle}, {Faisst}, {Kalinich},
  {Brodwin}, {Brescia}, \& {Cavuoti}}]{masters/etal:2015}
{Masters}, D., {Capak}, P., {Stern}, D., {et~al.} 2015, \apj, 813, 53

\bibitem[{{Masters} {et~al.}(2017){Masters}, {Stern}, {Cohen}, {Capak},
  {Rhodes}, {Castander}, \& {Paltani}}]{masters/etal:2017}
{Masters}, D.~C., {Stern}, D.~K., {Cohen}, J.~G., {et~al.} 2017, \apj, 841, 111

\bibitem[{{Masters} {et~al.}(2019){Masters}, {Stern}, {Cohen}, {Capak},
  {Stanford}, {Hernitschek}, {Galametz}, {Davidzon}, {Rhodes}, {Sand ers},
  {Mobasher}, {Castander}, {Pruett}, \& {Fotopoulou}}]{masters/etal:2019}
{Masters}, D.~C., {Stern}, D.~K., {Cohen}, J.~G., {et~al.} 2019, \apj, 877, 81

\bibitem[{{McCracken} {et~al.}(2012){McCracken}, {Milvang-Jensen}, {Dunlop},
  {Franx}, {Fynbo}, {Le F{\`e}vre}, {Holt}, {Caputi}, {Goranova}, {Buitrago},
  {Emerson}, {Freudling}, {Hudelot}, {L{\'o}pez-Sanjuan}, {Magnard}, {Mellier},
  {M{\o}ller}, {Nilsson}, {Sutherland}, {Tasca}, \&
  {Zabl}}]{mccracken/etal:2012}
{McCracken}, H.~J., {Milvang-Jensen}, B., {Dunlop}, J., {et~al.} 2012, \aap,
  544, A156

\bibitem[{{McQuinn} \& {White}(2013)}]{mcquinn/etal:2013}
{McQuinn}, M. \& {White}, M. 2013, \mnras, 433, 2857

\bibitem[{{Miller} {et~al.}(2007){Miller}, {Kitching}, {Heymans}, {Heavens}, \&
  {van Waerbeke}}]{miller/etal:2007}
{Miller}, L., {Kitching}, T.~D., {Heymans}, C., {Heavens}, A.~F., \& {van
  Waerbeke}, L. 2007, \mnras, 382, 315

\bibitem[{{Morrison} {et~al.}(2017){Morrison}, {Hildebrandt}, {Schmidt},
  {Baldry}, {Bilicki}, {Choi}, {Erben}, \& {Schneider}}]{morrison/etal:2017}
{Morrison}, C.~B., {Hildebrandt}, H., {Schmidt}, S.~J., {et~al.} 2017, \mnras,
  467, 3576

\bibitem[{{Naim} {et~al.}(1997){Naim}, {Ratnatunga}, \&
  {Griffiths}}]{naim/etal:1997}
{Naim}, A., {Ratnatunga}, K.~U., \& {Griffiths}, R.~E. 1997, arXiv e-prints,
  astro

\bibitem[{{Newman}(2008)}]{newman:2008}
{Newman}, J.~A. 2008, \apj, 684, 88

\bibitem[{{Newman} {et~al.}(2013){Newman}, {Cooper}, {Davis}, {Faber}, {Coil},
  {Guhathakurta}, {Koo}, {Phillips}, {Conroy}, {Dutton}, {Finkbeiner}, {Gerke},
  {Rosario}, {Weiner}, {Willmer}, {Yan}, {Harker}, {Kassin}, {Konidaris},
  {Lai}, {Madgwick}, {Noeske}, {Wirth}, {Connolly}, {Kaiser}, {Kirby},
  {Lemaux}, {Lin}, {Lotz}, {Luppino}, {Marinoni}, {Matthews}, {Metevier}, \&
  {Schiavon}}]{newman/etal:2013}
{Newman}, J.~A., {Cooper}, M.~C., {Davis}, M., {et~al.} 2013, \apjs, 208, 5

\bibitem[{{Planck Collaboration} {et~al.}(2018){Planck Collaboration},
  {Aghanim}, {Akrami}, {Ashdown}, {Aumont}, {Baccigalupi}, {Ballardini},
  {Banday}, {Barreiro}, {Bartolo}, {Basak}, {Battye}, {Benabed}, {Bernard},
  {Bersanelli}, {Bielewicz}, {Bock}, {Bond}, {Borrill}, {Bouchet}, {Boulanger},
  {Bucher}, {Burigana}, {Butler}, {Calabrese}, {Cardoso}, {Carron},
  {Challinor}, {Chiang}, {Chluba}, {Colombo}, {Combet}, {Contreras}, {Crill},
  {Cuttaia}, {de Bernardis}, {de Zotti}, {Delabrouille}, {Delouis}, {Di
  Valentino}, {Diego}, {Dor{\'e}}, {Douspis}, {Ducout}, {Dupac}, {Dusini},
  {Efstathiou}, {Elsner}, {En{\ss}lin}, {Eriksen}, {Fantaye}, {Farhang},
  {Fergusson}, {Fernandez-Cobos}, {Finelli}, {Forastieri}, {Frailis},
  {Franceschi}, {Frolov}, {Galeotta}, {Galli}, {Ganga}, {G{\'e}nova-Santos},
  {Gerbino}, {Ghosh}, {Gonz{\'a}lez-Nuevo}, {G{\'o}rski}, {Gratton},
  {Gruppuso}, {Gudmundsson}, {Hamann}, {Handley}, {Herranz}, {Hivon}, {Huang},
  {Jaffe}, {Jones}, {Karakci}, {Keih{\"a}nen}, {Keskitalo}, {Kiiveri}, {Kim},
  {Kisner}, {Knox}, {Krachmalnicoff}, {Kunz}, {Kurki-Suonio}, {Lagache},
  {Lamarre}, {Lasenby}, {Lattanzi}, {Lawrence}, {Le Jeune}, {Lemos},
  {Lesgourgues}, {Levrier}, {Lewis}, {Liguori}, {Lilje}, {Lilley}, {Lindholm},
  {L{\'o}pez-Caniego}, {Lubin}, {Ma}, {Mac{\'{\i}}as-P{\'e}rez}, {Maggio},
  {Maino}, {Mandolesi}, {Mangilli}, {Marcos-Caballero}, {Maris}, {Martin},
  {Martinelli}, {Mart{\'{\i}}nez-Gonz{\'a}lez}, {Matarrese}, {Mauri}, {McEwen},
  {Meinhold}, {Melchiorri}, {Mennella}, {Migliaccio}, {Millea}, {Mitra},
  {Miville-Desch{\^e}nes}, {Molinari}, {Montier}, {Morgante}, {Moss}, {Natoli},
  {N{\o}rgaard-Nielsen}, {Pagano}, {Paoletti}, {Partridge}, {Patanchon},
  {Peiris}, {Perrotta}, {Pettorino}, {Piacentini}, {Polastri}, {Polenta},
  {Puget}, {Rachen}, {Reinecke}, {Remazeilles}, {Renzi}, {Rocha}, {Rosset},
  {Roudier}, {Rubi{\~n}o-Mart{\'{\i}}n}, {Ruiz-Granados}, {Salvati}, {Sandri},
  {Savelainen}, {Scott}, {Shellard}, {Sirignano}, {Sirri}, {Spencer},
  {Sunyaev}, {Suur-Uski}, {Tauber}, {Tavagnacco}, {Tenti}, {Toffolatti},
  {Tomasi}, {Trombetti}, {Valenziano}, {Valiviita}, {Van Tent}, {Vibert},
  {Vielva}, {Villa}, {Vittorio}, {Wandelt}, {Wehus}, {White}, {White},
  {Zacchei}, \& {Zonca}}]{planck/cosmo:2018}
{Planck Collaboration}, {Aghanim}, N., {Akrami}, Y., {et~al.} 2018, ArXiv
  e-prints [\eprint[arXiv]{1807.06209}]

\bibitem[{{Popesso} {et~al.}(2009){Popesso}, {Dickinson}, {Nonino}, {Vanzella},
  {Daddi}, {Fosbury}, {Kuntschner}, {Mainieri}, {Cristiani}, {Cesarsky},
  {Giavalisco}, {Renzini}, \& {GOODS Team}}]{popesso/etal:2009}
{Popesso}, P., {Dickinson}, M., {Nonino}, M., {et~al.} 2009, \aap, 494, 443

\bibitem[{{R Core Team}(2015)}]{R}
{R Core Team}. 2015, R: A Language and Environment for Statistical Computing, R
  Foundation for Statistical Computing, Vienna, Austria

\bibitem[{{Schneider} {et~al.}(2006){Schneider}, {Knox}, {Zhan}, \&
  {Connolly}}]{schneider/etal:2006}
{Schneider}, M., {Knox}, L., {Zhan}, H., \& {Connolly}, A. 2006, \apj, 651, 14

\bibitem[{{Straatman} {et~al.}(2018){Straatman}, {van der Wel}, {Bezanson},
  {Pacifici}, {Gallazzi}, {Wu}, {Noeske}, {Bari{\v{s}}i{\'c}}, {Bell},
  {Brammer}, {Calhau}, {Chauke}, {Franx}, {van Houdt}, {Labb{\'e}}, {Maseda},
  {Mu{\~n}oz-Mateos}, {Muzzin}, {van de Sande}, {Sobral}, \&
  {Spilker}}]{straatman/etal:2018}
{Straatman}, C. M.~S., {van der Wel}, A., {Bezanson}, R., {et~al.} 2018, \apjs,
  239, 27

\bibitem[{{Tanaka} {et~al.}(2018){Tanaka}, {Coupon}, {Hsieh}, {Mineo},
  {Nishizawa}, {Speagle}, {Furusawa}, {Miyazaki}, \&
  {Murayama}}]{tanaka/etal:2018}
{Tanaka}, M., {Coupon}, J., {Hsieh}, B.-C., {et~al.} 2018, \pasj, 70, S9

\bibitem[{{Troxel} {et~al.}(2018{\natexlab{a}}){Troxel}, {Krause}, {Chang},
  {Eifler}, {Friedrich}, {Gruen}, {MacCrann}, {Chen}, {Davis}, {DeRose},
  {Dodelson}, {Gatti}, {Hoyle}, {Huterer}, {Jarvis}, {Lacasa}, {Lemos},
  {Peiris}, {Prat}, {Samuroff}, {S{\'a}nchez}, {Sheldon}, {Vielzeuf}, {Wang},
  {Zuntz}, {Lahav}, {Abdalla}, {Allam}, {Annis}, {Avila}, {Bertin}, {Brooks},
  {Burke}, {Carnero Rosell}, {Carrasco Kind}, {Carretero}, {Crocce}, {Cunha},
  {D'Andrea}, {da Costa}, {De Vicente}, {Diehl}, {Doel}, {Evrard}, {Flaugher},
  {Fosalba}, {Frieman}, {Garc{\'\i}a-Bellido}, {Gaztanaga}, {Gerdes},
  {Gruendl}, {Gschwend}, {Gutierrez}, {Hartley}, {Hollowood}, {Honscheid},
  {James}, {Kirk}, {Kuehn}, {Kuropatkin}, {Li}, {Lima}, {March}, {Menanteau},
  {Miquel}, {Mohr}, {Ogando}, {Plazas}, {Roodman}, {Sanchez}, {Scarpine},
  {Schindler}, {Sevilla-Noarbe}, {Smith}, {Soares-Santos}, {Sobreira},
  {Suchyta}, {Swanson}, {Thomas}, {Walker}, \& {Wechsler}}]{troxel/etal:2018b}
{Troxel}, M.~A., {Krause}, E., {Chang}, C., {et~al.} 2018{\natexlab{a}},
  \mnras, 479, 4998

\bibitem[{{Troxel} {et~al.}(2018{\natexlab{b}}){Troxel}, {MacCrann}, {Zuntz},
  {Eifler}, {Krause}, {Dodelson}, {Gruen}, {Blazek}, {Friedrich}, {Samuroff},
  {Prat}, {Secco}, {Davis}, {Fert{\'e}}, {DeRose}, {Alarcon}, {Amara},
  {Baxter}, {Becker}, {Bernstein}, {Bridle}, {Cawthon}, {Chang}, {Choi}, {De
  Vicente}, {Drlica-Wagner}, {Elvin-Poole}, {Frieman}, {Gatti}, {Hartley},
  {Honscheid}, {Hoyle}, {Huff}, {Huterer}, {Jain}, {Jarvis}, {Kacprzak},
  {Kirk}, {Kokron}, {Krawiec}, {Lahav}, {Liddle}, {Peacock}, {Rau},
  {Refregier}, {Rollins}, {Rozo}, {Rykoff}, {S{\'a}nchez}, {Sevilla-Noarbe},
  {Sheldon}, {Stebbins}, {Varga}, {Vielzeuf}, {Wang}, {Wechsler}, {Yanny},
  {Abbott}, {Abdalla}, {Allam}, {Annis}, {Bechtol}, {Benoit-L{\'e}vy},
  {Bertin}, {Brooks}, {Buckley-Geer}, {Burke}, {Carnero Rosell}, {Carrasco
  Kind}, {Carretero}, {Castander}, {Crocce}, {Cunha}, {D'Andrea}, {da Costa},
  {DePoy}, {Desai}, {Diehl}, {Dietrich}, {Doel}, {Fernandez}, {Flaugher},
  {Fosalba}, {Garc{\'{\i}}a-Bellido}, {Gaztanaga}, {Gerdes}, {Giannantonio},
  {Goldstein}, {Gruendl}, {Gschwend}, {Gutierrez}, {James}, {Jeltema},
  {Johnson}, {Johnson}, {Kent}, {Kuehn}, {Kuhlmann}, {Kuropatkin}, {Li},
  {Lima}, {Lin}, {Maia}, {March}, {Marshall}, {Martini}, {Melchior},
  {Menanteau}, {Miquel}, {Mohr}, {Neilsen}, {Nichol}, {Nord}, {Petravick},
  {Plazas}, {Romer}, {Roodman}, {Sako}, {Sanchez}, {Scarpine}, {Schindler},
  {Schubnell}, {Smith}, {Smith}, {Soares-Santos}, {Sobreira}, {Suchyta},
  {Swanson}, {Tarle}, {Thomas}, {Tucker}, {Vikram}, {Walker}, {Weller},
  {Zhang}, \& {DES Collaboration}}]{troxel/etal:2018}
{Troxel}, M.~A., {MacCrann}, N., {Zuntz}, J., {et~al.} 2018{\natexlab{b}},
  \prd, 98, 043528

\bibitem[{{van den Busch} {et~al.}(in prep.){van den Busch}, {Hildebrandt},
  {Wright}, \& et~al.}]{vandenbusch/etal:2019}
{van den Busch}, J., {Hildebrandt}, H., {Wright}, A.~H., \& et~al. in prep.

\bibitem[{{Vanzella} {et~al.}(2008){Vanzella}, {Cristiani}, {Dickinson},
  {Giavalisco}, {Kuntschner}, {Haase}, {Nonino}, {Rosati}, {Cesarsky},
  {Ferguson}, {Fosbury}, {Grazian}, {Moustakas}, {Rettura}, {Popesso},
  {Renzini}, {Stern}, \& {GOODS Team}}]{vanzella/etal:2008}
{Vanzella}, E., {Cristiani}, S., {Dickinson}, M., {et~al.} 2008, \aap, 478, 83

\bibitem[{{Venemans} {et~al.}(2015){Venemans}, {Verdoes Kleijn}, {Mwebaze},
  {Valentijn}, {Ba{\~n}ados}, {Decarli}, {de Jong}, {Findlay}, {Kuijken},
  {Barbera}, {McFarland}, {McMahon}, {Napolitano}, {Sikkema}, \&
  {Sutherland}}]{venemans/etal:2015}
{Venemans}, B.~P., {Verdoes Kleijn}, G.~A., {Mwebaze}, J., {et~al.} 2015,
  \mnras, 453, 2259

\bibitem[{{Ward}(1963)}]{ward/1963}
{Ward}, J. H.~J. 1963, Journal of the American Statistical Association, 58, 236

\bibitem[{Wehrens \& Buydens(2007)}]{wehrens/lutgarde:2007}
Wehrens, R. \& Buydens, L. M.~C. 2007, Journal of Statistical Software, 21, 1

\bibitem[{Wehrens \& Kruisselbrink(2018)}]{wehrens/kruisselbrink:2018}
Wehrens, R. \& Kruisselbrink, J. 2018, Journal of Statistical Software, 87, 1

\bibitem[{{Wright} {et~al.}(2018){Wright}, {Hildebrandt}, {Kuijken}, {Erben},
  {Blake}, {Buddelmeijer}, {Choi}, {Cross}, {de Jong}, {Edge},
  {Gonzalez-Fernandez}, {Gonz{\'a}lez Solares}, {Grado}, {Heymans}, {Irwin},
  {Kupcu Yoldas}, {Lewis}, {Mann}, {Napolitano}, {Radovich}, {Schneider},
  {Sif{\'o}n}, {Sutherland}, {Sutorius}, \& {Verdoes
  Kleijn}}]{wright/etal:2018b}
{Wright}, A.~H., {Hildebrandt}, H., {Kuijken}, K., {et~al.} 2018, arXiv
  e-prints, arXiv:1812.06077

\end{thebibliography}

\clearpage
\appendix 

\section{SOM Implementation}\label{sec: som implementation}
In this Appendix, we discuss the influence that the choice of SOM construction parameters and training data has on our
estimates of the spectroscopic representation of our dataset, discussed in Sect. \ref{sec: results I kv450}, using our
MICE2 simulations.  

\subsection{SOM construction parameters}
Here we detail the various SOM construction parameters that we must consider in order to generate a SOM 
for use in redshift calibration. There are undoubtedly more optional modifications that one can make to a SOM which 
are not discussed here, however we endeavour here to describe (albeit briefly) the parameters that are required for SOM
generation. We quantify the impact of the choice of each of these parameters in terms of the SOM spectroscopic
representation, which we quantify using the effective number density of the cosmic shear sample $n_{\rm eff}$ (Eqn.
\ref{eqn: neff}).

\subsubsection{Dimensionality \& Topology}
Two key choices related to SOM construction are those of the adopted  
dimensionality and topology. Dimensionality refers to jointly to the number and aspect ratio of individual cells that
make up the final manifold which we then project onto 2D. Topology refers to the choices regarding how those cells are
spread throughout the manifold. 

When considering dimensionality, previous studies have varied in their choices as to the optimum dimension that can
and/or should be used. In \cite{masters/etal:2015}, they implement a rectangular SOM with dimension $75\times105$
{ cells}, arguing that the asymmetry in the SOM manifold gives preferential direction to the principal manifold
component and thus improves convergence. Conversely, recent work by \cite{davidzon/etal:2019} implements a 
square SOM with dimension $80\times80$ {cells}. They present a simple method for determining their 
optimum SOM dimension using, jointly: the fraction of cells with significant representation (i.e. many
galaxies per cell) within the SOM, and the spread in the data about each weight vector relative to the source
photometric uncertainties. They optimise their SOM configuration using these parameters and conclude that the
$80\times80$ {cell} SOM is optimum for their application. 

We also investigate both the symmetric and asymmetric SOM construction cases. Unlike \cite{davidzon/etal:2019}, however,
we do not implement a range of SOMs with similar aspect ratio and different {cell} numbers. Instead, we use two
dimensionalities ($101\times101$ and $75\times150$), which we believe will exceed the maximum desirable fragmentation of
the data (for our purposes). We then implement a hierarchical clustering of the SOM {cells} to produce $n$ distinct
groupings of the data. The importance of the cluster number $n$ is discussed briefly below (Sect. \ref{sec: clust}) and
at length in Appendix \ref{sec: Clustering Results}. We argue that this mode of analysis is preferential to using the
SOM {cells} themselves to optimise the number of galaxies per group \citep[as was done in ][]{davidzon/etal:2019}, as this
separates the two somewhat different issues of optimising galaxy grouping and allowing the SOM due flexibility to
accurately map the manifold. 

Compounding the matter further, in addition to the raw number of {cells} in each SOM dimension, there are also
questions regarding what {cell} shape and surface topology is best for the SOM. {cell} shape becomes particularly relevant
in dense areas of the manifold, where different {cell} shapes can cause data to be differently distributed in the final
SOM (and thus impacting our grouping of like-with-like data). SOM topology, conversely, is most influenced in the 
sparser areas of the manifold. The choice of topology is typically either flat or toroidal; that is the edges of the SOM
manifold are either free or reconnect to form a continuous surface, respectively. 

Within our SOM implementation using the {\tt kohonen} package \citep{wehrens/kruisselbrink:2018,wehrens/lutgarde:2007},
it is trivial to generate SOMs (and SuperSOMs, where multiple layers are used) with arbitrary dimensionality, using
rectangular or hexagonal {cells}, and across a flat or toroidal manifold. We can therefore test the
influence of these construction choices on our final SOMs. 
\subsubsection{Clustering}\label{sec: clust}
On top of performing the training of the SOM using a particular dimensionality and topology, one can then refine
the grouping of the data within the SOM by using a hierarchical clustering on the final SOM optimisation tree. 
In this way, a high-resolution SOM grid can be adaptively lowered in its resolution after training, in an effort to
test the influence of the overall clustering to the conclusions. 

The choice of SOM clustering is particularly important to our SOM direct redshift calibration, as it dictates the number
of discrete weights which are available to reweight the spectroscopic data. In addition, the coarseness (or fineness) of
the SOM clustering will influence the width of the per-cluster \Nz\ distributions, possibly inducing biases if the
clustering level is too small. 

Due to its importance, we dedicate Appendix \ref{sec: Clustering Results} to the exploration of SOM clustering, and its
influence on our results. We find though, that the total number of clusters is irrelevant to our results beyond roughly  
$2000$ (see Appendix \ref{sec: Clustering Results}). 
For all tests in this work we use cluster numbers defined on the KV450 data, as described in Appendix \ref{sec:
Clustering Results}). 
\subsubsection{Training data}
Finally, the training data itself is naturally of great importance to the fidelity of the SOM in terms of spectroscopic
representation and redshift calibration. This includes the question of what
information (from a given sample) is relevant to parse to the SOM for training. Given 9-band photometric
data, the options are many-fold; we test a range of combinations of colours and magnitudes: 
\begin{itemize}
\item 9 magnitudes (0:9); 
\item 8 colours (8:0); 
\item 8 colours and 1 magnitude (8:1);  
\item 37 colours and 1 magnitude (37:1); or
\item 37 colours and 9 magnitudes (37:9).   
\end{itemize}
Each of these options could provide more information to the SOM, allowing the calibration to improve, or it may 
add too-much redundancy to the dataset and dilute the maps ability to extract colour-redshift information. 

In addition to which information to parse to the SOM during training, there is also an important question regarding
which data to use for the training; what sample do we parse for training, and how does the sample's cosmic variance,
photometric noise, and object selection influence the results. To investigate the importance of the sample itself we can
construct KV450-like spectroscopic samples, samples which are missing large and unique parts of the colour-redshift
space, and/or perfectly representative spectroscopic compilations. 
\subsection{Parameter Selection and Uncertainties}\label{sec: som opt results}
Determination of the SOM parameters is, at least to some degree, arbitrary. That is, we may make some
assumptions about the nature of the high-dimensional manifold and use this to influence our choices of dimension,
topology, clustering, and training data; but due to our inability to visualise greater than 3-dimensional space, 
these assumptions will always be imperfect. Indeed, this is part of the reason SOMs were developed in the first
instance. 

We extend the methodology of \cite{davidzon/etal:2019} to determine an optimal SOM construction, 
marginalising over arbitrarily chosen parameters in our SOM construction to produce an estimate of the 
method-induced uncertainty on our estimates of spectroscopic representation and redshift calibration accuracy.  
We perform a 2-stage process to a) determine the SOM parameters that are most
suitable for our analysis, and b) quantify and incorporate the uncertainty on our analysis introduced by these
somewhat arbitrary choices. To this end, we perform the SOM construction for our MICE2 simulations 
using the full grid of SOM options; namely:
\begin{itemize}
\item Dimensionality: 75x150 or 101x101; 
\item Topology: Toroidal or non-toroidal;  
\item {Cell} structure: Hexagonal or rectangular; 
\item Iteration steps: 100 or 1000;  
\item Training parameters: 0:9, 37:1, or 37:9; and 
\item Training data: KV450-like, noDEEP2, or perfectly sampled. 
\end{itemize}
We choose these options as we believe that they span the full gambit of possible options from the simplest 
(a square, non-toroidal SOM with rectangular {cells}, trained on 9-band photometry from a perfectly representative
spectroscopic dataset) to the most complex (a rectangular, toroidal SOM with hexagonal {cells}, trained on 37 colours and
9 magnitudes for a heavily biased spectroscopic sample).  
This creates a sample of $144$ individual SOM setups, which we run for each of our $5$ tomographic bins. 
Again, in order to maintain consistency between the different dimensionalities, we run these tests with the number of 
clusters set at $10,000$. 
These options are all equally valid. It is therefore our hope that the resulting 
redshift distributions from each of these SOMs are largely consistent. Or, inverting the situation, we can use this 
grid of optional and arbitrary parameter choices to generate an estimate of the uncertainty imprinted on the final 
SOM calibration related to our choice of parameters; essentially exploring the uncertainty on the SOM \Nz\ when 
marginalising over the arbitrarily selected SOM construction parameters. 

To quantify the influence of the SOM construction parameters, we use the measured representation of the spectral
catalogue, quantified using the $n^\prime_{\rm eff}/n_{\rm eff}$ statistic from Sect. \ref{sec: results I kv450}.  This
indicates the level to which the photometric data are represented by the spectroscopic catalogue.  

\begin{figure}
\centering
\includegraphics[width=\columnwidth]{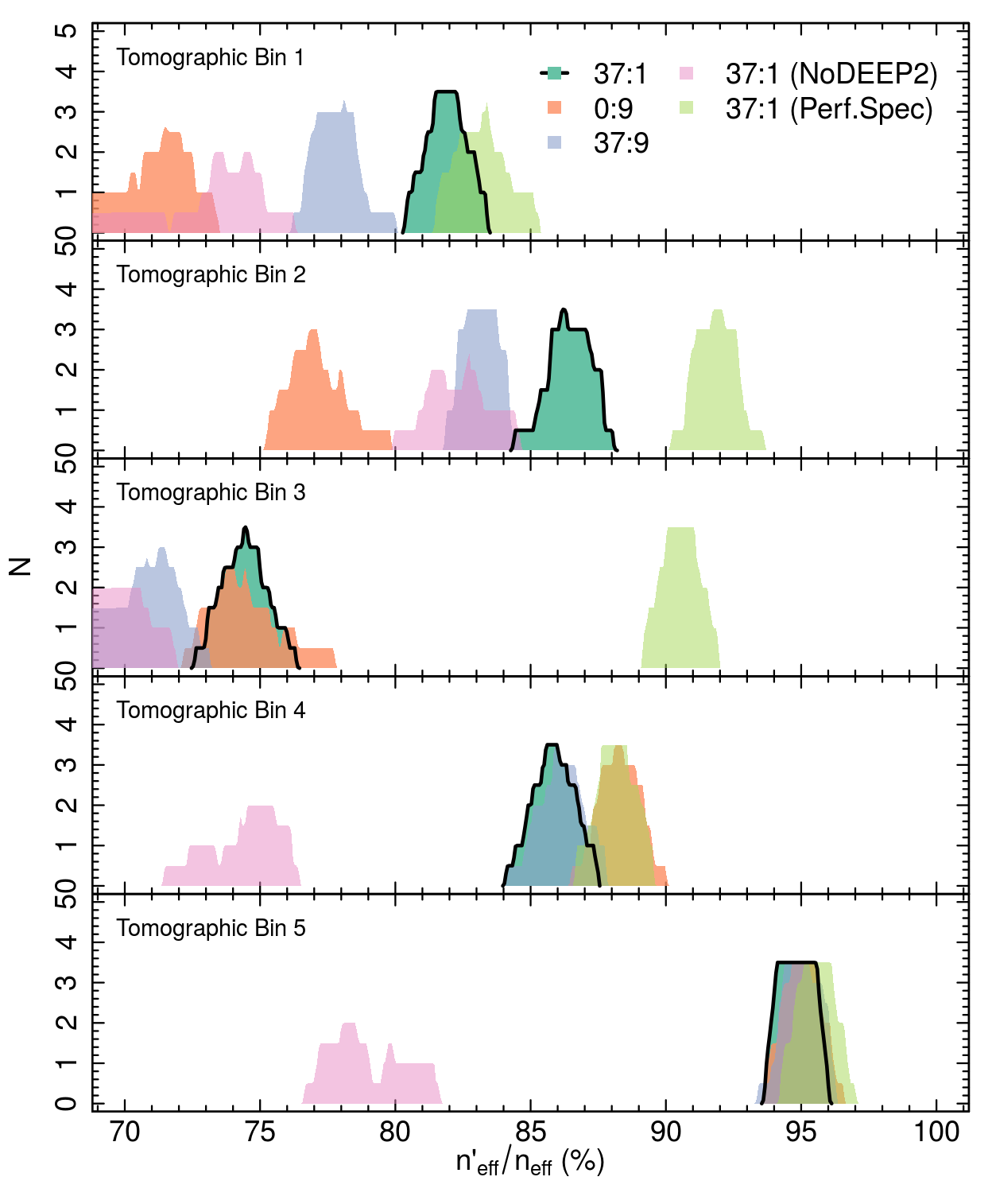}
\caption{The change in the effective number density of the MICE2 cosmic shear sample $n_{\rm eff}^\prime / n_{\rm eff}$, 
caused by the choice of SOM construction parameters and training samples.  Each panel
shows one tomographic bin from a single realisation of the MICE2 mock KiDS dataset. We separate, in particular, 5 option
selections that cause the most significant change to the observed representation in the SOMs: 3 different sets of
training inputs, and 3 different spectroscopic dataset constructions.  The legend indicates the spectroscopic
samples used (KV450-like fiducial setup, without DEEP2, and using a perfect sampling of the photometric data) and/or the
input training data (\#colours:\#magnitudes).  The remaining $16$ SOM constructions within each of these subsets are
shown as the variously coloured histograms. We can see that the construction of the SOM induces a $\sim$percent level
uncertainty on the representation fraction $n^\prime_{\rm eff}/n_{\rm eff}$; the results presented throughout the paper
are therefore robust to the construction of our SOM.   
}\label{fig: option pdfs}
\end{figure}

Figure \ref{fig: option pdfs} shows the resulting values of $n^\prime_{\rm eff}/n_{\rm eff}$ for our grid of SOM
construction parameters, grouped into 5 individual (dominant) factors: the choice of input training
parameters, and of training data. 

The choice of training parameters has a noticeable effect on the observed representation fractions, as the differing
photometric information allow the SOM to learn more about the nature of SEDs. While one may naively expect the SOM to
have equal success when provided with either  9 magnitudes or 37 colours and 1 magnitude (i.e. all magnitude
combinations), this is infact not the case. This can be seen in Figure \ref{fig: option pdfs} where, for example, the
first tomographic bin representation ($85\%$ for the fiducial 37:1 training) has a considerably lower spectroscopic
representation when trained on purely magnitude information alone (i.e. no colours; $76\%$). The interpretation of this
is that, when providing the SOM with purely magnitude based training data, the SOM must learn itself that it is the
magnitude combinations (i.e. the colours) which are important discriminants of the data. This learning appears to be
facilitated better when the SOM is provided with all of the colours directly,
especially in the lowest tomographic bins. The redundant information still has benefits in the training there.
Interestingly, however, the addition of the raw magnitudes back into the training set with the full combination of
colours has little effect (compare the 37:1 and 37:9 histograms in Figure \ref{fig: option pdfs}); the information from
the colours alone appears to saturate for KiDS-like data. 

Finally, we look at the influence of the different training samples. While discussed at length in Sect. \ref{sec:
results I kv450}, we note here that the conclusions we make about the representation of individual spectroscopic subsets
(such as zCOSMOS, DEEP2, and VVDS) are not impacted by the construction of our SOM. This can be seen by the noDEEP2 and
perfect spectroscopic compilation histograms in Figure \ref{fig: option pdfs}. They show that the construction only
causes a $\sim 1\%$ uncertainty on the overall representation statistics, regardless of the choice of training sample.
Said differently, we conclude that the results presented in Sect. \ref{sec: results I kv450} are robust to the choices
made in our SOM construction. 

\section{Influence of Cluster Number}\label{sec: Clustering Results}
With our SOM-based direct photometric redshift calibration method (Sect. \ref{sec: som method}), the weights which are applied
to each spectroscopic galaxy are estimated using the groupings of self-similar photometric and spectroscopic data, as
determined by the SOM. This creates a fundamental link between the number of groupings made and the flexibility of our
method to reweight the spectroscopic data: increasing the number of groups allows for a more flexible weighting scheme.
However there is also a counter-effect: with increased fragmentation of the datasets comes increased fractions of the
photometric dataset which are no longer represented by the spectroscopy. These lost photometric sources decrease the
signal-to-noise of cosmic shear, negatively impacting cosmic shear science. Therefore, a careful consideration of the interplay between
the grouping of galaxies and redshift distribution estimation is crucial.  In this appendix, we explore the interplay
between the number of SOM groupings, the accuracy of the calibrated redshift distributions, and the number of sources
which are represented by spectroscopy and therefore make it into our photometric `gold sample'. 

First, we discuss the process of grouping galaxies within the SOM. 
In the simplest case, the grouping of galaxies can be determined by the SOM {cells} themselves. This necessitates
adapting the number of SOM {cells} to the point where one is happy with the typical number of sources per group; such a
process places a conflict between group number and SOM resolution. 

SOM resolution is important in determining the ability of the SOM to map complex areas of the hyper-surface accurately.
Therefore, instead of restricting the accuracy of the SOM surface mapping for the sake of increasing the number of
galaxies per grouping, it would be optimal to generate the surface mapping using a high-dimensional grid and then
group these SOM {cells} together post-facto, to produce the desired number of galaxies per grouping. 

This is the process that we invoke here. {To do this we utilize the native ``hclust'' 
function within {\sc R} \citep{R}. This function performs a hierarchical cluster analysis using a set
of dissimilarities for the $n$ objects being clustered. We opt to use a simple Euclidean distance matrix between SOM codebook
vectors (\ie the set of $n$ vectors which jointly describe where a {cell} sits in
$n$-dimensions) as our clustering basis. Initially, each object is assigned to its own cluster and then the algorithm proceeds
iteratively, at each stage joining the two most similar clusters, and continuing until there is just a single cluster.  At
each stage distances between clusters are recomputed by the Lance-Williams dissimilarity update formula according to the
particular clustering method being used. A number of different clustering methods are provided. We implement the
complete linkage method of clustering, which finds similar clusters. This is preferable (for our purposes) compared to,
for example, Ward's minimum variance method \citep[which aims at finding compact, spherical clusters in N
dimensions;][]{ward/1963} or the
single linkage method (which is closely related to the minimal spanning tree and adopts a ‘friends of friends’
clustering strategy). Details of these algorithms and their implementation can be found in
\cite{everitt/1974,hartigan/1975}, and in the extensive {\sc R} documentation.} 
Using this clustering methodology, can group a SOM consisting of $\sim10000$ {cells} into $3000$ groupings, each with a
different number of component {cells} (and therefore galaxies). We are thus able to generate arbitrary numbers of
groupings without sacrificing SOM resolution.

\begin{figure}
\centering
\includegraphics[width=\columnwidth]{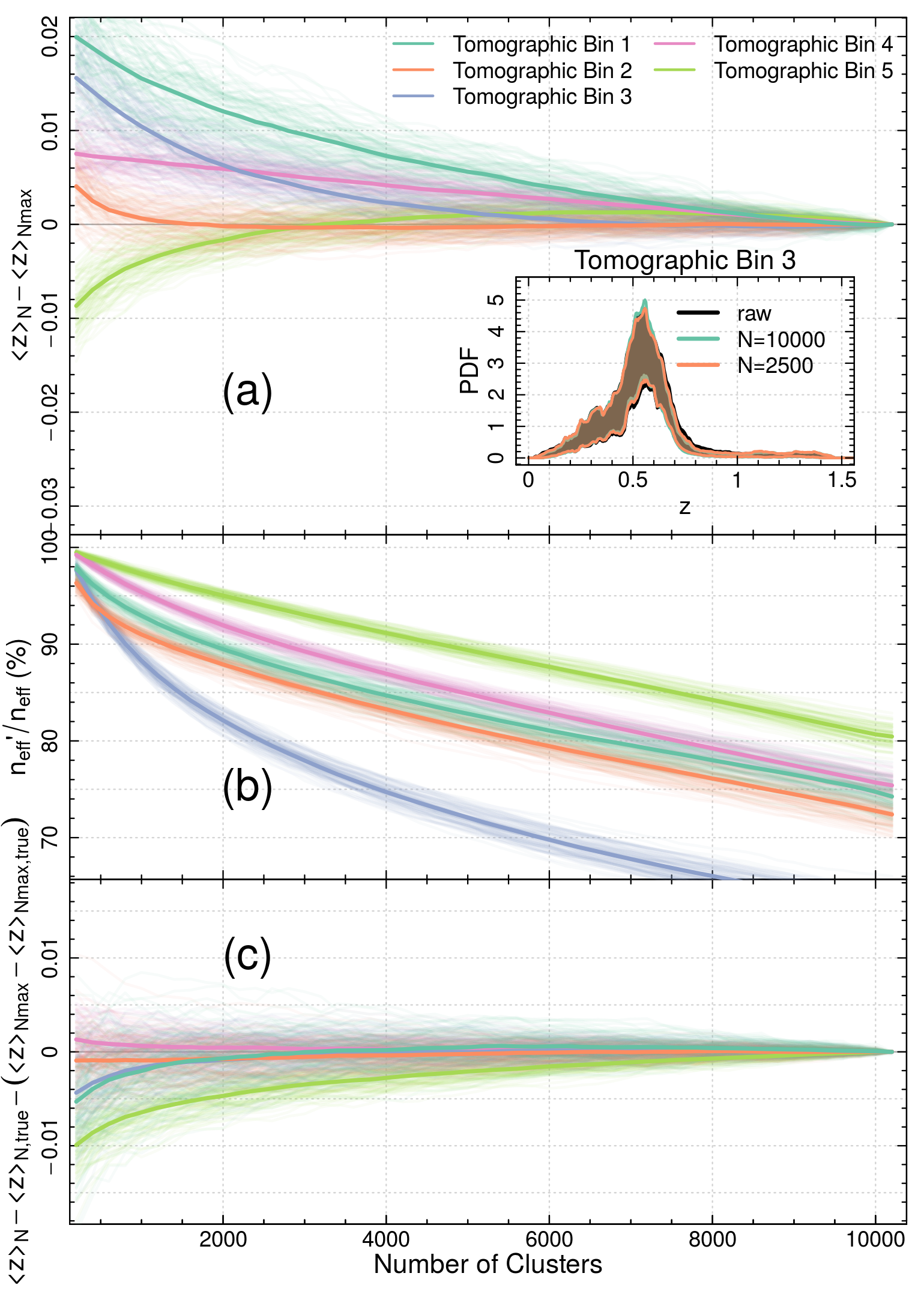}
\caption{The influence of the number of SOM clusters on the resulting tomographic bin redshifts and the recovered
photometric misrepresentation for the MICE2 simulation dataset. We show three statistics as a function of cluster
number. {\em Panel a:} the change in the tomographic mean estimate. {\em Panel b:} the change in the
representation of the spectroscopic catalogue. {\em Panel c:} the change in the tomographic mean bias.  
The inset of panel (a) shows the distribution of PDFs estimated for the third tomographic bin with $10000$ and $2500$
clusters. The black curve shows the distribution of the unweighted spec-z sample.  
Panel (a) demonstrates that, with increasing cluster number, our estimate of the tomographic redshift distribution
mean changes significantly. The middle panel demonstrates why: with more clusters there is more fragmentation of the photometric
data and significant decreases in the spectroscopic representation, such that the `gold' sample decreases in size. 
However, crucially, panel (c) demonstrates
that the loss of representation does not change the low level of bias in the recovered redshift distribution. 
}\label{fig: MICE2 clusters} 
\end{figure}

Figure \ref{fig: MICE2 clusters} demonstrates the influence of the number of SOM hierarchical clusters on the resulting tomographic
bin redshift means, the recovered photometric misrepresentation, and the redshift bin biases for the $100$ line-of-sight
realisations of our 
MICE2 simulated dataset. We compare the mean of the estimated tomographic
redshift distribution with the final (most highly clustered) estimate, after the removal of sources that are classified
as being unrepresented in the spectroscopic compilation (panel a). This allows us to explore how the mean estimates
themselves are influenced by the choice of cluster number. Photometric misrepresentation is defined in using the same
$n_{\rm eff}$ statistics as used in Sect. \ref{sec: results I kv450} (panel b). Finally the redshift distribution biases
(panel c) are measured by comparing the estimated and mean true redshift distributions for each number of clusters. 
We show the bias of the individual redshift distribution estimates also after subtracting off the residual bias 
from the highest cluster number estimate ($\langle z \rangle_{{\rm max}}$). Again, this allows us to remove the effects of noise and
sample variance, which induce a maximal scatter and bias of $\sigma_{\Delta\langle z \rangle}\sim 0.006$ and $|\Delta\langle z \rangle|\leq0.01$
respectively (see Sect. \ref{sec: results II}).

Panel (a) of Fig. \ref{fig: MICE2 clusters} shows that the mean estimates appear quite sensitive to the choice of cluster number.
The mean estimate for tomographic bin 1 varies by $|\langle z \rangle_{n}-\langle z \rangle_{{\rm max}}| \approx 0.02$ over the probed range of
clusters. Simultaneously, as can be seen in panel (b), the representation is also changing significantly. This is an
indication of the loss of (typically high-redshift) sources that we experience when we overly fragment our SOM
groupings. While this will considerably impact the cosmic shear signal for the final gold sample, it is important to 
recognise that these two effects, however, work in tandem to ensure that the final {\em bias} on each of the tomographic
bin means remains negligible over essentially the entire range of cluster choices (panel c). 

Overall, Fig. \ref{fig: MICE2 clusters} demonstrates that, beyond the regime where the clusters are large (and so the
reweighting flexibility is small), the accuracy of the tomographic redshift calibration is essentially insensitive to
the choice of the number of clusters. This is certainly not the case for the representation statistics, however, and
therefore which we seek to maximise for our goal of cosmic shear science. 

We typically observe a joint decrease in representation and bin average redshift with increasing cluster number. 
This indicates that the data being removed are preferentially at high-redshift, likely being expelled due to
fluctuations in photometric noise; the spectroscopy scatter between SOM cells where the similarly-noisy photometric data
reside. 

This loss of high-redshift sources is relevant for both cosmic shear science and the representation statistics presented
in Sects. \ref{sec: results I kv450} and \ref{sec: results I mice2}. We can see from panel (b) of Fig. \ref{fig: MICE2
clusters} that, if we were to use only $10$ clusters, we would infer that the spectroscopic catalogue was perfectly
representative. However naturally this is a false interpretation, because hidden within this clustering are groups of
{ cells}
with vastly different true \Nz\ distributions. Similarly, using the maximal cluster number will produce a very high
fidelity gold sample, and a very accurate calibrated redshift distribution. However this distribution will be severely
lacking in the highest-redshift sources; those which carry the most cosmic shear signal. 
We must therefore decide on the criteria to determine which number of clusters is adequate/necessary for the correct
description of the data, without causing sources to be unnecessarily lost from our gold sample. 

\begin{figure}
\centering
\includegraphics[scale=0.5]{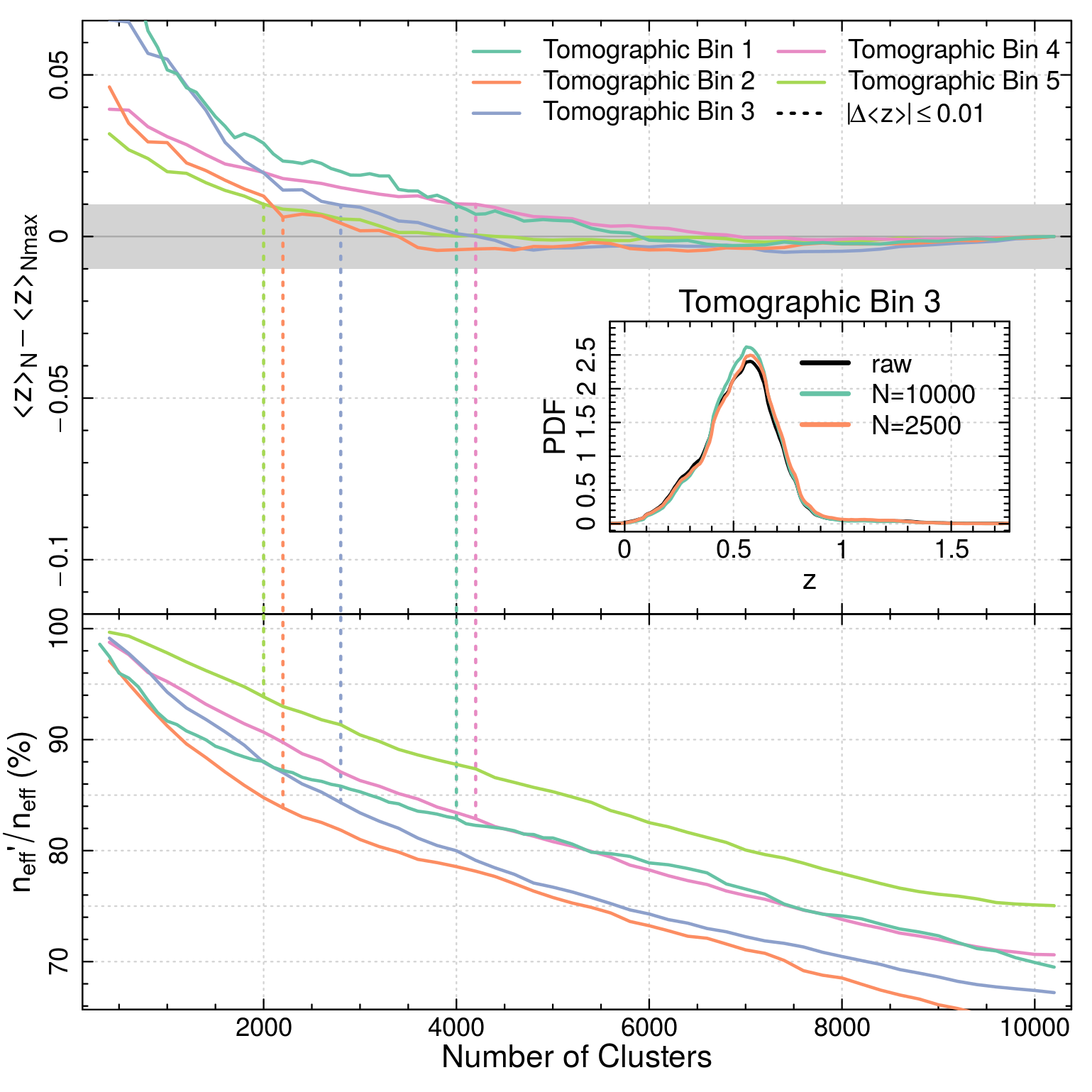}
\caption{The influence of the number of SOM clusters on the resulting tomographic bin redshifts and the recovered photometric 
misrepresentation for the full KV450 photometric and spectroscopic datasets. The figure is constructed in the same way
as Fig. \protect \ref{fig: MICE2 clusters}. Here we additionally show the point at which the tomographic bin mean
redshifts cross the $|\langle z \rangle_{n}-\langle z \rangle_{{\rm max}}| \leq 0.01$ boundary (shown with grey shading). This criteria
designates our per-tomographic-bin cluster-number choice for this work. }\label{fig: KV450 clusters}
\end{figure}

We can tie this decision back to our overall goal of photometric redshift calibration. Using the
simulations, we can determine the point below which there are insufficient clusters to accurately recover the true
redshift distribution, and use this as our cluster number for the simulations. 
Indeed, such an estimate would require only a simple threshold to be applied in panel (c) of Fig. \ref{fig: MICE2
clusters}.  However naturally such a determination cannot be made on the data (as the truth is not known) and the
application of values estimated with the simulations to the data may risk bias: we know that the MICE2 simulations
truncate abruptly at $z=1.4$, whereas the universe thankfully does not. Instead, we can define the clustering using a
quantity that is measurable on both the simulations and the data: the change in the mean redshift (for a particular
number of clusters $n$; $\langle z \rangle_{n}$) compared to the maximally clustered data (\ie\ where every {cell} is an individual
cluster; $\langle z \rangle_{{\rm max}}$). This is the statistic shown in panel (a) of Fig. \ref{fig: MICE2 clusters}, and now we
also compute this statistic for the real KV450 data in Fig.  \ref{fig: KV450 clusters}. Using this figure, we can apply
a simple threshold and determine the acceptable number of clusters to use in our representation estimation. We opt to
use a value of $|\langle z \rangle_{n}-\langle z \rangle_{{\rm max}}| \leq 0.01$, and the resulting cluster numbers are shown using the dashed
lines in the figure.  Figure \ref{fig: KV450 clusters} shows that the mean estimates do vary, and the strong function of
representativeness as a function of cluster number is also evident in data, as with the simulations. Again, though, it
is important to note that this value of $|\langle z \rangle_{n}-\langle z \rangle_{{\rm max}}| \leq 0.01$ is not equivalent to a bias of
$|\Delta \langle z \rangle| \leq 0.01$, as was shown in Fig. \ref{fig: MICE2 clusters}, as the `gold' sample for the maximally
clustered case is very different from the optimal sample. 

We use Fig. \ref{fig: KV450 clusters} to define the cluster number which we use for the entirety of this work, for both
simulations and data, per tomographic bin. We determine the cluster number which satisfies $|\langle z
\rangle_{n}-\langle z \rangle_{{\rm
max}}| \leq 0.01$, and annotate these points as dashed lines within the figure. We round these points to the nearest
$100$ clusters, and in this way define the number of clusters used to determine the representation statistics for the
KV450 spectroscopic compilation, and determine which sources make up our photometric `gold sample'.  The numbers for
bins one to five are, respectively: $n_{\rm clust} = {6200,2700,3500,4400,2200}$.

\section{Spectroscopic vs Photometric Training}\label{sec: spec training}
Previous works such as the C3R2 project aim to determine the sampling of spectroscopic data by constructing a SOM 
trained on the photometric data; essentially performing the reverse of our implemented method. The C3R2 project 
find a considerable number of cells (in Euclid-like data) which are lacking spectroscopic information. This 
raises the concern that our training on spectroscopic data may miss bubbles of $n$-dimensional colour space, 
and that photometric data in these bubbles would then be randomly scattered throughout our SOM, creating biases. 

This is a fair concern, and is grounded in the idea that the SOM is unable to learn about parts of colour-colour 
space that are simply not present in the spectroscopic training data. Such an effect will be successfully mitigated 
in our methodology provided that the SOM is able to correctly partition sources from untrained parts of colour-colour 
space into specific {cells}; that the SOM has learned that there is a gap in its training set. We can test the robustness
the influence of our choice of training in two ways. Firstly, we can simply perform our training using the photometric
dataset, assess the quality of the resulting SOM, and estimate the robustness of our results to this alternative
training. Secondly, we can perform our fiducial training on a spectroscopic compilation that we know is catastrophically
under-representative in colour-redshift space, and determine the robustness of our results in this case. 

\subsection{Photometric Training}
We test how our method performs when training on the photometric dataset, rather than the spectroscopic. All other
aspects of the process are the same (\ie\ the bands used, SOM construction, cluster determination, etc). 
We start by comparing the suitability of the photometrically trained SOM for use in our redshift distribution
calibration method by exploring firstly the per-cluster \Nz\ distributions, and secondly the coverage statistics
described in Sect. \ref{sec: results I kv450}.   

A primary assumption of our SOM photometric redshift calibration method is that the \Nz\ distributions assigned
to each group within the spectroscopic catalogue be narrow; ideally delta-function like. We can estimate the typical
spread of each cluster \Nz\ using the distribution of nMAD values for the spec-$z$ within each cluster. 
In our spectroscopically trained SOM, the $75^{\rm th}$ percentile of the cluster \Nz\ width
distribution (per tomographic bin) is $\Delta z_{\rm clust}/(1+z_{\rm clust}) \in \{0.043,0.048,0.060,0.048,0.086\}$ (we
show the $75^{\rm th}$ percentile as the median nMAD is 0 for the first two tomographic bins). 
This indicates that the majority of {cells} (agnostic to the photometric representation therein) have typical spreads of 
$\Delta z_{\rm clust}/(1+z_{\rm clust}) \lesssim 0.08$ in all tomographic bins. 
Comparatively, the photometrically trained SOM has an average cluster \Nz\ width of 
$\Delta z_{\rm clust} / (1+z_{\rm clust}) \in \{0.050,0.040,0.075,0.055,0.093\}$; $1-2 \%$ poorer than the spectroscopic training in
all but the second bin. The same conclusion holds when weighting the clusters by their contribution to the tomographic
effective number density (\ie\ when weighting each cluster nMAD estimate by the photometric shear-measurement weights). 

This demonstrates that, for the purposes of our redshift distribution calibration, the spectroscopic training yields
groupings that more explicitly trace the colour-redshift relation than our photometric training. Examination of the
photometric representation statistics gives an indication as to why this may be the case. We compute the representation
statistic $n^\prime_{\rm eff}/n_{\rm eff}$ (shown in Table \ref{tab: coverage}), but now for
our SOM trained on the photometric data. The most dramatic change seen is in the fraction of {cells} which contain
spectra ($f_{\rm pix}$). In our spectroscopically trained SOM this was $f_{\rm pix} = 91.9\%$. For the photometrically
trained SOM this value drops significantly, to $f_{\rm pix} = 66.9\%$. However this is not accompanied by a
corresponding catastrophic drop in the overall representation of the photometry: $n^\prime_{\rm eff}/n_{\rm eff} = 94.1 \%$ for our
photometrically trained SOM. This indicates that considerable area ($>20\%$) within our photometrically trained SOM is being
allocated to photometric data which contribute just $6\%$ of the shear measurement weight. As a result, the
spectroscopic data are assigned to fewer SOM {cells} and therefore have greater per-cluster widths. 
Otherwise, the photometrically trained SOM demonstrates the same trends as seen in Table \ref{tab: coverage}; DEEP2
remains the most unique and influential of our spectroscopic subsamples, and different samples dominate
the information contained in different tomographic bins. Finally, the representation of the photometric data is poorer
in the first three tomographic bins when training on the photometric data, $n^\prime_{\rm eff}/n_{\rm eff} \in
{75.6,71.6,78.6}\%$, with the second tomographic bin being the most heavily affected. In the highest two tomographic
bins, however, the representation statistics remain unchanged under photometric training; $n^\prime_{\rm eff}/n_{\rm
eff} \in {82.3,93.7} \%$. 

\subsection{Catastrophically unrepresentative training} 
We can further test whether the SOM methodology is robust to the influence of such a 
catastrophic absence of data in the training sample. The simplest way to do this is to construct two trained SOMs: 
one based on the full spectroscopic compilation, and one based on the spectroscopic compilation without DEEP2 (our 
most influential dataset). We have already demonstrated that DEEP2 occupies a unique area of the colour-manifold (see 
Sect. \ref{sec: results I kv450}), and so we know that the SOM trained without DEEP2 will be missing a large, unique, part of 
the colour-colour training set. We can then populate each of these SOMs with the full spectroscopic and photometric
datasets, and determine:
\begin{itemize}
\item how the photometric misrepresentation of the full photometric dataset differs between the two SOMs; and 
\item how the DEEP2 data are distributed within the two different SOMs. 
\end{itemize}

\begin{figure*}
\centering
\includegraphics[scale=0.7,trim=0 30 0 50,clip]{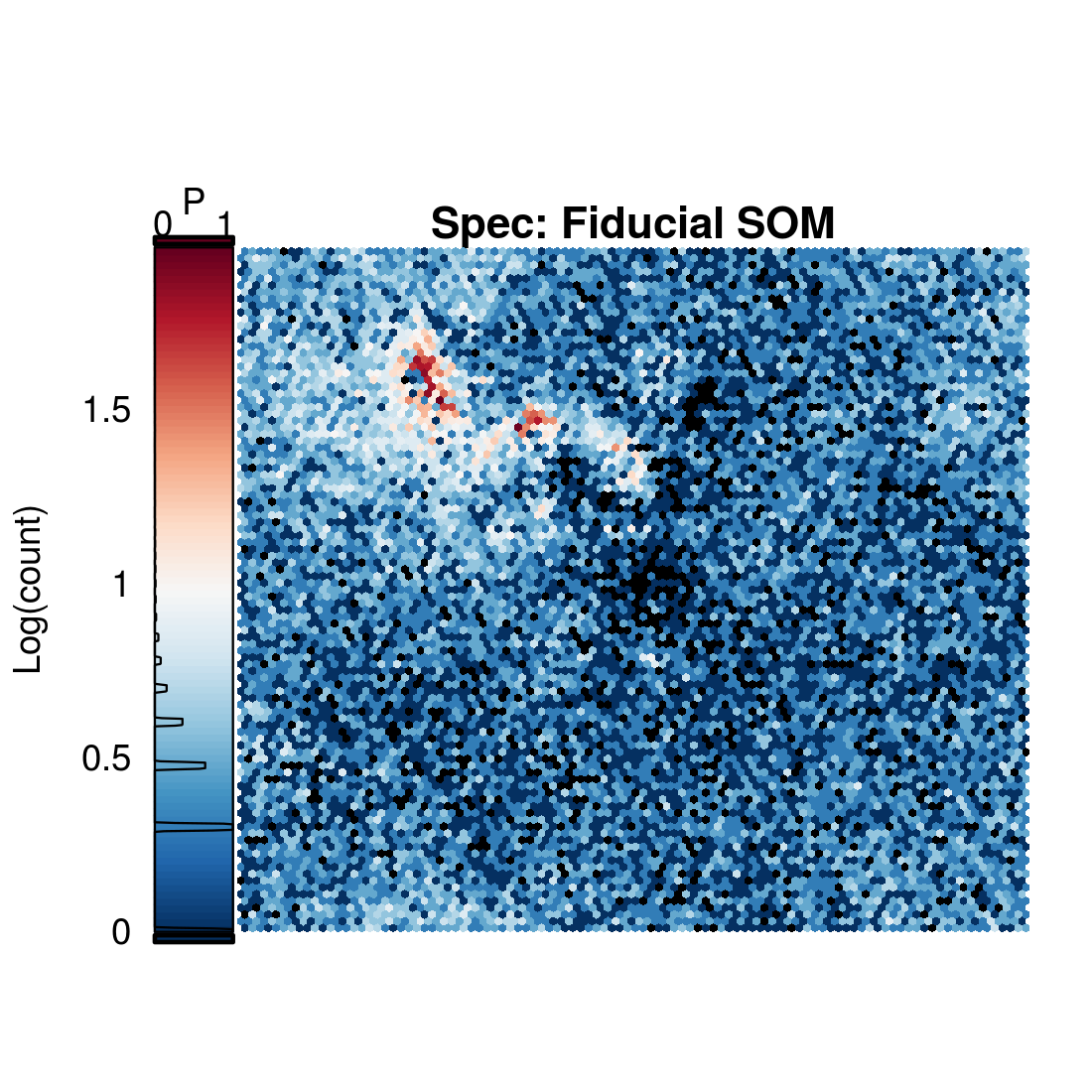}\includegraphics[scale=0.7,trim=0 30 0 50,clip]{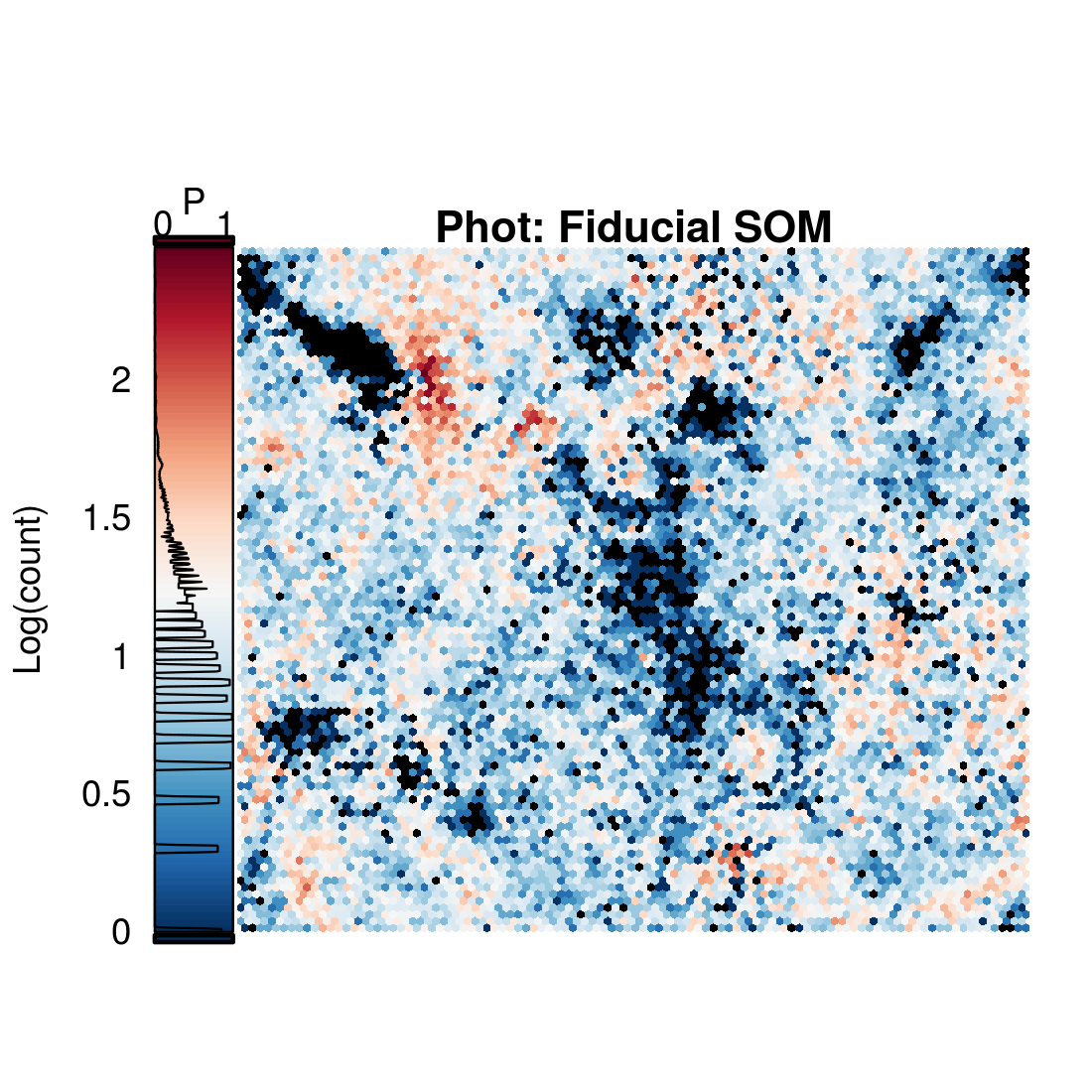}
\includegraphics[scale=0.7,trim=0 30 0 50,clip]{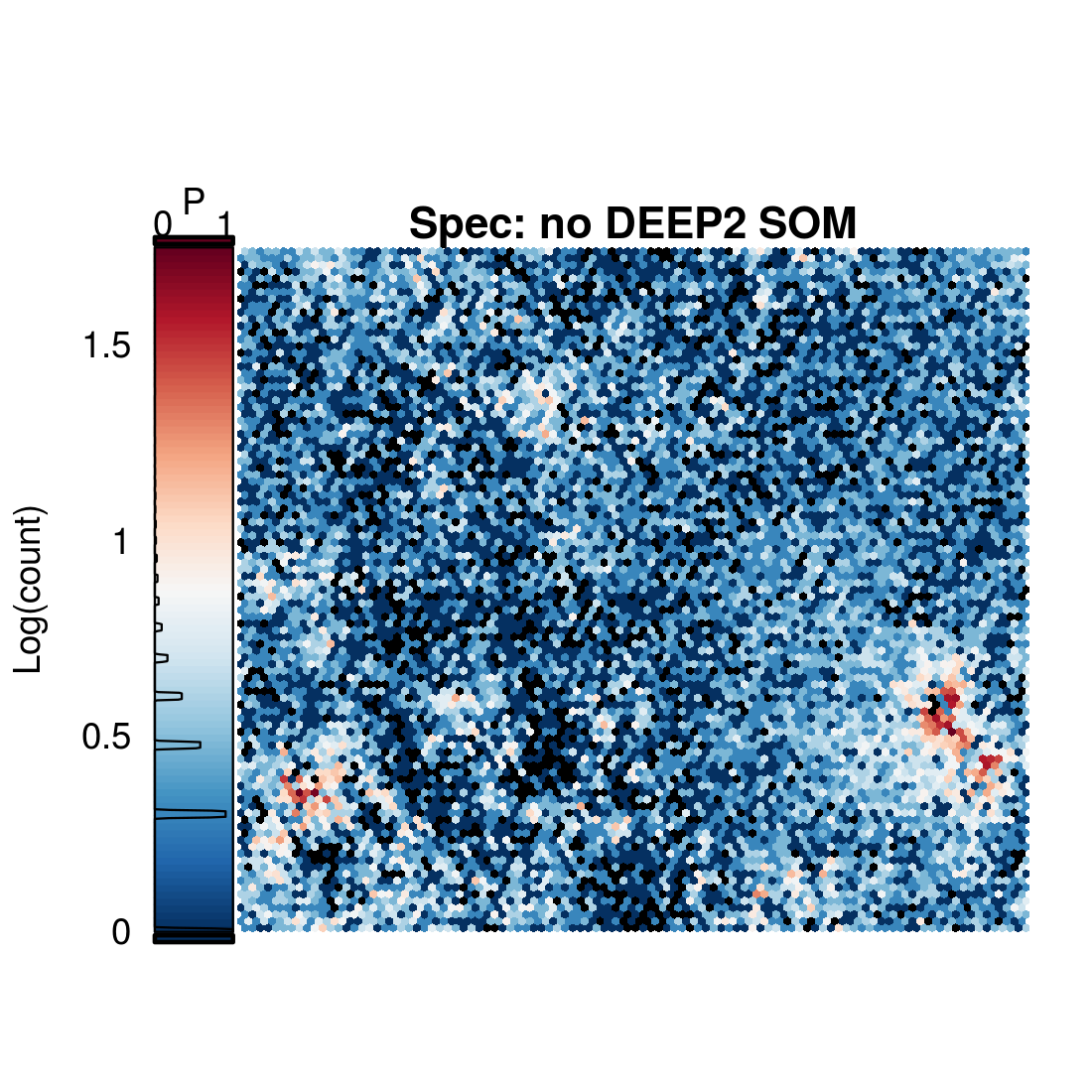}\includegraphics[scale=0.7,trim=0 30 0 50,clip]{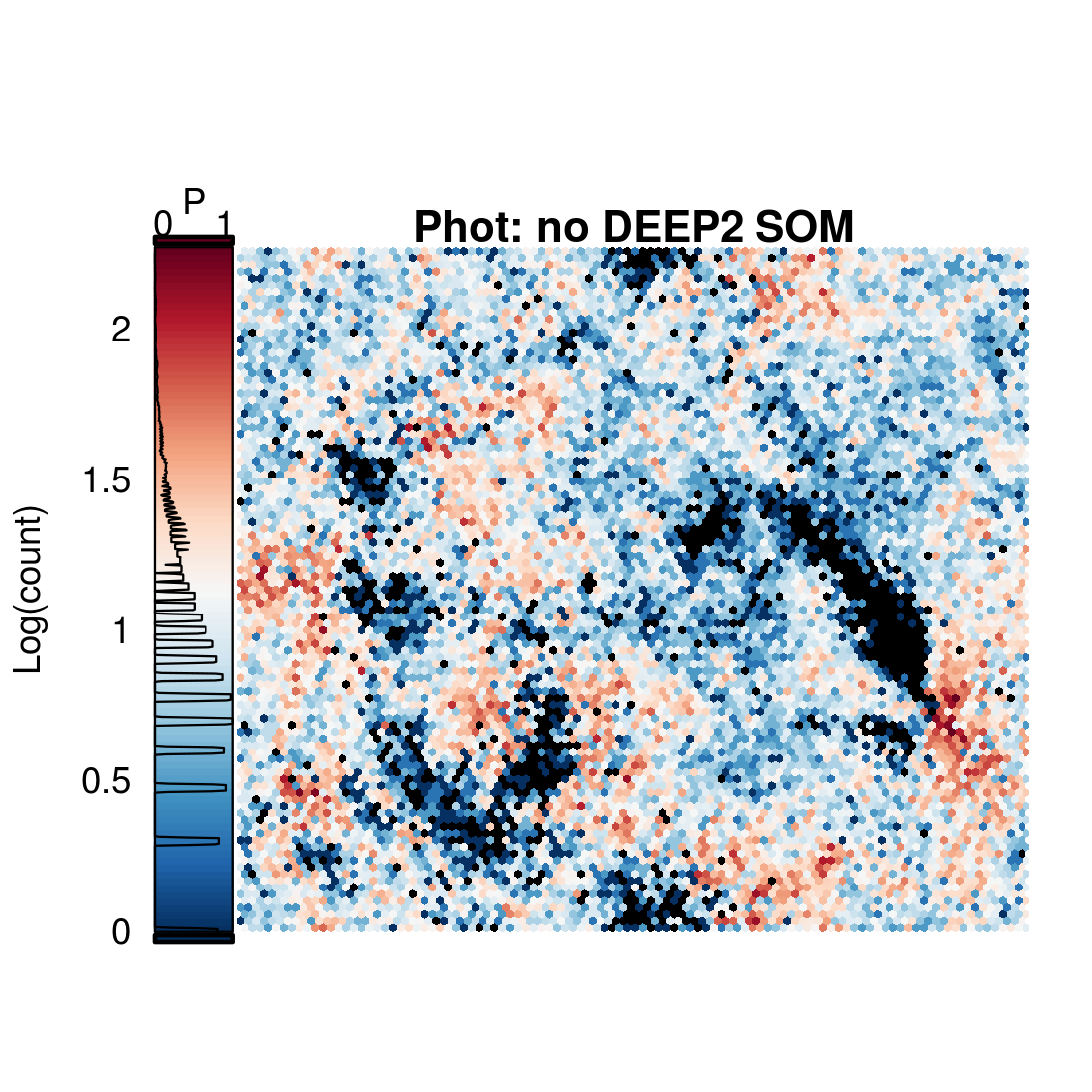}
\caption{The distribution of spectroscopic (left column) and photometric (right column) data, within our SOMs 
trained on the full spectroscopic compilation (top row) and the without-DEEP2 compilation (bottom row). The 
photometric data have similar structures (the missing bright-magnitude wedge can be used as a reference in each 
of these panels). Despite the different distribution of spectral and photometric data in the two SOMs, the
representation statistics remain essentially unchanged: in both cases $\sim 5\%$ of the photometric data 
occupy {cells} not containing spectra.}\label{fig: training test 1}
\end{figure*}

\begin{figure*}
\centering
\includegraphics[scale=0.7,trim=0 30 0 20,clip]{training_test_2.png}\includegraphics[scale=0.7,trim=0 30 0 20,clip]{training_test_3.png}
\includegraphics[scale=0.7,trim=0 30 0 20,clip]{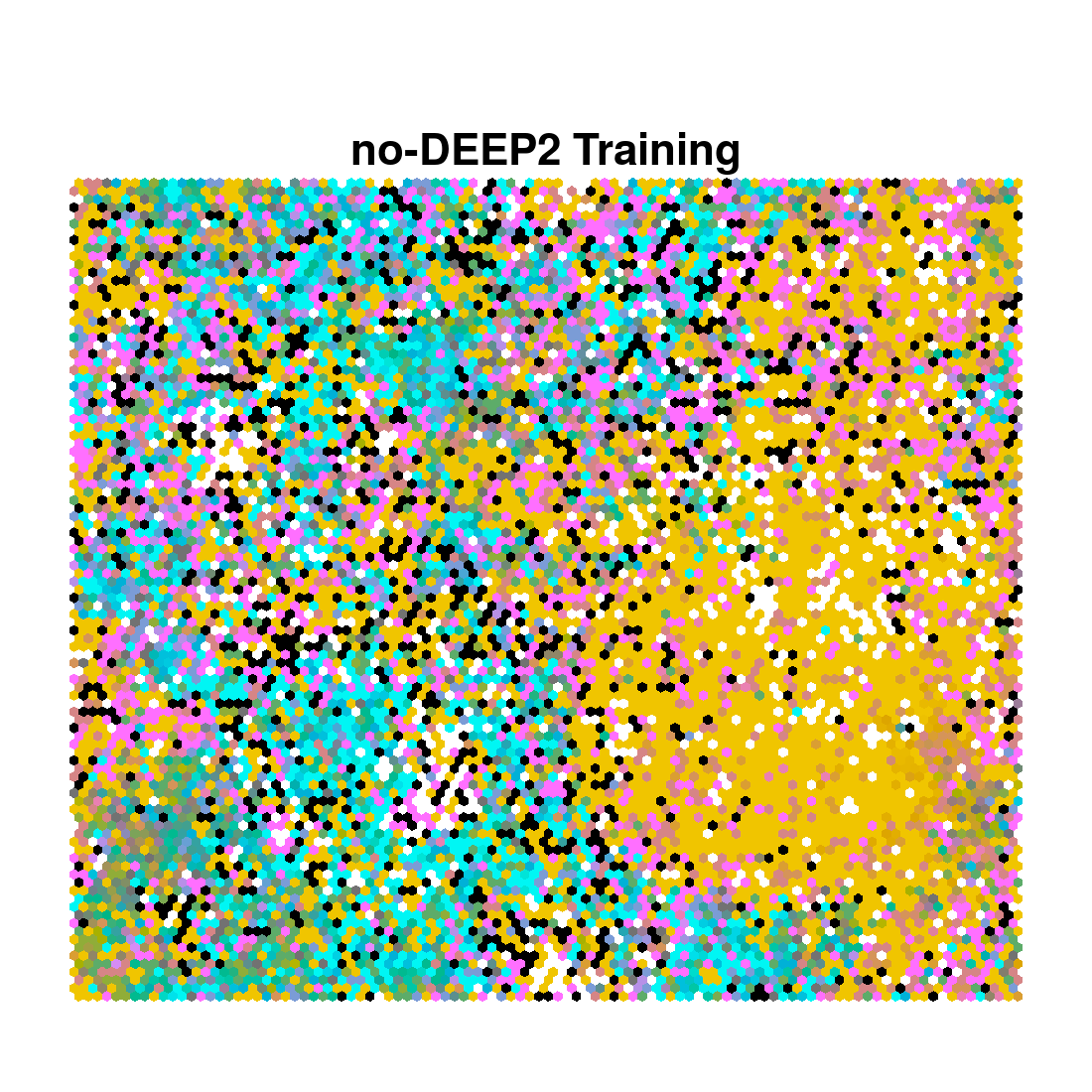}\includegraphics[scale=0.7,trim=0 30 0 20,clip]{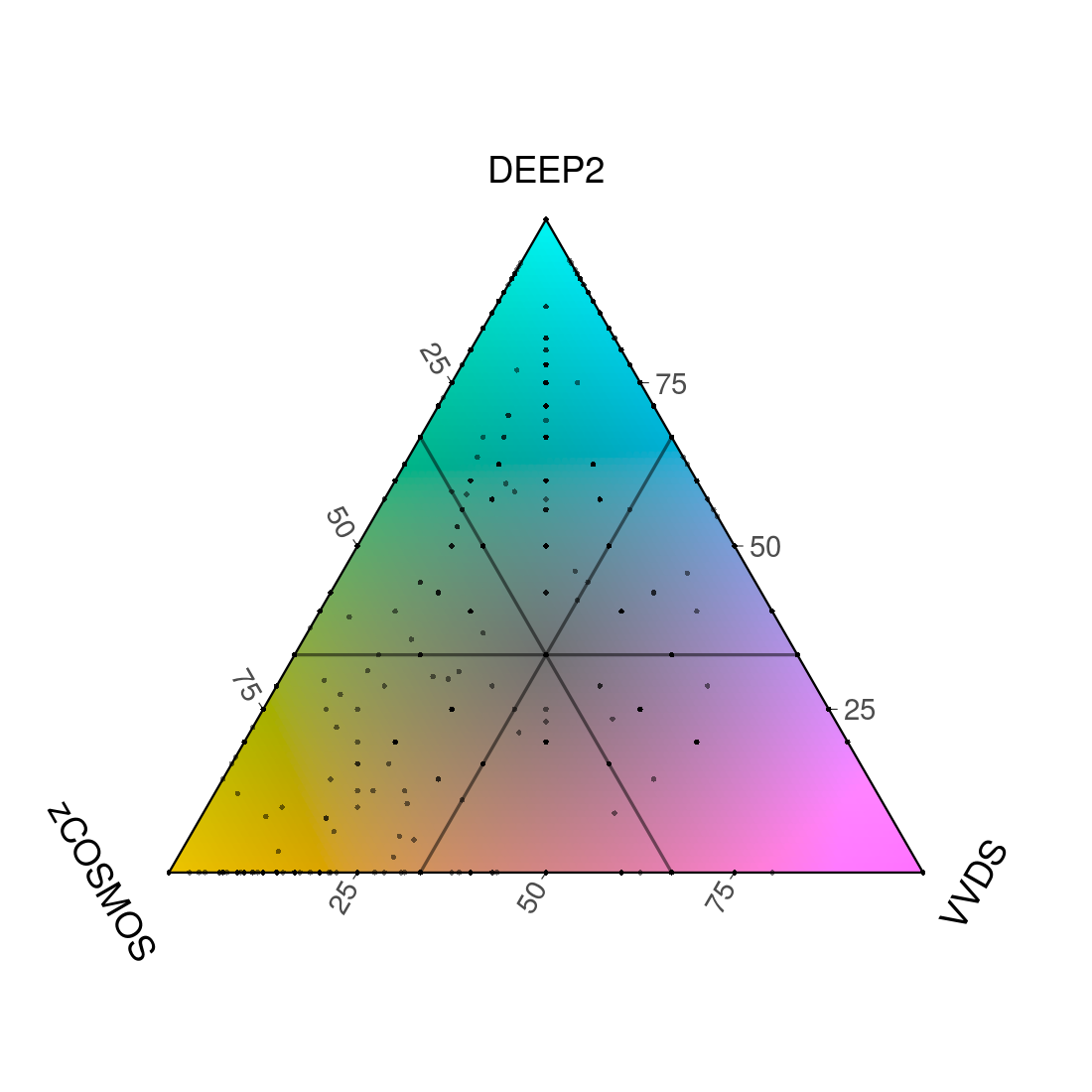}
\caption{The fractional occupation per SOM {cell} of each of the DEEP2, zCOSMOS, and VVDS samples. The upper row  
repeats the fiducial training shown in Fig. \protect \ref{fig: som samples}, which can be compared to the no-DEEP2 training in the
lower row. DEEP2 is assigned far fewer {cells} in the SOM trained on the spectroscopic dataset excluding DEEP2
($\sim11\%$ fewer).}\label{fig: training test 2}
\end{figure*}

Figure \ref{fig: training test 1} shows our two trained SOMs, coloured by the counts of the spectroscopic and
photometric datasets. The figure shows that the two SOMs distribute the photometric and spectroscopic data somewhat 
differently under the two trainings. Using the wedge of empty {cells} in the photometric SOMs for
orientation\footnote{Remember that these SOMs have toroidal geometry. That is, the SOM obeys pacman rules: what goes out
the top comes in the bottom, what goes out the left comes in the right.}
(these are the bright-galaxy {cells} which exist in the spectroscopic compilation but not in the lensing sample), 
we can see that the no-DEEP2 training is dispersing the spectral and photometric data over a greater area within 
the SOM manifold. Put differently, the manifold {cells} have adapted to allocate fewer {cells} to the area of
colour-colour space that houses DEEP2 and is missing from the training set. However, importantly, that space has still
been mapped and allocated space within our SOM. This can be seen in 
Figure \ref{fig: training test 2}, which shows the same two SOMs now coloured by the occupation 
statistics of DEEP2, zCOSMOS, and CDFS (as in Figure \ref{fig: som samples}). The visual effect is clear; there are
far fewer exclusively-DEEP2 (blue) {cells} in the no-DEEP2 trained SOM. Indeed, there is a $\sim 11\%$ decrease in the
total number of {cells} assigned to DEEP2: from $42.8\%$ in the fiducial case to $31.8\%$ in the no-DEEP2 training. 
This decrease in real estate, however, is not accompanied by a significant change in the estimated misrepresentation 
of the photometric data; a change of only $1\%$ is seen. 

So, the training without DEEP2 does cause a significant change in the SOM area allocated to the missing spectroscopic
data. Crucially, however, the spectroscopic and photometric data which belong to DEEP2 are still correctly allocated to
the same {cells}. The SOM has recognised that there is an area of colour-colour space missing, but naturally does not 
assign it the same weight in the final manifold mapping. Therefore, we conclude that, even in the case of
catastrophically missing spectral data, we will not see extensive contamination of the redshift calibration by sources
arbitrarily distributed within the SOM space. 


\section{\Nz\ biasing via sample selection}\label{sec: SOM Bias}
In the SOM photometric redshift calibration method we assume that redshift distributions per SOM cell/cluster are
narrow, and therefore that each cell traces a single population of galaxies in colour-redshift space.  When this
assumption does not hold, the method can become sensitive to differences between population distributions in the
photometric and spectroscopic samples. To combat this effect we implement tomographic binning of both the photometric
and spectroscopic samples, in contrast to previous direct calibration implementations \citep[see for example][]{hildebrandt/etal:2017,hildebrandt/etal:2018}. In this appendix, we demonstrate the importance of ensuring that
photometric and spectroscopic samples have consistent redshift-dependent selection functions applied, and how the
invocation of a SOM based direct redshift calibration {\rm without} consistent selection functions can lead to
considerable bias in the recovered redshift distributions. 

\begin{figure*} \centering
\includegraphics[scale=0.5]{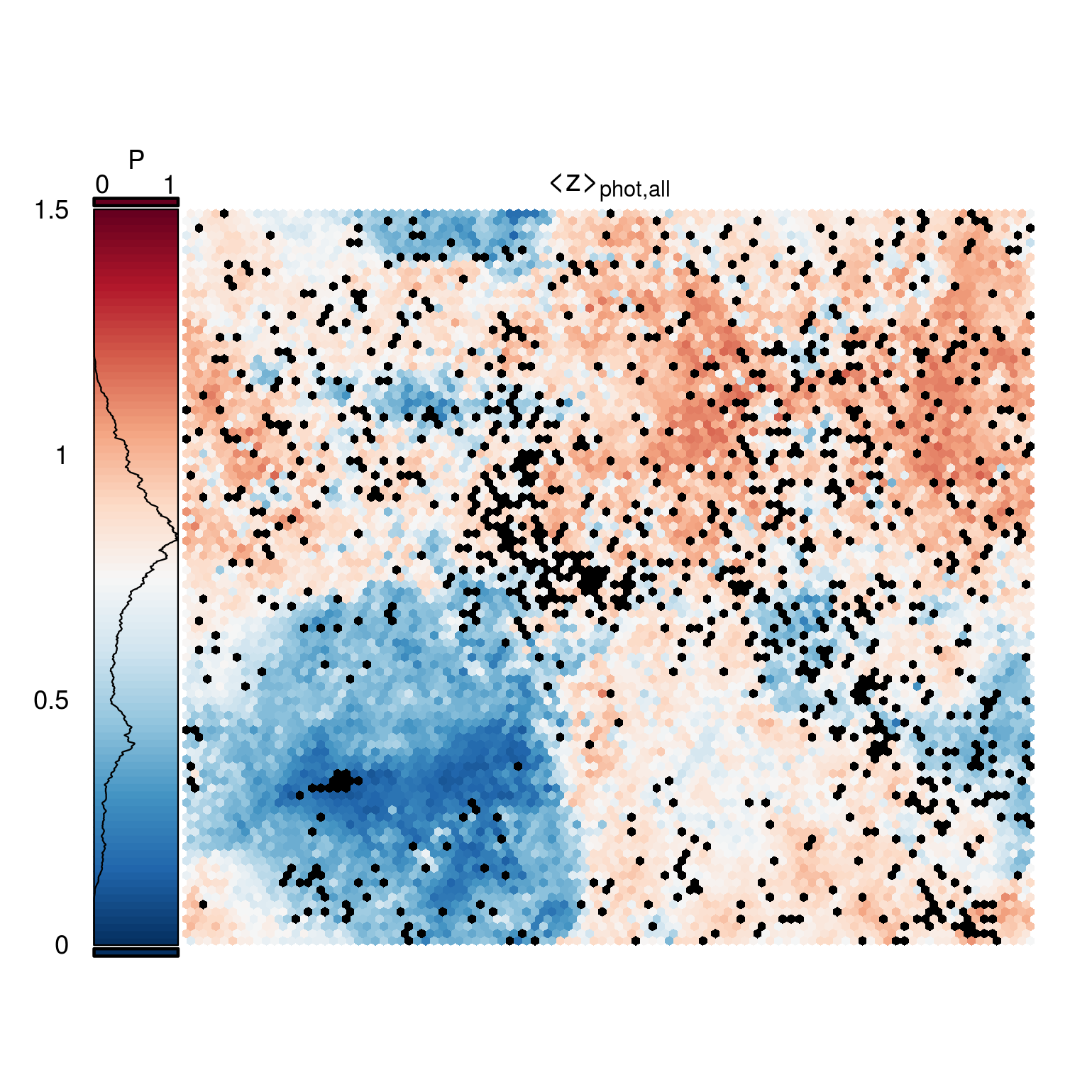}
\includegraphics[scale=0.5]{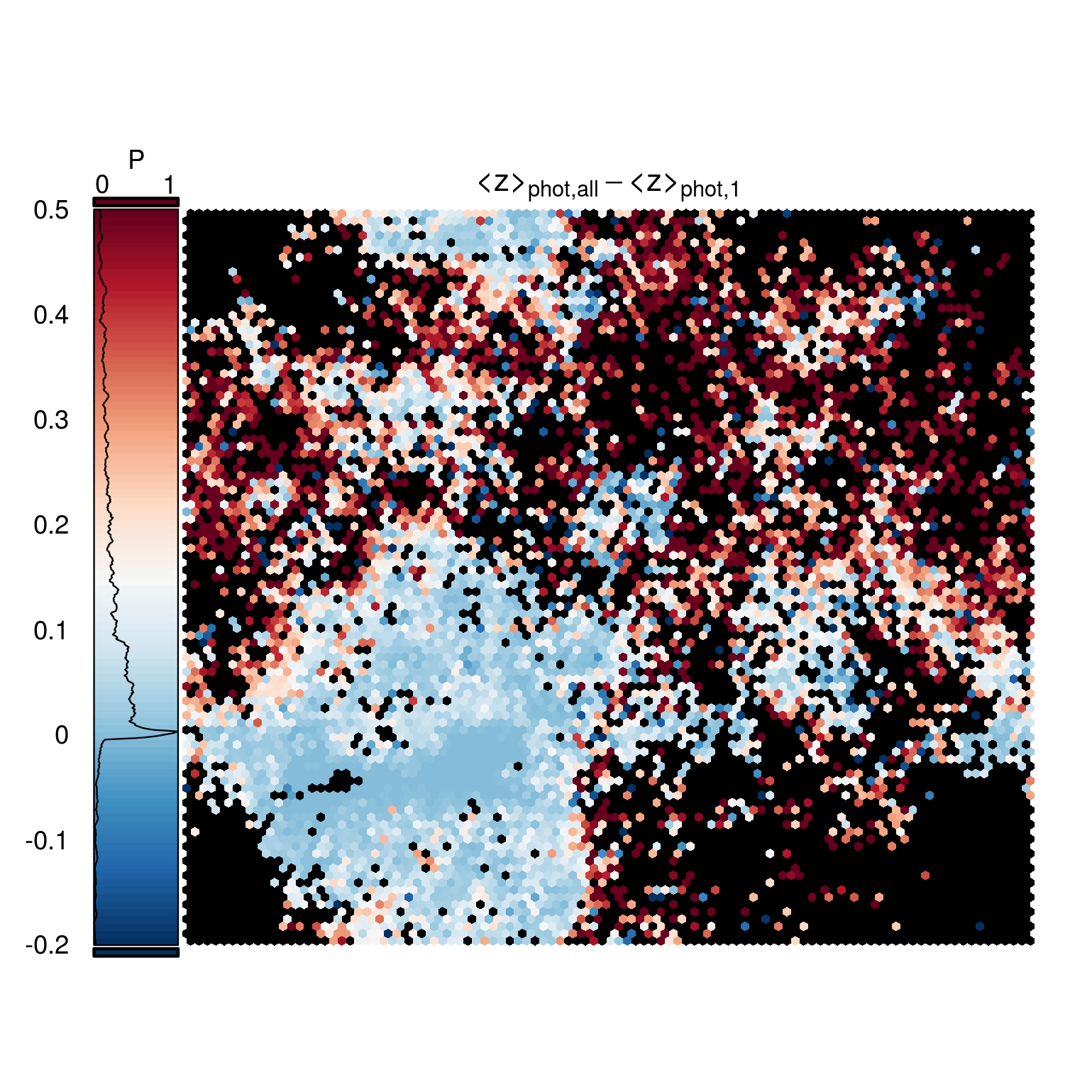}
\caption{{\em Right:} The distribution of mean redshift as a function of SOM cell for the fiducial MICE2 simulation.
{\em Right:} the distribution of the difference in SOM cell mean redshifts for the full photometric sample compared to
the first tomographic bin. The figure demonstrates that the \Nz\ of the high-$z$ SOM cells in the lowest tomographic bin
is systematically different to those cells in the full sample. This effect creates the bias observed in the SOM direct
calibration. }\label{fig: tomo1 dz} \end{figure*}

We first consider a trained SOM which maps the full colour-redshift space of our MICE2 simulations. Figure \ref{fig:
tomo1 dz} shows the distribution of mean (simulated/true) redshifts within individual SOM cells ($\langle z \rangle_{\rm
phot,all}$; left panel). 
The cells map the colour-redshift distribution of the training sample, and we can see from the PDF of the
$\langle z \rangle_{\rm phot,all}$ values (shown within the colour-bar) that the number of cells with a particular mean redshift,
closely traces the redshift distribution of the spectroscopic data (see Fig. \ref{fig: specz dist} for comparison). 
 cells in the SOM generally have a narrow spectroscopic redshift distribution ($\langle\sigma_z\rangle = 0.08$), however
photometric noise and colour redshift degeneracies mean that cells can have low-significance wings in the
redshift distribution. Such wings need not be a significant source of uncertainty/bias in redshift distribution
calibration, provided that the spectroscopic and photometric data are similarly affected. The photometric and
spectroscopic data must share the same noise properties over the manifold, which causes galaxies to scatter around the map in
a consistent manner. 

This requirement can, however, be easily violated. The right panel in Fig. \ref{fig: tomo1 dz} shows the difference
between SOM cell mean redshifts for the full photometric sample ($\langle z \rangle_{\rm phot,all}$, as in the left panel) and the
 cell means of the same data after selecting only sources within the first tomographic bin ($\langle z \rangle_{\rm phot,1}$).
Firstly we can see that the photometric galaxies in the first tomographic bin preferentially trace the low-redshift SOM
 cells; photometric redshift strongly correlates with SOM cell redshift. However due to the considerable size of the
photometric catalogue, there are many nominally high-redshift cells which contain photometric galaxies that are
exclusively at low-redshift, thus creating a considerable population of highly biased cells ($\langle z \rangle_{\rm
phot,all}-\langle z \rangle_{\rm phot,1}>0.1$). It is clear from the figure that the high-$z$
portion of sources in the first tomographic bin has a systematically biased per-cell \Nz\ compared to the overall
sample. This in turn causes the reweighting to over-represent the high-$z$ data in the log-$z$ tomographic bins, leading to
a bias in the estimated redshift distributions.  
 
The conclusion is thus: tomographic binning imposes a strong redshift dependent selection on the photometric catalogue.
This selection modifies the true \Nz\ per SOM cell for the photometric catalogue, thereby introducing a bias in the
redshift calibration. Interestingly this effect cannot be mitigated by pre-selecting cells
based on their spread in spectroscopic (or even simulated) redshifts. The asymmetric sampling of the \Nz\ affects all
 cells to some degree, and down-selecting cells based on their initial spread fails to remove the bias at any
significant level. For this reason, in our analysis we apply tomographic binning to both the spectroscopic and
photometric catalogues.  

The observations here have a more fundamental impact, though, than simply that of dictating the method of implementing
tomographic binning in our SOM. The result acts as a cautionary tale for all works utilising direct comparisons between
SOM-based groupings of training and analysis data. It suggests that {\em any} form of systematic selection which 
differs between the training and analysis groupings used in SOM analysis has the ability to introduce complex biases. 
Should this bias be noise-dependent then the effect increases, as inter-cell scatter also varies between the two
datasets. 

Such an observation may be of importance to other SOM-based clustering studies, such as those of \cite{davidzon/etal:2019}
and \cite{buchs/etal:2019}: should there exist even subtle noise and selection differences between the training and
analysis datasets used in any SOM clustering study, it is possible that non-trivial biases in inference can be
introduced. 


\section{Catastrophic spec-$z$ Failures}\label{sec: catastrophic specz}
{ In \cite{hildebrandt/etal:2018} the authors tested the impact of calibrating KiDS cosmic shear redshift
distributions using the high-fidelity photometric redshift compilation of \citep{lilly/etal:2009}, finding that it
caused a shift in their estimate of ${S}_8=\sigma_8\sqrt{\Omega_{\rm m}/0.3}$. 
It has been hypothesised that this observed shift could be due (at least
in part) to the presence of catastrophically misidentified spectroscopic redshifts in the spectroscopic compilation
used for direct photometric redshift calibration.  
One such form of catastrophic} misidentification was reported recently in \cite{laigle/etal:2019}, where they explored outliers in the 
COSMOS15 \citep{laigle/etal:2016} photo-$z$ vs. spec-$z$ distribution and found $\sim35\%$ of outliers (which equates to roughly 1.3\% of all sources) 
had evidence of contamination within the spectroscopic slit. The implication of this study being that $\sim 35\%$ of outliers in the
COSMOS15 photo-$z$ vs. spec-$z$ distribution are not failures of the photo-$z$, but rather are
caused by failures in the spectroscopy. {Should such a population of failed spectroscopic redshifts also be present in
the spectroscopic compilation used for direct calibration, considerable increases in reconstruction bias could occur.} 

We therefore explore the effect of introducing catastrophic spec-$z$ failures into the spectroscopic compilation. 
To simulate spectroscopic failures {of the kind explored by \cite{laigle/etal:2019}}, we make the conservative assumption that the observed fraction of catastrophic
spectroscopic failures in the COSMOS15 observations of \cite{laigle/etal:2019} apply equally to all samples within our
spectroscopic dataset, even though this is known to be untrue. {DEEP2, for example, is analysed such that any
spectra which show features from multiple redshifts are always given low confidence (and are not included in the
direct calibration compilation). Additionally, some spectra in DEEP2 are observed twice with different slit orientations, allowing
estimation of the fraction of slit-orientation-induced catastrophic redshift failures. Their analysis indicates that
repeated observations produce repeated reliable (nQ$\geq 3$) redshifts $86\%$ of the time, and that these discrepant
$14\%$ are dominated by technical failures or simply low SNR in one of the observations. Crucially, they find that pairs
of high-confidence (expected $95\%$ reliability; nQ$=3$) and certain (expected $99.5\%$ reliability; nQ$=4$) sources
have redshifts that differ by more than $500$km/s only $0.68\%$ and $0.24\%$ of the time, respectively
\citep{newman/etal:2013}. Therefore catastrophic failures of the form seen in \cite{laigle/etal:2019} are, at minimum,
greatly suppressed in DEEP2.} 

To simulate the effect of catastrophic misidentification {of this type}, we first identify the appropriate number of spectra to
contaminate as a function of $i$-band magnitude. We do this by taking the observed photo-$z$ vs. spec-$z$ outlier 
rates, as a function of $i$-band magnitude, from Table $2$ of \cite{laigle/etal:2019}: 
${1.7,6.7,10.2,22.0}\%$ for the magnitude bins $i \in {(-\infty,23),[23,24),[24,25),[25,\infty)}$. 
Again to be conservative, we have allowed the upper and lower magnitude bins to be open-ended; 
the brightest bin extends to all bright magnitudes and the faintest bin extends to all fainter magnitudes.
{ The faint end choice, however, is of little consequence; there are very few spectroscopic sources in our
simulation with $i$-band fluxes fainter than $24^{\rm th}$ magnitude.}

We use the estimated outlier fractions and observed fraction of outlier spec-$z$ failures to generate the relevant 
fraction of catastrophic mis-identifications in our MICE2 mock KiDS spectroscopic dataset. This equates to 
${0.6,2.3,3.6,7.7}\%$ spec-$z$ failures within each of the $i$-band magnitude bins. {Weighting each of these bins by the
number of spectra therein produces an expected catastrophic failure rate for our simulations of $1.07\%$. This is
slightly lower than the same calculated on the real KV450 spectroscopic compilation ($1.27\%$), due to the simulated 
spectroscopic data being slightly shallower than the full spectroscopic dataset. In order to account for this 
difference, we artificially inflate the rate of catastrophic failures across the full magnitude range by $25\%$, to 
$43.7\%$ of observed outliers. This produces a simulated catastrophic outlier fraction of $1.33\%$, which we use for the 
following tests.} We simulate the production of catastrophic failures in 3 ways. 

1. We assume that the misidentified spec-$z$ must
come from a source with {an observed $i$-band magnitude} at least as bright as the source that it is replacing.
{ This failure mode implies that the brightest of two sources within the slit will be the one for which a redshift
is reported, and produces catastrophic failures that preferentially move to lower redshifts than the original source. }

2. We allow the catastrophic
redshift to be sampled arbitrarily from the distribution of all {spectra. This mode of failure replicates the
circumstance where the observed (secure) redshift distribution of a spectroscopic survey is determined exclusively by
the true redshift distribution of the galaxies within the survey volume and the secure-redshift success function
(dictated predominantly by the telescope and spectrograph design: wavelength range, resolution, spectrum SNR, etc); i.e.
that there is no significant redshift-dependant spectroscopic pre-selection. If this is the case, catastrophic failure
can only occur when the contaminating galaxy has a true redshift which lies within the secure-redshift success window.
Therefore, catastrophic failure in this mode can be simply drawn from the observed distribution of galaxies with
successful redshift estimates.  This mode of failure will be increasingly unrealistic with increasingly heavy
spectroscopic colour-preselection, however. 

3. We allow catastrophic failures to be sampled uniformly in the
simulation redshift range.} For each of these three cases we generate such a contaminated sample 100 times, and
calculate the resulting tomographic redshift distributions each time. 

The results of our contaminations on each of the tomographic bins is shown in  Table \ref{tab: catastrophic}. {We
find that the maximally induced bias is at the $\Delta \langle z \rangle < 0.005$ level}, and so is of little concern to
our analysis\footnote{Recall that this estimate invokes multiple maximally-biasing assumptions and in reality the
influence of such contaminants will likely be smaller than presented here.} This is true for all three forms of
catastrophic failure in our MICE2 simulations. The maximal bias is also seen always in the highest tomographic bin,
which is dominated by the DEEP2 dataset and therefore will have significantly better contamination properties than is
assumed here. We therefore conclude that the catastrophic misidentification of spec-$z$, even for a greater fraction of
spectra than is known to be possibly affected, cannot cause significant systematic biases in the estimated SOM redshift
distributions for KiDS-like surveys. 

\begin{table*}
\centering
\caption{The effect of catastrophic spec-$z$ misidentification on the estimated SOM tomographic bin mean 
redshifts. The simulations here all assume a catastrophic  
failure fraction of $35\%$. {Combining this rate with the observed outlier rate in COSMOS and the spectroscopic
number counts in KiDS gives the expected failure rate: $1.30\%$. The simulation is slightly shallower than the KiDS
data, giving an outlier rate of $1.03\%$. We simulate our outliers with this rate (Fiducial) and with two artificially
inflated rates: increasing the instance of failures over all magnitudes, and increasing failures at faint magnitudes. 
The three rows in each set of rates show the results when forcing failures to be
brighter than the target galaxy ($z\sim {\rm N}(z|i\leq i_0)$; top), arbitrarily sampled from the full spec-$z$ distribution 
($z\sim {\rm N}(z)$; middle), or uniformly
sampled from the redshift range ($z\sim {\rm U}(z)$; bottom). }
In all cases the results show shifts of less than $0.005$ in the tomographic bin means. Given the magnitude of these biases 
is small, we conclude that catastrophic spectral mis-identifications cannot cause significant biasing of the SOM tomographic redshift 
distributions for KiDS-like analyses. }\label{tab: catastrophic}
\begin{tabular}{cc|ccccc}
\hline
\hline
\multicolumn{2}{c}{Catastrophic} & \multicolumn{5}{c}{$\Delta \langle z \rangle$ $(\langle z \rangle_{\rm 0}-\langle z\rangle_{\rm c})$} \\
Rate         & Type & Bin 1 & Bin 2 & Bin 3 & Bin 4 & Bin 5   \\
\hline
          &$z\sim {\rm N}(z|i\leq i_0)$&$ 0.0023 \pm 0.0011 $ & $ 0.0018 \pm 0.0008 $ & $ 0.0008 \pm 0.0007 $ & $ -0.0010 \pm 0.0006 $ & $ -0.0031 \pm 0.0006 $ \\
 Fiducial &$z\sim {\rm N}(z)$          &$ 0.0029 \pm 0.0010 $ & $ 0.0022 \pm 0.0007 $ & $ 0.0011 \pm 0.0007 $ & $ -0.0007 \pm 0.0006 $ & $ -0.0028 \pm 0.0006 $ \\
          &$z\sim {\rm U}(z)$          &$ 0.0029 \pm 0.0013 $ & $ 0.0020 \pm 0.0009 $ & $ 0.0011 \pm 0.0009 $ & $ -0.0009 \pm 0.0008 $ & $ -0.0030 \pm 0.0009 $ \\
\hline
 Higher   &$z\sim {\rm N}(z|i\leq i_0)$&$ 0.0029 \pm 0.0011 $ & $ 0.0021 \pm 0.0009 $ & $ 0.0011 \pm 0.0008 $ & $ -0.0013 \pm 0.0006 $ & $ -0.0037 \pm 0.0008 $ \\
 overall  &$z\sim {\rm N}(z)$          &$ 0.0036 \pm 0.0011 $ & $ 0.0027 \pm 0.0010 $ & $ 0.0015 \pm 0.0009 $ & $ -0.0010 \pm 0.0006 $ & $ -0.0034 \pm 0.0008 $ \\
 contam   &$z\sim {\rm U}(z)$          &$ 0.0034 \pm 0.0013 $ & $ 0.0025 \pm 0.0011 $ & $ 0.0014 \pm 0.0011 $ & $ -0.0010 \pm 0.0007 $ & $ -0.0038 \pm 0.0010 $ \\
\hline
 Higher   &$z\sim {\rm N}(z|i\leq i_0)$&$ 0.0023 \pm 0.0010 $ & $ 0.0021 \pm 0.0008 $ & $ 0.0011 \pm 0.0009 $ & $ -0.0014 \pm 0.0006 $ & $ -0.0042 \pm 0.0008 $ \\
 faint    &$z\sim {\rm N}(z)$          &$ 0.0029 \pm 0.0010 $ & $ 0.0025 \pm 0.0010 $ & $ 0.0013 \pm 0.0010 $ & $ -0.0008 \pm 0.0005 $ & $ -0.0037 \pm 0.0008 $ \\
 contam   &$z\sim {\rm U}(z)$          &$ 0.0028 \pm 0.0012 $ & $ 0.0023 \pm 0.0011 $ & $ 0.0012 \pm 0.0012 $ & $ -0.0012 \pm 0.0009 $ & $ -0.0042 \pm 0.0010 $ \\
\hline
\hline
\end{tabular}
\end{table*}


\end{document}